\documentclass[11pt]{article}
\usepackage[utf8]{inputenc}
\usepackage{caption}
\usepackage{braket,slashed,bm}
\usepackage{jheppub}
\usepackage{simplewick}
\usepackage{array,multirow}
\usepackage{amsmath} \DeclareMathAlphabet{\mathpzc}{OT1}{pzc}{m}{it}
\usepackage[normalem]{ulem}
\usepackage{calligra}
\usepackage{floatrow}
\DeclareMathAlphabet{\mathpzc}{OT1}{pzc}{m}{it}
\usepackage{mathrsfs}
\usepackage{tikz}

\usepackage{subcaption}
\usepackage{lscape}
\usepackage{hyperref}
\usepackage{graphicx}
\usepackage{booktabs}
\newfloatcommand{capbtabbox}{table}[][\FBwidth]

\graphicspath{{./Figures/}}

\def\beq{\begin{equation}}
\def\eeq{\end{equation}}
\def\bea{\begin{eqnarray}}
\def\eea{\end{eqnarray}}
\def\nn{\nonumber \\}
\def\hyp{\mathsf{y}}

\newcommand{\Lagr}{\mathcal{L}}

\begin{document}
\title{EWPD in the SMEFT to dimension eight}

\author[a]{Tyler Corbett,}

\author[b]{Andreas Helset,}

\author[c]{Adam Martin,}

\author[a]{and Michael Trott}

\affiliation[a]{Niels Bohr Institute, University of Copenhagen,
Blegdamsvej 17, DK-2100, Copenhagen, Denmark}

\affiliation[b]{Walter Burke Institute for Theoretical Physics, California Institute of Technology,
Pasadena, CA 91125, USA}

\affiliation[c]{Department of Physics, University of Notre Dame, Notre Dame, IN, 46556, USA}

\abstract{We calculate the ${\cal{O}}(\langle H^\dagger H \rangle^2/\Lambda^4)$ corrections to LEP electroweak precision data
using the geometric formulation of the Standard Model Effective Field Theory (SMEFT).
We report our results in simple-to-use interpolation tables that allow the
interpretation of this data set to dimension eight for the first time.
We demonstrate the impact of these previously unknown terms in the case of a general analysis in the SMEFT,
and also in the cases of two distinct models matched to dimension eight.
Neglecting such dimension-eight corrections to LEP observables introduces a theoretical error in SMEFT studies.
We report some preliminary studies defining such a theory error,
explicitly demonstrating the effect of previously unknown
dimension-eight SMEFT corrections on LEP observables.}

\subheader{\normalsize
        \vspace*{-3.7em}
        \begin{flushright}
	CALT-TH/2021-007
        \end{flushright}
}

\maketitle
\section{Introduction}
Interpreting current experimental results while allowing for the Standard Model (SM)
to break down at higher energies in future experimental studies is a key task in particle physics. This can be done in a way that is agnostic about
new physics at higher energies by using an effective field theory (EFT). In this approach we are Taylor expanding
in the low-energy measurement scale(s) divided by the scale of some new physics effects. This defines the ``power counting" of the EFT.
When combined with the assumed symmetries, the low-energy field content,
and the representations of the fields under these symmetries,
this defines an EFT.
The power of EFT studies of data sets
resides in the fact that such an approach is systematically improvable with quantum loop corrections,
and corrections that are higher order in the power counting without knowledge of the UV completion of the EFT.

When the particle spectrum includes an $\rm SU(2)_L$ scalar doublet ($H$), and the mass scale of heavy new physics is parametrically
separated from the electroweak scale, the Standard Model
Effective Field Theory (SMEFT) is the appropriate EFT for data
with a measurement scale proximate to the electroweak scale.\footnote{For a review on EFT and the SMEFT, see Ref.~\cite{Brivio:2017vri}.
Subtleties in mixing of heavy and light states can potentially lead to the HEFT \cite{Feruglio:1992wf,Grinstein:2007iv,Barbieri:2007bh,Buchalla:2012qq,Alonso:2012px}.
These subtleties do not change our conclusions; see Refs.~\cite{Corbett:2015lfa,Buchalla:2016bse,Brivio:2017vri,Passarino:2019yjx,Chang:2019vez,Cohen:2020xca} for related scientific discussions.}

The Large Electron Positron (LEP) collider
provided a series of precise measurements on the properties of the SM states interacting at energies proximate to the $\mathcal{Z}$ mass
\cite{ALEPH:2005ab}.
Over a decade after the conclusion of the LEP experimental program,
no consistent and complete analysis of this data in terms of the SMEFT, extended to sub-leading order in the power counting,
had been developed
until the geometric formulation of the SMEFT defined the relevant formalism not only to sub-leading order,
but to all orders in the expansion in $\sqrt{\langle H^\dagger H \rangle}/\Lambda$ in
Refs.~\cite{Helset:2018fgq,Corbett:2019cwl,Helset:2020yio,Hays:2020scx}.\footnote{Here $\sqrt{\langle H^\dagger H \rangle} \equiv v_T$ is defined to be the vacuum expection
	value (vev) of the Higgs field in the SMEFT. In this paper we generally do not
draw a distinction between the notation $\bar{v}_T$ and $v$, using the latter for notational brevity at times.
An exception to this rule, where the distinction between the SM classical vev $v$ and the minimum of the potential in the SMEFT
$\bar{v}_T$ is important, is discussed in Section \ref{matchingsection}.}
To develop this approach the SMEFT was reformulated geometrically
as the Higgs field space geometry plays a key role in this EFT
\cite{Burgess:2010zq,Alonso:2015fsp,Alonso:2016btr,Alonso:2016oah}. Equally important was the
development and use of Hilbert series techniques in Refs.~\cite{Lehman:2015via,Lehman:2015coa,Henning:2015daa,Henning:2015alf}.

In this paper, we present a complete set of explicit results that allow the study of LEP Electroweak Precision Data (EWPD) constraints
to dimension eight for the first time.
We also study the effect of previously unknown and neglected dimension-eight
corrections when interpreting LEP data.\footnote{The results in this work extend previous
results \cite{Grinstein:1991cd,Han:2004az,Alonso:2013hga,Corbett:2017qgl,Brivio:2017bnu} in a consistent Effective Field Theory
extension of the SM (nowadays called the SMEFT). See also the related
Refs.~\cite{Falkowski:2014tna,Efrati:2015eaa,Falkowski:2019hvp}. We also note that the first work to stress
the need to characterise a theory error due to the neglect of dimension-eight operators
when interpreting EWPD is Ref.~\cite{Berthier:2015oma}.
This point has also been stressed in several recent studies on other observables,
see Refs.~\cite{Alte:2017pme,Keilmann:2019cbp}.}

\section{SMEFT and geoSMEFT}\label{setup}

The SMEFT Lagrangian is
\begin{align}
	\Lagr_{\textrm{SMEFT}} &= \Lagr_{\textrm{SM}} + \Lagr^{(d)}, &   \Lagr^{(d)} &= \sum_i \frac{C_i^{(d)}}{\Lambda^{d-4}}\mathcal{Q}_i^{(d)}
	\quad \textrm{ for } d>4.
\end{align}
The higher-dimensional operators $\mathcal{Q}_i^{(d)}$ are constructed out of the SM fields.
The particle spectrum includes an $\rm SU(2)_L$ scalar doublet ($H$) with hypercharge
$\hyp_h = 1/2$. The operators $\mathcal{Q}_i^{(d)}$ are labelled with a mass dimension $d$ superscript
and multiply unknown Wilson coefficients $C_i^{(d)}$.
We define $\tilde{C}^{(d)}_i \equiv C^{(d)}_i \bar{v}_T^{d-4}/\Lambda^{d-4}$ and use the Warsaw basis~\cite{Grzadkowski:2010es} for
$\mathcal{L}^{(6)}$ and Refs.~\cite{Helset:2020yio,Hays:2018zze} for $\mathcal{L}^{(8)}$ results.
Our remaining notation is defined in Refs.~\cite{Alonso:2013hga,Brivio:2017vri}.

The geometric formulation of the SMEFT (geoSMEFT~\cite{Helset:2018fgq,Corbett:2019cwl,Helset:2020yio,Hays:2020scx})
organizes the theory in terms of field-space connections.
This approach builds on the results reported in Refs.~\cite{Vilkovisky:1984st,Burgess:2010zq,Alonso:2015fsp,Alonso:2016btr,Alonso:2016oah,Misiak:2018gvl}.
Using this formulation, the SMEFT was consistently formulated at all orders in the expansion in ${\cal{O}}(v/\Lambda)$ for
two- and three-point functions. In particular, the theory was consistently formulated for these $n$-point functions to ${\cal{O}}(v^4/\Lambda^4)$
in Ref.~\cite{Hays:2020scx}, including input parameter shifts. This is sufficient to examine the effect of heretofore
unknown dimension-eight corrections on EWPD observables.
\begin{center}
\begin{table}[t!]
\centering
\tabcolsep 8pt
\begin{tabular}{|>{$}c<{$}c|*3{r@{ $\pm$ }l|}}
\toprule
\multicolumn{2}{|c|}{Observable} & \multicolumn{2}{c|}{$\{\hat{\alpha}, \hat{m}_Z,\hat{G}_F\}$ inputs} & \multicolumn{2}{c|}{$\{\hat{m}_W, \hat{m}_Z,\hat{G}_F\}$ inputs} & \multicolumn{2}{c|}{Exp. result \cite{ALEPH:2005ab}}
\\ \midrule
\Gamma_{e,\mu}& [\rm MeV] & 83.978 & 0.013  & 84.003 & 0.018 & 83.92 & 0.12 \\
\Gamma_{\tau}& [\rm MeV] & 83.788 & 0.013  & 83.813 & 0.018 & 84.08 & 0.22 \\
\Gamma_{\nu}&  [\rm MeV] & 167.166& 0.015  & 167.168 & 0.015 & 166.333 & 0.5 \\
\Gamma_{u}& [\rm MeV]& 299.91 & 0.16 & 300.18 & 0.20  & \multicolumn{2}{c|}{-}\\
\Gamma_{c}&  [\rm MeV] & 299.84 & 0.16  & 300.10 & 0.20 & 300.5 & 5.3 \\
\Gamma_{d,s}&  [\rm MeV] & 382.77 & 0.13  & 383.01 & 0.17 & \multicolumn{2}{c|}{-}\\
\Gamma_{b}&  [\rm MeV] & 375.88 & 0.13  & 376.12 & 0.17 & 377.6 & 1.3 \\
\Gamma_{Z} & [\rm MeV] & 2494.4 & 0.7  & 2495.7 & 1.0 & 2495.2 & 2.3 \\
R_{\ell} && 20.749 & 0.007  & 20.758 & 0.008 & 20.767 & 0.025 \\
R_{c} &&  0.17221 & 0.00002  & 0.17223 & 0.00003 & 0.1721 & 0.003 \\
R_{b} &&  0.21588 & 0.00003  & 0.21586 & 0.00003 & 0.21619 & 0.00066 \\
A_{FB}^{\ell} &&
0.01632  & 0.00022  & 0.01718 & 0.00037 & 0.0171 & 0.0010 \\
A_{FB}^{c} &&
0.07370 & 0.00070  & 0.07583 & 0.00117 & 0.0707 & 0.0035 \\
A_{FB}^{b} &&  0.10341  & 0.00097 & 0.10615 & 0.00162 & 0.0992 & 0.0016  \\
\sigma_{\rm Had}^0&[\rm pb]&41,491&5&41,489&5& 41,488 &6\\
\bottomrule
\end{tabular}
\caption{Predictions for LEPI observables in the two input parameter schemes. The $\{\hat{m}_W, \hat{m}_Z,\hat{G}_F\}$ scheme
results are derived using \cite{Awramik:2006uz,Dubovyk:2019szj,Freitas:2014hra,Awramik:2003rn}.
In particular $A_{FB}^{c}$ is derived using Ref.~\cite{Awramik:2006uz}. We have compared the results
for $A_{FB}^{\ell}$ using Refs.~\cite{Awramik:2006uz,Dubovyk:2019szj} and the results agree within
quoted errors.
\label{schemeresults}}
\end{table}
\end{center}
\section{EWPD observables}\label{observables}
We seek to interpret the results in Table \ref{schemeresults} in the SMEFT consistently to ${\cal{O}}(v^4/\Lambda^4)$.
For the predictions of these measurements, we need numerical values of Lagrangian parameters. These are defined
in an input parameter scheme. Such a scheme is a free choice, and two choices are in common use in the literature.
These are the $\{\hat{m}_W, \hat{m}_Z,\hat{G}_F\}$ and $\{\hat{\alpha}, \hat{m}_Z,\hat{G}_F\}$ schemes.
We report results in both of these schemes, and use the numerical input parameter results in
Table \ref{tab:inputs} to fix values of Lagrangian parameters to this end.
For the SM results for the observables in each scheme, we update the theoretical predictions.
For partial widths and ratios of partial widths we update the results beyond those quoted in Ref.~\cite{Brivio:2017bnu}.
These numerical values were determined using the interpolation formula in Refs.~\cite{Freitas:2014hra,Awramik:2003rn}.
In addition, we use the expansion formula in Refs.~\cite{Awramik:2006uz,Dubovyk:2019szj} to determine up-to-date numerical values of
the $A_{FB}^i$ pseudo-observables in the $\{\hat{m}_W, \hat{m}_Z,\hat{G}_F\}$ scheme, leading to Table \ref{schemeresults}.\footnote{We once again
thank A. Freitas for helpful comments and advice on using the results of Refs.~\cite{Freitas:2014hra,Awramik:2003rn,Awramik:2006uz,Dubovyk:2019szj}.}

We present our results for SMEFT corrections normalized to SM predictions.
Our results can be modified to take into account new SM predictions
by multiplying by the ratio of the SM prediction in Table \ref{schemeresults} divided by the new SM prediction.
\begin{center}
\begin{table}
\centering
\tabcolsep 8pt
\begin{tabular}{|c|c|c|}
\hline
Input parameters&Value&Ref.\\
\hline
$\hat m_Z$ [GeV]&$91.1876\pm0.0021$&\cite{Zyla:2020zbs}\\
$\hat m_W$ [GeV]&$80.387\pm0.016$&\cite{Aaltonen:2013iut}\\
$\hat m_h$ [GeV] &$125.10\pm0.14$&\cite{Zyla:2020zbs}\\
$\hat m_t$ [GeV] &$172.4\pm0.7$&\cite{Zyla:2020zbs}\\
$\hat{m}_b$ [GeV]&     $4.18\pm0.03$ & \cite{Olive:2016xmw}\\
$\hat{m}_c$ [GeV]&     $1.27\pm0.02$   & \cite{Olive:2016xmw}\\
$\hat{m}_\tau$ [GeV]&  $1.77686\pm0.00012$   & \cite{Olive:2016xmw}\\
$\hat{G}_F$ [GeV$^{-2}$] & 1.1663787 $\cdot 10^{-5}$&  \cite{Olive:2016xmw,Mohr:2012tt} \\
$\hat \alpha_{EW}$&1/137.03599084(21)&\cite{Zyla:2020zbs}\\
$\Delta\alpha$&$0.0590\pm0.0005$&\cite{Dubovyk:2019szj}\\
$\hat \alpha_s$&$0.1179\pm0.0010$&\cite{Zyla:2020zbs}\\
\hline
$m_W^{\hat\alpha}$ [GeV]&$80.36\pm0.01$&--\\
$\Delta\alpha^{\hat m_W}$&$0.0576\pm0.0008$&--\\
\hline
\end{tabular}
\caption{Input parameter values used to predict EWPD theory predictions for both schemes.
$m_W^{\hat\alpha}$ is the value of $m_W$ inferred in the $\{\hat\alpha,\hat m_Z,\hat G_F\}$ scheme
using the interpolation formula of Refs.~\cite{Freitas:2014hra,Awramik:2003rn,Awramik:2006uz,Dubovyk:2019szj},
which includes SM loop corrections, while $\Delta\alpha^{\hat m_W}$ is the shift in the
value of alpha due to hadronic effects for the $\{\hat m_W,\hat m_Z,\hat G_F\}$ scheme. For an introductory
discussion on the use of $\Delta\alpha$ relating low scale measurements of $\hat\alpha$
and higher scale values above the hadronic resonance region
see Ref.~\cite{Wells:2005vk}. Note that we use a tree level value of $m_W$, not $m_W^{\hat\alpha}$,
in calculating the numerical coefficients for the shifts due to the SMEFT in EWPD.}
\label{tab:inputs}
\end{table}
\end{center}
The observables are
\begin{align}
\bar{\Gamma}_i &= \frac{\hat{m}_Z \, N^i_c}{24 \, \pi} \left(|g_{eff}^{\mathcal{Z},i_L}|^2 + |g_{eff}^{\mathcal{Z},i_R}|^2 \right)(1 - \frac{4\hat{m}_i^2}{\hat{m}_Z^2})^{3/2},& \quad
\bar{\Gamma}_{had} &= \bar{\Gamma}_{u} + \bar{\Gamma}_{d} +\bar{\Gamma}_{c} + \bar{\Gamma}_{s} + \bar{\Gamma}_{b}, \\
\bar{R}_{c,b} &= \frac{\bar{\Gamma}_{c,b}}{\bar{\Gamma}_{had}}, & \quad
\bar{R}_{\ell} &= \frac{\bar{\Gamma}_{had}}{\bar{\Gamma}_{\ell}}, \\
\bar{\sigma}_{had}^{0} &= \frac{12 \, \pi}{\hat{m}_Z^2} \, \frac{\bar{\Gamma}_e \, \bar{\Gamma}_{had}}{\bar{\Gamma}_Z^2},
& \quad \bar{A}_{FB}^{0,f} &= \frac{3}{4} \, \bar{A}_\ell \, \bar{A}_f,
\end{align}
where
\bea
\bar{A}_i= \frac{(g_{eff}^{\mathcal{Z},i_L}-g_{eff}^{\mathcal{Z},i_R})(g_{eff}^{\mathcal{Z},i_L}+g_{eff}^{\mathcal{Z},i_R})}{(g_{eff}^{\mathcal{Z},i_L})^2+(g_{eff}^{\mathcal{Z},i_R})^2}.
\eea
The bar notation indicates theoretical predictions in a canonically normalized SMEFT
to mass dimension $d$. A hat indicates an experimentally measured quantity,
or a numerically defined quantity using measured input parameters.
$g_{eff}^{\mathcal{Z},i_L}$ are defined in Eq.~\eqref{effcouplings}.

\section{EWPD results}
\label{ewpdresults}
The results of Refs.~\cite{Helset:2020yio,Hays:2020scx}
allow EWPD to be studied to ${\cal{O}}(v^4/\Lambda^4)$ in the SMEFT.
We neglect in our results corrections further suppressed by SM masses, and proportional
to the small decay
width of the $\mathcal{Z}$ compared to its mass, as LEP data is strongly peaked at $p^2 \sim m^2_\mathcal{Z}$. We also neglect self-interference effects in the decay due to dipole operators
squared. Both of these corrections in the SMEFT are calculable and neglected here largely for brevity
of presentation. These extra effects only further support our main point,
calling for a cautious interpretation of LEP constraints in the SMEFT.\footnote{For high enough $\Lambda$ these neglected effects, along with loop corrections, might be on the same order or larger than the dimension-eight corrections. In that case a more comperehensive analysis of EWPD is called for.}

In both input parameter schemes,
$\{\hat{m}_Z,\hat{G}_F\}$ are used to fix the dimensions, so the observables are defined in terms of these dimensionful parameters,
with shifts due to corrections to the input observables conventionally included in the shifts
to the effective $\mathcal{Z}$ couplings $g_{\rm eff,pr}^{\mathcal{Z},\psi}$.
Here $p,r$ are flavor labels.
For the observables we study, the energy scale is fixed to be $p^2 \simeq \hat{m}_{\mathcal{Z}}^2$
and the SMEFT corrections scale as ${\cal{O}}(v^{2n}/\Lambda^{2n})$.
Once the corrections to the effective $\mathcal{Z}$ couplings in an input scheme are known to an order in this expansion, EWPD
can be analyzed to the same order.

The effective couplings are defined at all orders in $v/\Lambda$ to be \cite{Helset:2020yio,Hays:2020scx}
\bea\label{effcouplings}
g_{\rm eff,pr}^{\mathcal{Z},\psi} &=& \frac{\bar{g}_Z}{2} \left[(2 s_{\theta_Z}^2 \, Q_\psi -\sigma_3)\delta_{pr}+ \bar{v}_T \langle L_{3,4}^{\psi,pr} \rangle + \sigma_3
\bar{v}_T \langle L_{3,3}^{\psi,pr} \rangle  \right] \nn
&=& \langle g_{\rm SM,pr}^{\mathcal{Z},\psi} \rangle
+ \langle g_{\rm eff,pr}^{\mathcal{Z},\psi} \rangle_{\mathcal{O}(v^2/\Lambda^2)}
+ \langle g_{\rm eff,pr}^{\mathcal{Z},\psi} \rangle_{\mathcal{O}(v^4/\Lambda^4)} + \dots
\eea
Here $\psi_L = \{q_L,\ell_L\}$,
while $\psi_R =\{u_R,d_R,e_R\}$ and $\sigma_3 = 1$ for $u_L, \nu_L$ while $\sigma_3 = -1$ for $d_L, e_L$.
$L_{3,4},L_{3,3}$ are geoSMEFT field space connections defined in Refs.~\cite{Helset:2020yio,Hays:2020scx}.
\subsection{SMEFT to $\mathcal{L}^{(6)}$}
It is straightforward to derive
\bea
\langle \bar{\Gamma}^{\textrm{ SMEFT}}_i \rangle = \hat{\Gamma}^{\textrm{SM}}_i + \langle{\bar{\Gamma}}_i\rangle_{\mathcal{O}(v^2/\Lambda^2)} + \dots
\eea
for each EWPD observable in the SMEFT.\footnote{See Refs.~\cite{Han:2004az,Grinstein:1991cd,Berthier:2015oma,Brivio:2017bnu,Ellis:2020unq} for past analyses consistent
with these results.} By Taylor expanding the predictions to linear order in the corrections to the partial widths
via the effective couplings, we have that
\begin{align}
\frac{\langle \bar{\Gamma}_{\mathcal{Z} \rightarrow \bar{\psi}_p \psi_p}\rangle_{{\cal{O}}(v^2/\Lambda^2)}}{\hat{\Gamma}^{\rm SM}_{\mathcal{Z} \rightarrow \bar{\psi}_p \psi_p}}
&=  2  \frac{\rm{Re} \, \left[\langle g_{\rm SM,pp}^{\mathcal{Z},\psi_L}\rangle \, \langle g_{\rm eff,pp}^{\mathcal{Z},\psi_L}\rangle_{{\cal{O}}(v^2/\Lambda^2)}
\right]}
{|\langle g_{\rm SM,pp}^{\mathcal{Z},\psi_L}\rangle|^2+ |\langle g_{\rm SM,pp}^{\mathcal{Z},\psi_R}\rangle|^2}
+ 2  \frac{\rm{Re} \, \left[\langle g_{\rm SM,pp}^{\mathcal{Z},\psi_R}\rangle \, \langle g_{\rm eff,pp}^{\mathcal{Z},\psi_R}\rangle_{{\cal{O}}(v^2/\Lambda^2)}
\right]}
{|\langle g_{\rm SM,pp}^{\mathcal{Z},\psi_L}\rangle|^2+ |\langle g_{\rm SM,pp}^{\mathcal{Z},\psi_R}\rangle|^2} \nn
&= N^{\psi_R} \langle g_{\rm eff,pp}^{\mathcal{Z},\psi_R}\rangle_{{\cal{O}}(v^2/\Lambda^2)}
+N^{\psi_L} \langle g_{\rm eff,pp}^{\mathcal{Z},\psi_L}\rangle_{{\cal{O}}(v^2/\Lambda^2)}.
\end{align}
The $N^{\psi_{R/L}}$ are numerical coefficients that are reported in Table \ref{effectivecoupling1}.
For example, for $\Gamma_{\mathcal{Z} \rightarrow \bar{u} u}$, $N^{u_{R}} = 2.66$ while $N^{u_{L}} = -6.29$ in the
$\{\hat{m}_W, \hat{m}_Z,\hat{G}_F\}$ input parameter scheme.

Each partial width $\bar{\Gamma}_i$,
and the sum of partial widths $\bar{\Gamma}_{had}$, $\bar{\Gamma}_{Z}$, are defined at linear order
in SMEFT perturbations via Table \ref{effectivecoupling1}.
Linear perturbations in the partial widths then define $\bar{R}_{c,b,\ell}$ and $\bar{\sigma}_{had}^{0}$ via
\bea
	\frac{\bar{R}_{c,b}}{\hat{R}^{\textrm{SM}}_{c,b}} &=& 1 +
	\frac{\langle{\bar{\Gamma}}_{c,b}\rangle_{\mathcal{O}(v^2/\Lambda^2)}}{\hat{\Gamma}^{\textrm{SM}}_{c,b}}
	-\frac{\langle{\bar{\Gamma}}_{had}\rangle_{\mathcal{O}(v^2/\Lambda^2)}}{\hat{\Gamma}^{\textrm{SM}}_{had}}  + \dots \\
	\frac{\bar{R}_{\ell}}{\hat{R}^{\textrm{SM}}_{\ell}} &=& 1 +
	\frac{\langle{\bar{\Gamma}}_{had}\rangle_{\mathcal{O}(v^2/\Lambda^2)}}{\hat{\Gamma}^{\textrm{SM}}_{had}} -
	\frac{\langle{\bar{\Gamma}}_{\ell}\rangle_{\mathcal{O}(v^2/\Lambda^2)}}{{\hat{\Gamma}^{\textrm{SM}}_{\ell}}} + \dots \\
	\frac{\bar{\sigma}_{had}^{0,\textrm{SMEFT}}}{\hat{\sigma}_{had}^{0,\textrm{SM}}} &=& 1
	+\frac{\langle{\bar{\Gamma}}_{e}\rangle_{\mathcal{O}(v^2/\Lambda^2)}}{\hat{\Gamma}^{\textrm{SM}}_{e}}
	+\frac{\langle{\bar{\Gamma}}_{had}\rangle_{\mathcal{O}(v^2/\Lambda^2)}}{\hat{\Gamma}^{\textrm{SM}}_{had}}
	- 2 \frac{\langle{\bar{\Gamma}}_{Z}\rangle_{\mathcal{O}(v^2/\Lambda^2)}}{\hat{\Gamma}^{\textrm{SM}}_{Z}} + \dots
\eea
The remaining observables, $\bar{A}_{FB}^{0,f}$ for $f = \{\ell,c,b\}$
have the leading SMEFT perturbation
\begin{align}
	\frac{\langle \bar{A}_i\rangle_{\mathcal{O}(v^2/\Lambda^2)}}{\hat{A}_i^{\textrm{SM}}} &=&
	\frac{4 \, \langle g_{\textrm{SM}}^{\mathcal{Z},i_L} \rangle\, \langle g_{\textrm{SM}}^{\mathcal{Z},i_R}\rangle}{\langle g_{\textrm{SM}}^{\mathcal{Z},i_L}\rangle^4-\langle g_{\textrm{SM}}^{\mathcal{Z},i_R}\rangle^4}
	\left[\langle g_{\textrm{SM}}^{\mathcal{Z},i_R} \rangle \langle g_{eff}^{\mathcal{Z},i_L}\rangle_{\mathcal{O}(v^2/\Lambda^2)}
		-\langle g_{\textrm{SM}}^{\mathcal{Z},i_L} \rangle \langle g_{eff}^{\mathcal{Z},i_R}\rangle_{\mathcal{O}(v^2/\Lambda^2)}
\right].
\end{align}
The required numerical coefficients to construct these observables are given in Table \ref{Aiexpansion}.
For example, for the bottom quark in the $\{\hat{m}_W, \hat{m}_Z,\hat{G}_F\}$ scheme we find that
\bea
	\frac{\langle \bar{A}_b\rangle_{\mathcal{O}(v^2/\Lambda^2)}}{\hat{A}_b^{\textrm{SM}}} &=&
 2.2 \langle g_{eff}^{\mathcal{Z},d_R}\rangle_{\mathcal{O}(v^2/\Lambda^2)} + 0.39 \langle g_{eff}^{\mathcal{Z},d_L}\rangle_{\mathcal{O}(v^2/\Lambda^2)} .
\eea
The $\bar{A}_{FB}^{0,f}$ follow directly via
\bea
\frac{(\bar{A}_{FB}^{0,f})^{\textrm{SMEFT}}}{(\hat{A}_{FB}^{0,f})^{\textrm{SM}}} = 1 + \frac{\langle \bar{A}_e\rangle_{\mathcal{O}(v^2/\Lambda^2)}}{\hat{A}_e^{\textrm{SM}}}
+ \frac{\langle \bar{A}_f\rangle_{\mathcal{O}(v^2/\Lambda^2)}}{\hat{A}_f^{\textrm{SM}}} + \dots
\eea
Each of the effective couplings is expanded into SMEFT Wilson coefficients in Table \ref{effectivecouplingtoWC}.

\subsection{SMEFT to $\mathcal{L}^{(8)}$}
Defining EWPD to dimension eight in the SMEFT requires an expansion of the observables to $\mathcal{O}(v^4/\Lambda^4)$,
and the definition of the effective couplings to $\mathcal{O}(v^4/\Lambda^4)$.
The latter is defined in Tables \ref{effectivecoupling2} and \ref{effectivecoupling3}.
Expressing the results compactly, we build upon the presentation in Ref.~\cite{Hays:2020scx}.
Expanding to second order the partial widths
\bea
\langle \bar{\Gamma}^{\textrm{SMEFT}}_i \rangle = \hat{\Gamma}^{\textrm{SM}}_i + \langle{\bar{\Gamma}}_i\rangle_{\mathcal{O}(v^2/\Lambda^2)}
+ \langle{\bar{\Gamma}}_i\rangle_{\mathcal{O}(v^4/\Lambda^4)} + \dots
\eea
where
\begin{align}
\frac{\langle \bar{\Gamma}^{\rm SMEFT}_{\mathcal{Z} \rightarrow \bar{\psi}_p \psi_p}\rangle_{{\cal{O}}(v^4/\Lambda^4)}}{\hat{\Gamma}^{\rm SM}_{\mathcal{Z} \rightarrow \bar{\psi}_p \psi_p}}
&=  2  \frac{\rm{Re} \, \left[\langle g_{\rm SM,pp}^{\mathcal{Z},\psi_L}\rangle \, \langle g_{\rm eff,pp}^{\mathcal{Z},\psi_L}\rangle_{{\cal{O}}(v^4/\Lambda^4)}
\right]}
{|\langle g_{\rm SM,pp}^{\mathcal{Z},\psi_L}\rangle|^2+ |\langle g_{\rm SM,pp}^{\mathcal{Z},\psi_R}\rangle|^2}
+ 2  \frac{\rm{Re} \, \left[\langle g_{\rm SM,pp}^{\mathcal{Z},\psi_R}\rangle \, \langle g_{\rm eff,pp}^{\mathcal{Z},\psi_R}\rangle_{{\cal{O}}(v^4/\Lambda^4)}
\right]}
{|\langle g_{\rm SM,pp}^{\mathcal{Z},\psi_L}\rangle|^2+ |\langle g_{\rm SM,pp}^{\mathcal{Z},\psi_R}\rangle|^2} \nonumber \\
&+ \frac{|\langle g_{\rm eff,pp}^{\mathcal{Z},\psi_L}\rangle_{{\cal{O}}(v^2/\Lambda^2)}|^2+ |\langle g_{\rm eff,pp}^{\mathcal{Z},\psi_R}\rangle_{{\cal{O}}(v^2/\Lambda^2)}|^2}{|\langle g_{\rm SM,pp}^{\mathcal{Z},\psi_L}\rangle|^2+ |\langle g_{\rm SM,pp}^{\mathcal{Z},\psi_R}\rangle|^2}.
\end{align}
There is dependence on the $\langle g_{\rm eff,pp}^{\mathcal{Z},\psi_{L/R}} \rangle_{{\cal{O}}(v^{2n}/\Lambda^{2n})}$ at each order in the expansion $n$.
The (pseudo)-observables also have dependence on the squared dimension-six effective couplings.
The required numerical coefficients are given in Table \ref{effectivecoupling1}.
As an example, for decays to up quarks in the $\{\hat{m}_W, \hat{m}_Z,\hat{G}_F\}$ scheme we have that
\begin{align}
\frac{\langle \bar{\Gamma}^{\rm SMEFT}_{\mathcal{Z} \rightarrow \bar{u}_p u_p}\rangle_{{\cal{O}}(v^4/\Lambda^4)}}{\hat{\Gamma}^{\rm SM}_{\mathcal{Z} \rightarrow \bar{u}_p u_p}}
&= -6.29 \langle g_{\rm eff,pp}^{\mathcal{Z},u_L}\rangle_{{\cal{O}}(v^4/\Lambda^4)} +
 2.66 \langle g_{\rm eff,pp}^{\mathcal{Z},\psi_R}\rangle_{{\cal{O}}(v^4/\Lambda^4)}
+ 12.1 |\langle g_{\rm eff,pp}^{\mathcal{Z},\psi_L}\rangle_{{\cal{O}}(v^2/\Lambda^2)}|^2 \nn
&+ 12.1|\langle g_{\rm eff,pp}^{\mathcal{Z},\psi_R}\rangle_{{\cal{O}}(v^2/\Lambda^2)}|^2
\end{align}
The observables $\bar{R}_{c,b}, \bar{R}_{\ell}, \bar{\sigma}_{had}^{0}$ are determined from the
expansion of the $\Gamma_i$ directly via
\bea
\frac{(\bar{\Gamma}_i/\bar{\Gamma}_j)^{\textrm{SMEFT}}}{(\hat{\Gamma}_i/\hat{\Gamma}_j)^{\textrm{SM}}} =
	1 &+& \frac{\langle{\bar{\Gamma}}_i\rangle_{\mathcal{O}(v^2/\Lambda^2)}}{\hat{\Gamma}_i^{\textrm{SM}}}
	- \frac{\langle{\bar{\Gamma}}_j\rangle_{\mathcal{O}(v^2/\Lambda^2)}}{\hat{\Gamma}_j^{\textrm{SM}}} +\left(\frac{\langle{\bar{\Gamma}}_j\rangle_{\mathcal{O}(v^2/\Lambda^2)}}{\hat{\Gamma}_j^{\textrm{SM}}}\right)^2 \\
	  &\, & + \frac{\langle{\bar{\Gamma}}_i\rangle_{\mathcal{O}(v^4/\Lambda^4)}}{\hat{\Gamma}_i^{\textrm{SM}}}
	  - \frac{\langle{\bar{\Gamma}}_j\rangle_{\mathcal{O}(v^4/\Lambda^4)}}{\hat{\Gamma}_j^{\textrm{SM}}}
	  -\frac{\langle{\bar{\Gamma}}_i\rangle_{\mathcal{O}(v^2/\Lambda^2)} \, \langle{\bar{\Gamma}}_j\rangle_{\mathcal{O}(v^2/\Lambda^2)}}{\hat{\Gamma}_i^{\textrm{SM}} \, \hat{\Gamma}_j^{\textrm{SM}}}.
 \nonumber
\eea
 For $\bar{A}_{FB}^{0,f}$ we expand directly in terms of the effective couplings via
 \bea
	\frac{\langle \bar{A}^{\textrm{SMEFT}}_i\rangle_{\mathcal{O}(v^4/\Lambda^4)}}{\hat{A}_i^{\textrm{SM}}} &=&
	\frac{2 \, \langle g_{\textrm{SM}}^{\mathcal{Z},i_R} \rangle^2 \, \langle g_{\rm eff,pp}^{\mathcal{Z},i_L}\rangle^2_{{\cal{O}}(v^2/\Lambda^2)}}
	{[\langle g_{\textrm{SM}}^{\mathcal{Z},i_L} \rangle^2+\langle g_{\textrm{SM}}^{\mathcal{Z},i_R} \rangle^2]^2}\,
\left(
	\frac{\langle g_{\textrm{SM}}^{\mathcal{Z},i_R} \rangle^2 - 3 \langle g_{\textrm{SM}}^{\mathcal{Z},i_L} \rangle^2}
	{\langle g_{\textrm{SM}}^{\mathcal{Z},i_L} \rangle^2-\langle g_{\textrm{SM}}^{\mathcal{Z},i_R} \rangle^2}
\right) \nn
 &-&
 \frac{2 \, \langle g_{\textrm{SM}}^{\mathcal{Z},i_L} \rangle^2 \, \langle g_{\rm eff,pp}^{\mathcal{Z},i_R}\rangle^2_{{\cal{O}}(v^2/\Lambda^2)}}
 {[\langle g_{\textrm{SM}}^{\mathcal{Z},i_L} \rangle^2+\langle g_{\textrm{SM}}^{\mathcal{Z},i_R} \rangle^2]^2}\,
 \left(
	 \frac{\langle g_{\textrm{SM}}^{\mathcal{Z},i_L} \rangle^2 - 3 \langle g_{\textrm{SM}}^{\mathcal{Z},i_R} \rangle^2}
	 {(\langle g_{\textrm{SM}}^{\mathcal{Z},i_L} \rangle^2-\langle g_{\textrm{SM}}^{\mathcal{Z},i_R} \rangle^2)}
 \right) \nn
 &+& 8 \, \langle g_{\rm eff,pp}^{\mathcal{Z},i_R}\rangle_{{\cal{O}}(v^2/\Lambda^2)} \langle g_{\rm eff,pp}^{\mathcal{Z},i_L}\rangle_{{\cal{O}}(v^2/\Lambda^2)}
 \left(\frac{\langle g_{\textrm{SM}}^{\mathcal{Z},i_L} \rangle \, \langle g_{\textrm{SM}}^{\mathcal{Z},i_R} \rangle}{(\langle g_{\textrm{SM}}^{\mathcal{Z},i_L} \rangle^2+\langle g_{\textrm{SM}}^{\mathcal{Z},i_R} \rangle^2)^2}\right) \\
 &+& 4 \, \frac{\langle g_{\textrm{SM}}^{\mathcal{Z},i_L} \rangle\langle g_{\rm eff,pp}^{\mathcal{Z},i_L}\rangle_{{\cal{O}}(v^4/\Lambda^4)}}
 {\langle g_{\textrm{SM}}^{\mathcal{Z},i_L} \rangle^4
 -\langle g_{\textrm{SM}}^{\mathcal{Z},i_R} \rangle^4}
 \langle g_{\textrm{SM}}^{\mathcal{Z},i_R} \rangle^2
 -4 \, \frac{\langle g_{\textrm{SM}}^{\mathcal{Z},i_R} \rangle \langle g_{\rm eff,pp}^{\mathcal{Z},i_R}\rangle_{{\cal{O}}(v^4/\Lambda^4)}}
 {\langle g_{\textrm{SM}}^{\mathcal{Z},i_L} \rangle^4
 -\langle g_{\textrm{SM}}^{\mathcal{Z},i_R} \rangle^4}
 \langle g_{\textrm{SM}}^{\mathcal{Z},i_L} \rangle^2. \nonumber
  \eea
The numerical dependence of EWPD
observables on the SMEFT induced effective coupling in the $\Gamma_i$
is largely scheme independent and is given in Table \ref{effectivecoupling1}.
The numerical dependence on the SMEFT induced effective couplings
for the $\bar{A}^{\textrm{SMEFT}}_i$ are given in Table \ref{Aiexpansion}.

The numerical dependence of the effective couplings on the Wilson coefficients
are reported at ${\cal{O}}(v^2/\Lambda^2),{\cal{O}}(v^4/\Lambda^4)$ in
Tables \ref{effectivecouplingtoWC}, \ref{effectivecoupling2}, and \ref{effectivecoupling3}. This expansion of the effective couplings in terms of the individual
Wilson coefficients carries a significant SMEFT input parameter scheme dependence, which increases at higher orders in
the $v^n/\Lambda^n$ expansion \cite{Hays:2020scx}.
This is expected due to the decoupling theorem and represents the effect of new physics being
absorbed into the lower energy measured input parameters.
This is a more significant issue for the SMEFT compared to many EFTs, due to the presence of a Higgs field.
\begin{table}\centering
\renewcommand{\arraystretch}{1.2}
\begin{center}
\begin{tabular}{|c|c|c|c|c|c|c|}
\hline
\multicolumn{7}{|c|}{Numerical dependence of $\langle \bar{\Gamma}_i \rangle/\hat{\Gamma}^{\textrm{SM}}_i$
 in the $\{\hat{m}_W, \hat{m}_Z,\hat{G}_F\}/\{\hat{\alpha}, \hat{m}_Z,\hat{G}_F\}$ schemes} \\
 \hline
 ${\cal{O}}(v^2/\Lambda^2)$  & $\langle \bar{\Gamma}_{u} \rangle/\hat{\Gamma}^{\textrm{SM}}_{u}$ & $\langle \bar{\Gamma}_{\nu} \rangle/\hat{\Gamma}^{\textrm{SM}}_{\nu}$ &
 $\langle \bar{\Gamma}_{\ell} \rangle/\hat{\Gamma}^{\textrm{SM}}_{\ell}$ & $\langle \bar{\Gamma}_{d,b} \rangle/\hat{\Gamma}^{\textrm{SM}}_{d,b}$
									 & $\langle \bar{\Gamma}_{Z} \rangle/\hat{\Gamma}^{\textrm{SM}}_{Z}$ & $\langle \bar{\Gamma}_{had} \rangle/\hat{\Gamma}^{\textrm{SM}}_{had}$ \\
\hline
$\langle g_{\rm eff,pp}^{\mathcal{Z},u_R}\rangle$ & 2.66/2.76 & & & & 2(0.320/0.331) & 2(0.458/0.475) \\
$\langle g_{\rm eff,pp}^{\mathcal{Z},d_R}\rangle$ & & & & -1.04/-1.08 & 3(-0.160/-0.166) & 3(-0.229/-0.236)\\
$\langle g_{\rm eff,pp}^{\mathcal{Z},\ell_R}\rangle$ &  & & -4.75/-4.93& &3(-0.160/-0.166) & \\
 $\langle g_{\rm eff,pp}^{\mathcal{Z},u_L}\rangle$ & -6.29/-6.19 & & & & 2(-0.756/-0.745) & 2(-1.08/-1.07)\\
 $\langle g_{\rm eff,pp}^{\mathcal{Z},d_L}\rangle$ & & & & 5.97/5.94  & 3(0.917/0.911) & 3(1.31/1.29) \\
 $\langle g_{\rm eff,pp}^{\mathcal{Z},\ell_L}\rangle$ &  & & 5.91/5.74 & & 3(0.199/0.193) & \\
 $\langle g_{\rm eff,pp}^{\mathcal{Z},\nu_L}\rangle$ & & -5.36/-5.36 & & & 3(-0.359/-0.359) & \\
 \hline
 ${\cal{O}}(v^4/\Lambda^4)$  & $\langle \bar{\Gamma}_{u} \rangle/\hat{\Gamma}^{\textrm{SM}}_{u}$ & $\langle \bar{\Gamma}_{u} \rangle/\hat{\Gamma}^{\textrm{SM}}_{\nu}$ &
 $\langle \bar{\Gamma}_{\ell} \rangle/\hat{\Gamma}^{\textrm{SM}}_{\ell}$ & $\langle \bar{\Gamma}_{d,b} \rangle/\hat{\Gamma}^{\textrm{SM}}_{d,b}$
									 & $\langle \bar{\Gamma}_{Z} \rangle/\hat{\Gamma}^{\textrm{SM}}_{Z}$ & $\langle \bar{\Gamma}_{had} \rangle/\hat{\Gamma}^{\textrm{SM}}_{had}$  \\
\hline
$\langle g_{\rm eff,pp}^{\mathcal{Z},u_R}\rangle$ & 2.66/2.76 & & & & 2(0.320/0.331) & 2(0.458/0.475) \\
$\langle g_{\rm eff,pp}^{\mathcal{Z},d_R}\rangle$ &  & & & -1.04/-1.08 & 3(-0.160/-0.166) & 3(-0.229/-0.236) \\
$\langle g_{\rm eff,pp}^{\mathcal{Z},\ell_R}\rangle$ &  & & -4.75/-4.93 & & 3(-0.160/-0.166) &  \\
 $\langle g_{\rm eff,pp}^{\mathcal{Z},u_L}\rangle$ & -6.29/-6.19 & & & & 2(-0.756/-0.745) & 2(-1.08/-1.07) \\
 $\langle g_{\rm eff,pp}^{\mathcal{Z},d_L}\rangle$ &  & & & 5.97/5.94 & 3(0.917/0.911) & 3(1.31/1.29) \\
 $\langle g_{\rm eff,pp}^{\mathcal{Z},\ell_L}\rangle$ &  & & 5.91/5.74  & & 3(0.199/0.193) &  \\
 $\langle g_{\rm eff,pp}^{\mathcal{Z},\nu_L}\rangle$ & & -5.36/-5.36 & & & 3(-0.359/-0.359) & \\
 \hline
 $\langle g_{\rm eff,pp}^{\mathcal{Z},u_R}\rangle^2$ & 12.1/12.1 & & & & 2(1.45/1.45) & 2(2.08/2.08) \\
 $\langle g_{\rm eff,pp}^{\mathcal{Z},d_R}\rangle^2$ & & & & 9.47/9.47 & 3(1.45/1.45) & 3(2.08/2.07) \\
 $\langle g_{\rm eff,pp}^{\mathcal{Z},\ell_R}\rangle^2$ &  & & 14.4/14.4 & & 3(0.485/0.485) &  \\
  $\langle g_{\rm eff,pp}^{\mathcal{Z},u_L}\rangle^2$ & 12.1/12.1  & & & & 2(1.45/1.45) & 2(2.08/2.08) \\
  $\langle g_{\rm eff,pp}^{\mathcal{Z},d_L}\rangle^2$ & & & & 9.47/9.47 & 3(1.45/1.45) & 3(2.08/2.07)  \\
  $\langle g_{\rm eff,pp}^{\mathcal{Z},\ell_L}\rangle^2$ & & & 14.4/14.4 & & 3(0.485/0.485) &  \\
  $\langle g_{\rm eff,pp}^{\mathcal{Z},\nu_L}\rangle^2$ & & 7.24/7.24 & & & 3(0.485/0.485) & \\
 \hline
\end{tabular}
\caption{Dependence of the partial widths on the effective couplings, scaled to the SM prediction of the partial width.
For the columns $\bar{\Gamma}_{\ell},\bar{\Gamma}_{u},\bar{\Gamma}_{d,b}$ the individual partial widths are reported.
The sum over flavors is explicit in the contribution to $\bar{\Gamma}_{had},\bar{\Gamma}_{Z}$.
The top section of the Table reports the dependence on $\langle g_{\rm eff,pp}^{\mathcal{Z},\psi}\rangle_{\mathcal{O}(v^2/\Lambda^2)}$.
The middle section of the Table reports the dependence on $\langle g_{\rm eff,pp}^{\mathcal{Z},\psi}\rangle_{\mathcal{O}(v^4/\Lambda^4)}$,
while the bottom section is the dependence on  $\langle g_{\rm eff,pp}^{\mathcal{Z},\psi}\rangle_{\mathcal{O}(v^2/\Lambda^2)}^2$. In the quoted results, $\langle \bar{\Gamma}_{d,b} \rangle/\hat{\Gamma}^{\textrm{SM}}_{d,b}$
was determined using numerical values of light quarks $d,s$ for the partial width. $\hat{\Gamma}^{\textrm{SM}}_{d,s}/\hat{\Gamma}^{\textrm{SM}}_{b}$
differs at the percent level in the SM. This leads to numerical differences, when combined with rounding effects,
in the results quoted that should be incorporated as a simple rescaling based on Table \ref{schemeresults}.
An empty entry indicates no dependence on the relevant effective coupling.}\label{effectivecoupling1}
\end{center}
\end{table}
\begin{table}\centering
\renewcommand{\arraystretch}{1.2}
\begin{center}
\begin{tabular}{|c|c|c|c|}
\hline
\multicolumn{4}{|c|}{$\langle \bar{A}_i\rangle/\hat{A}_i^{\textrm{SM}}$
 in the $\{\hat{m}_W, \hat{m}_Z,\hat{G}_F\}/\{\hat{\alpha}, \hat{m}_Z,\hat{G}_F\}$ schemes} \\
 \hline
 ${\cal{O}}(v^2/\Lambda^2)$  & $\langle \bar{A}_\ell\rangle/\hat{A}_\ell^{\textrm{SM}}$ &
 $\langle \bar{A}_c\rangle/\hat{A}_c^{\textrm{SM}}$ &
 $\langle \bar{A}_b\rangle/\hat{A}_b^{\textrm{SM}}$ \\
\hline
$\langle g_{\rm eff,pp}^{\mathcal{Z},u_R}\rangle$ & & -6.71/-7.23  &   \\
$\langle g_{\rm eff,pp}^{\mathcal{Z},d_R}\rangle$ &  & & 2.22/2.33  \\
$\langle g_{\rm eff,pp}^{\mathcal{Z},\ell_R}\rangle$ & 26.9/37.7  & &   \\
$\langle g_{\rm eff,pp}^{\mathcal{Z},u_L}\rangle$ & &-2.84/-3.22  &   \\
$\langle g_{\rm eff,pp}^{\mathcal{Z},d_L}\rangle$ &  & &  0.387/0.423 \\
$\langle g_{\rm eff,pp}^{\mathcal{Z},\ell_L}\rangle$ & 21.7/32.4 & &   \\
\hline
${\cal{O}}(v^4/\Lambda^4)$  & $\langle \bar{A}_\ell\rangle/\hat{A}_\ell^{\textrm{SM}}$ &
$\langle \bar{A}_c\rangle/\hat{A}_c^{\textrm{SM}}$ &
$\langle \bar{A}_b\rangle/\hat{A}_b^{\textrm{SM}}$ \\
\hline
$\langle g_{\rm eff,pp}^{\mathcal{Z},u_R}\rangle^2$ & & -12.0/-10.7  &   \\
$\langle g_{\rm eff,pp}^{\mathcal{Z},d_R}\rangle^2$ &  & & -17.8/-17.8  \\
$\langle g_{\rm eff,pp}^{\mathcal{Z},\ell_R}\rangle^2$ & 46.6/76.9  & &   \\
$\langle g_{\rm eff,pp}^{\mathcal{Z},u_L}\rangle^2$ & &-13.0/-14.7  &   \\
$\langle g_{\rm eff,pp}^{\mathcal{Z},d_L}\rangle^2$ &  & &  -1.77/-1.94 \\
$\langle g_{\rm eff,pp}^{\mathcal{Z},\ell_L}\rangle^2$ & -75.4/-106 & &   \\
\hline
$\langle g_{\rm eff,pp}^{\mathcal{Z},u_R}\rangle \langle g_{\rm eff,pp}^{\mathcal{Z},u_L}\rangle$ &  & -35.9/-37.8 &  \\
$\langle g_{\rm eff,pp}^{\mathcal{Z},d_R}\rangle \langle g_{\rm eff,pp}^{\mathcal{Z},d_L}\rangle$ &  & &-13.2/-13.9 \\
$\langle g_{\rm eff,pp}^{\mathcal{Z},\ell_R}\rangle \langle g_{\rm eff,pp}^{\mathcal{Z},\ell_L}\rangle$ & -56.3/-57.3 & &  \\
\hline
$\langle g_{\rm eff,pp}^{\mathcal{Z},u_R}\rangle$ & & -6.71/-7.23  &   \\
$\langle g_{\rm eff,pp}^{\mathcal{Z},d_R}\rangle$ &  & & 2.22/2.33  \\
$\langle g_{\rm eff,pp}^{\mathcal{Z},\ell_R}\rangle$ & 26.9/37.7  & &   \\
$\langle g_{\rm eff,pp}^{\mathcal{Z},u_L}\rangle$ & &-2.84/-3.22  &   \\
$\langle g_{\rm eff,pp}^{\mathcal{Z},d_L}\rangle$ &  & &  0.387/0.423 \\
$\langle g_{\rm eff,pp}^{\mathcal{Z},\ell_L}\rangle$ & 21.7/32.4 & &   \\
\hline
\end{tabular}
\caption{Numerical coefficients defining the dependence on the SMEFT effective couplings
in forward backward asymmetries. The expressions are normalized to the tree level SM values in each
input parameter
scheme: $\hat{A}_c^{\textrm{SM}} = 0.70/0.70$, $\hat{A}_b^{\textrm{SM}} = 0.94/0.94$,
$\hat{A}_\ell^{\textrm{SM}} = 0.21/0.15$. The significant scheme dependence of $\hat{A}_\ell^{\textrm{SM}}$ follows from
the accidental numerical  suppression of the value of the vectorial leptonic coupling,
rendering it more sensitive to scheme dependence.}\label{Aiexpansion}
\end{center}
\end{table}

\begin{table}\centering
\renewcommand{\arraystretch}{1.2}
\begin{center}
\begin{tabular}{|c|c|c|c|c|c|c|c|}
\hline
 \multicolumn{8}{|c|}{SMEFT corrections in the $\{\hat{m}_W, \hat{m}_Z,\hat{G}_F\}/\{\hat{\alpha}, \hat{m}_Z,\hat{G}_F\}$ scheme} \\
 \hline
${{\cal{O}}(\frac{v^2}{\Lambda^2})}$ & $\langle g_{\rm eff,pp}^{\mathcal{Z},u_R}\rangle$ & $\langle g_{\rm eff,pp}^{\mathcal{Z},d_R}\rangle$ & $\langle g_{\rm eff,pp}^{\mathcal{Z},\ell_R}\rangle$ &
 $\langle g_{\rm eff,pp}^{\mathcal{Z},u_L}\rangle$ & $\langle g_{\rm eff,pp}^{\mathcal{Z},d_L}\rangle$ & $\langle g_{\rm eff,pp}^{\mathcal{Z},\ell_L}\rangle$ & $\langle g_{\rm eff,pp}^{\mathcal{Z},\nu_L}\rangle$  \\
\hline
$\delta G_F^{(6)}$ & -0.08/0.15 &  0.04/-0.07 & 0.12/-0.22  & 0.18/0.41 & -0.22/-0.34 &  -0.15/-0.49 & 0.26/0.26 \\
$\tilde{C}_{HD}^{(6)}$ & -0.22/0.05 & 0.11/-0.03 & 0.33/-0.08 &-0.13/0.15 & 0.02/-0.12 & 0.24/-0.17 & 0.09/0.09 \\
$\tilde{C}_{HWB}^{(6)}$ & -0.21/0.39 &  0.10/-0.19 & 0.31/-0.58 & -0.21/0.39 & 0.10/-0.19 & 0.31/-0.58 & \\
$\tilde{C}_{H\psi}^{(6)}$ & 0.37/0.37 & 0.37/0.37 & 0.37/0.37 &0.37/0.37 & 0.37/0.37 & 0.37/0.37 & 0.37/0.37 \\
$\tilde{C}_{H\psi}^{3,(6)}$ &  & & & -0.37/-0.37  & 0.37/0.37 & 0.37/0.37 & -0.37/-0.37\\
\hline
\end{tabular}
\caption{The effective couplings expanded to ${{\cal{O}}(v^2/\Lambda^2)}$ in each input parameter scheme. $\delta G_F^{(6)}$
is defined in Ref.~\cite{Hays:2020scx}. Reported is the numerical coefficient multiplying each SMEFT correction.  $p=\{1,2,3\}$
is a flavor index. The operator subscript labels $\psi$ take on the values $\{u_R,d_R,\ell_R,q_L,\ell_L\}$, with the effective coupling $\psi$
label dictating the value of $\psi$.}\label{effectivecouplingtoWC}
\end{center}
\end{table}

\begin{table}\centering
\renewcommand{\arraystretch}{1.1}
\begin{center}
\begin{tabular}{|c|c|c|c|c|c|c|c|}
\hline
 \multicolumn{4}{|c|}{SMEFT corrections in $\{\hat{m}_W, \hat{m}_Z,\hat{G}_F\}/\{\hat{\alpha}, \hat{m}_Z,\hat{G}_F\}$ scheme} \\
 \hline
${{\cal{O}}(\frac{v^4}{\Lambda^4})}$ & $\langle g_{\rm eff,pp}^{\mathcal{Z},u_R}\rangle$ & $\langle g_{\rm eff,pp}^{\mathcal{Z},d_R}\rangle$ & $\langle g_{\rm eff,pp}^{\mathcal{Z},\ell_R}\rangle$  \\
\hline
$\langle g_{\rm eff}^{\mathcal{Z},\psi}\rangle^2$ & 14/5.5 & -27/-11 &-9.1/-3.6 \\
$\tilde{C}_{HB} \, \tilde{C}_{HWB}$ & -0.21/0.39 & 0.10/-0.19 & 0.31/-0.58 \\
$\tilde{C}_{HD}^2$ & 0.28/-0.026 & -0.14/0.013 & -0.42/0.040  \\
$\tilde{C}_{HD} \, \tilde{C}_{H\psi}^{(6)}$ & -0.83/-0.19 & -0.83/-0.19 & -0.83/-0.19  \\
$\tilde{C}_{HD} \, \tilde{C}_{HWB}$ & 0.59/-0.19 & -0.29/0.097 & -0.88/0.29  \\
$\tilde{C}_{HD} \langle g_{\rm eff}^{\mathcal{Z},\psi}\rangle$ & 4.0/0.50 & 4.0/0.50 & 4.0/0.50 \\
$(\tilde{C}_{H\psi}^{(6)})^2$ & 0.62/1.4 & -1.2/-2.8 & -0.42/-0.93 \\
$\tilde{C}_{HWB} \, \tilde{C}_{H\psi}^{(6)}$ & -0.69/0.58 & -0.69/0.58 &  -0.69/0.58 \\
$\tilde{C}_{H\psi}^{(6)} \langle g_{\rm eff}^{\mathcal{Z},\psi}\rangle$ & -6.7/-5.8 & 13/12 & 4.5/3.9  \\
$\tilde{C}_{HWB} \, \langle g_{\rm eff}^{\mathcal{Z},\psi}\rangle$ & 3.7/0.26 & 3.7/0.26 & 3.7/0.26  \\
$\tilde{C}_{HW} \, \tilde{C}_{HWB}$ & -0.21/0.39 & 0.10/-0.19 & 0.31/-0.58  \\
$\tilde{C}_{HD}^{(8)}$ &-0.014/0.026 & 0.0069/-0.013 &  0.021/-0.040  \\
$\tilde{C}_{HD,2}^{(8)}$ & -0.21/0.026 & 0.10/-0.013 &  0.31/-0.040  \\
$\tilde{C}_{H\psi}^{(8)}$ & 0.19/0.19 & 0.19/0.19 & 0.19/0.19 \\
$\tilde{C}_{HW,2}^{(8)}$ & -0.38/0 & 0.19/0 & 0.58/0  \\
$\tilde{C}_{HWB}^{(8)}$ & -0.10/0.19 & 0.051/-0.097 & 0.15/-0.29  \\
$\delta G_F^{(8)}$ & -0.078/0.15 & 0.039/-0.075 & 0.12/-0.22  \\
$(\tilde{C}_{HWB}^{(6)})^2$ & 0.19/-0.35 &  -0.096/0.18 & -0.29/0.53  \\
\hline
\end{tabular}
\caption{The effective couplings expanded to ${{\cal{O}}(v^4/\Lambda^4)}$ in each input parameter scheme. $\delta G_F^{(8)}$
and the remaining operator forms are defined in Ref.~\cite{Helset:2020yio,Hays:2020scx}.
$\langle g_{\rm eff,pp}^{\mathcal{Z},\psi}\rangle$ is understood to be $\langle g_{\rm eff,pp}^{\mathcal{Z},\psi}\rangle_{{\cal{O}}(v^2/\Lambda^2)}$
in the left most column. Reported is the numerical coefficient multiplying each SMEFT correction.  $p=\{1,2,3\}$
is a flavour index. We have eliminated $\delta G_F^{(6)}$ in favor of introducing $\langle g_{\rm eff,pp}^{\mathcal{Z},\psi}\rangle_{\mathcal{O}(v^2/\Lambda^2)}^2$
in these expressions.}\label{effectivecoupling2}
\end{center}
\end{table}

\begin{table}\centering
\renewcommand{\arraystretch}{1.1}
\begin{center}
\begin{tabular}{|c|c|c|c|c|c|c|c|}
\hline
 \multicolumn{5}{|c|}{SMEFT corrections in the $\{\hat{m}_W, \hat{m}_Z,\hat{G}_F\}/\{\hat{\alpha}, \hat{m}_Z,\hat{G}_F\}$ scheme} \\
 \hline
${{\cal{O}}(\frac{v^4}{\Lambda^4})}$ &
 $\langle g_{\rm eff,pp}^{\mathcal{Z},u_L}\rangle$ & $\langle g_{\rm eff,pp}^{\mathcal{Z},d_L}\rangle$ & $\langle g_{\rm eff,pp}^{\mathcal{Z},\ell_L}\rangle$ & $\langle g_{\rm eff,pp}^{\mathcal{Z},\nu_L}\rangle$  \\
\hline
$\langle g_{\rm eff}^{\mathcal{Z},\psi}\rangle^2$ & -5.8/-0.92 & 4.8/1.9 & 7.3/0.40 & -4.1/-4.1\\
$\tilde{C}_{HB} \, \tilde{C}_{HWB}$  & -0.21/0.39 & 0.10/-0.19 & 0.31/-0.58 & \\
$\tilde{C}_{HD}^2$ & -0.0073/-0.073 & 0.0060/0.060 & 0.084/0.086 & -0.046/-0.046 \\
$\tilde{C}_{HD} \, \tilde{C}_{H\psi}^{(6)}$  & 0.088/-0.19 & -0.072/-0.19 & 0.33/-0.19 & -0.19/-0.19 \\
$\tilde{C}_{HD} \, \tilde{C}_{HWB}$ & 0.079/-0.19 & -0.084/0.097 & 0.086/0.29 & \\
$\tilde{C}_{HD} \langle g_{\rm eff}^{\mathcal{Z},\psi}\rangle$ & -0.97/0.50 & -0.11/0.50 & -2.3/0.50 & 0.50/0.50 \\
$(\tilde{C}_{H\psi}^{(6)})^2$ &-0.26/0.11 & 0.22/-0.026 & 0.33/-0.14 & -0.19/-0.19 \\
$\tilde{C}_{HWB} \, \tilde{C}_{H\psi}^{(6)}$ & 0.29/-0.62 & 0.12/-0.49 & 0.56/-0.62 & \\
$\tilde{C}_{H\psi}^{(6)} \langle g_{\rm eff}^{\mathcal{Z},\psi}\rangle$ & 2.8/0.042 & -2.3/-0.64 & -3.6/0.24 & 2.0/2.0 \\
$\tilde{C}_{HWB} \, \langle g_{\rm eff}^{\mathcal{Z},\psi}\rangle$  & -1.6/2.3 & -0.65/1.7 & -3.0/2.5 & \\
$\tilde{C}_{HW} \, \tilde{C}_{HWB}$  &-0.21/0.39 & 0.10/-0.19 & 0.31/-0.58 & \\
$\tilde{C}_{HD}^{(8)}$ & 0.033/0.073 & -0.039/-0.060 & -0.026/-0.086 & 0.046/0.046 \\
$\tilde{C}_{HD,2}^{(8)}$ & -0.16/0.073 & 0.057/-0.060 & 0.26/-0.086 & 0.046/0.046 \\
$\tilde{C}_{H\psi}^{(8)}$  & 0.19/0.19 & 0.19/0.19 & 0.19/0.19 & 0.19/0.19\\
$\tilde{C}_{HW,2}^{(8)}$ &  -0.38/0 & 0.19/0 & 0.58/0 & \\
$\tilde{C}_{HWB}^{(8)}$  & -0.10/0.19 & 0.051/-0.097 & 0.15/-0.29 & \\
$\delta G_F^{(8)}$  & 0.18/0.41 & -0.22/-0.34 & -0.15/-0.49 & 0.26/0.26 \\
$(\tilde{C}_{HWB}^{(6)})^2$ & -0.081/-0.20 & 0.017/-0.017 & 0.23/0.49 & \\
$\tilde{C}_{HD}^{(6)} \, \tilde{C}_{H\psi}^{3,(6)}$ &-0.088/0.19 &-0.072/-0.19 & 0.33/-0.19 & 0.19/0.19 \\
$\tilde{C}_{H\psi,2}^{(8)}$ & -0.19/-0.19 & 0.19/0.19 & 0.19/0.19 & -0.19/-0.19\\
$(\tilde{C}_{H\psi}^{3,(6)})^2$ & -0.26/0.11 & 0.22/-0.026 & 0.33/-0.14 & -0.19/-0.19\\
$\tilde{C}_{H\psi}^{(6)}\, \tilde{C}_{H\psi}^{3,(6)}$ & 0.53/-0.22 & 0.43/-0.052 & 0.67/-0.29 & 0.37/0.37 \\
$\tilde{C}_{HWB}^{(6)} \, \tilde{C}_{H\psi}^{3,(6)}$ & -0.29/0.62 & 0.12/-0.49 & 0.56/-0.62 & \\
$\tilde{C}_{H\psi}^{3,(6)} \, \langle g_{\rm eff}^{\mathcal{Z},\psi}\rangle$ & -2.8/-0.042 & -2.3/-0.64 & -3.6/0.24 & -2.0/-2.0 \\
$\tilde{C}_{H\psi}^{3,(8)}$ & -0.19/-0.19 & 0.19/0.19 & 0.19/0.19 & -0.19/-0.19 \\
\hline
\end{tabular}
\caption{The effective couplings expanded to ${{\cal{O}}(v^4/\Lambda^4)}$ in each input parameter scheme. $\delta G_F^{(8)}$
and the remaining operator forms are defined in Ref.~\cite{Helset:2020yio,Hays:2020scx}.
$\langle g_{\rm eff,pp}^{\mathcal{Z},\psi}\rangle$ is understood to be $\langle g_{\rm eff,pp}^{\mathcal{Z},\psi}\rangle_{\cal{O}}(v^2/\Lambda^2)$
in the left most column. Reported is the numerical coefficient multiplying each SMEFT correction.  $p=\{1,2,3\}$
is a flavour index. We have eliminated $\delta G_F^{(6)}$ in favor of introducing $\langle g_{\rm eff,pp}^{\mathcal{Z},\psi}\rangle_{\mathcal{O}(v^2/\Lambda^2)}^2$
in these expressions.}\label{effectivecoupling3}
\end{center}
\end{table}

\subsection{Expansion of $\bar{R}_\ell$}
We illustrate the use of the formula we present defining EWPD to ${\cal{O}}(v^4/\Lambda^4)$
using the example of $\bar{R}_\ell$. First we use the result for the expansion of this observable to second order
\bea
	\frac{\bar{R}^{\textrm{SMEFT}}_\ell}{\hat{R}^{\textrm{SM}}_\ell} &=&
	\frac{(\hat{\Gamma}_{had}/\hat{\Gamma}_\ell)^{\textrm{SMEFT}}}{(\hat{\Gamma}_{had}/\hat{\Gamma}_{\ell})^{\textrm{SM}}} \\
&=&
1 + \frac{\langle{\Gamma}^{\textrm{SMEFT}}_{had}\rangle_{\mathcal{O}(v^2/\Lambda^2)}}{\hat{\Gamma}_{had}^{\textrm{SM}}}
- \frac{\langle{\Gamma}^{\textrm{SMEFT}}_\ell\rangle_{\mathcal{O}(v^2/\Lambda^2)}}{\hat{\Gamma}_\ell^{\textrm{SM}}} +\left(\frac{\langle{\Gamma}^{\textrm{SMEFT}}_\ell\rangle_{\mathcal{O}(v^2/\Lambda^2)}}{\hat{\Gamma}_\ell^{\textrm{SM}}}\right)^2  \nonumber \\
&\, & + \frac{\langle{\Gamma}^{\textrm{SMEFT}}_{had}\rangle_{\mathcal{O}(v^4/\Lambda^4)}}{\hat{\Gamma}_{had}^{\textrm{SM}}}
- \frac{\langle{\Gamma}^{\textrm{SMEFT}}_\ell\rangle_{\mathcal{O}(v^4/\Lambda^4)}}{\hat{\Gamma}_\ell^{\textrm{SM}}}
-\frac{\langle{\Gamma}^{\textrm{SMEFT}}_{had}\rangle_{\mathcal{O}(v^2/\Lambda^2)} \, \langle{\Gamma}^{\textrm{SMEFT}}_\ell\rangle_{\mathcal{O}(v^2/\Lambda^2)}}{\hat{\Gamma}_{had}^{\textrm{SM}} \, \hat{\Gamma}_\ell^{\textrm{SM}}}.
 \nonumber
\eea
Then we substitute the values from Table \ref{effectivecoupling1} into this expression, using the $\{\hat{m}_W, \hat{m}_Z,\hat{G}_F\}$
input parameter scheme results, finding
\bea
	\bar{R}_\ell/\hat{R}^{\textrm{SM}}_\ell &=&
1+  \left[3.9 \langle g_{eff}^{\mathcal{Z},d_L}\rangle - 0.69 \langle g_{eff}^{\mathcal{Z},d_R}\rangle - 5.9 \langle g_{eff}^{\mathcal{Z},\ell_L}\rangle
+ 4.8 \langle g_{eff}^{\mathcal{Z},\ell_R}\rangle - 2.2 \langle g_{eff}^{\mathcal{Z},u_L}\rangle + 0.92 \langle g_{eff}^{\mathcal{Z},u_R}\rangle \right] \nonumber \\
&+& \left[3.9 \langle g_{eff}^{\mathcal{Z},d_L}\rangle -0.69 \langle g_{eff}^{\mathcal{Z},d_R}\rangle -5.9 \langle g_{eff}^{\mathcal{Z},\ell_L}\rangle
+ 4.8 \langle g_{eff}^{\mathcal{Z},\ell_R}\rangle -2.2 \langle g_{eff}^{\mathcal{Z},u_L}\rangle + 0.92 \langle g_{eff}^{\mathcal{Z},u_R}\rangle \right]_{\mathcal{O}(v^4/\Lambda^4)}  \nonumber \\
&+& \left[6.2 \langle g_{eff}^{\mathcal{Z},d_L}\rangle^2 +6.2 \langle g_{eff}^{\mathcal{Z},d_R}\rangle^2 +21 \langle g_{eff}^{\mathcal{Z},\ell_L}\rangle^2
+8.2 \langle g_{eff}^{\mathcal{Z},\ell_R}\rangle^2 +4.2 \langle g_{eff}^{\mathcal{Z},u_L}\rangle^2 +4.2 \langle g_{eff}^{\mathcal{Z},u_R}\rangle^2 \right] \nonumber \\
&+& \left[-23 \langle g_{eff}^{\mathcal{Z},d_L}\rangle \,  \langle g_{eff}^{\mathcal{Z},\ell_L}\rangle
+19 \langle g_{eff}^{\mathcal{Z},d_L}\rangle \,  \langle g_{eff}^{\mathcal{Z},\ell_R}\rangle
+ 4.1 \langle g_{eff}^{\mathcal{Z},d_R}\rangle \,  \langle g_{eff}^{\mathcal{Z},\ell_L}\rangle
-3.3 \langle g_{eff}^{\mathcal{Z},d_R}\rangle \,  \langle g_{eff}^{\mathcal{Z},\ell_R}\rangle\right]  \nn
&+&\left[- 56 \langle g_{eff}^{\mathcal{Z},\ell_L}\rangle \,  \langle g_{eff}^{\mathcal{Z},\ell_R}\rangle
+ 13 \langle g_{eff}^{\mathcal{Z},\ell_L}\rangle \,  \langle g_{eff}^{\mathcal{Z},u_L}\rangle
- 5.4 \langle g_{eff}^{\mathcal{Z},\ell_L}\rangle \,  \langle g_{eff}^{\mathcal{Z},u_R}\rangle
- 10 \langle g_{eff}^{\mathcal{Z},\ell_R}\rangle \,  \langle g_{eff}^{\mathcal{Z},u_L}\rangle  \right]  \nn
&+&\left[4.4 \langle g_{eff}^{\mathcal{Z},\ell_R}\rangle \,  \langle g_{eff}^{\mathcal{Z},u_R}\rangle\right]. \label{exampleRell}
\eea
All of the $g_{eff}$
appearing in Eq.~\eqref{exampleRell}, excepting those on the second line, are
$\langle g_{eff}^{\mathcal{Z},\psi}\rangle_{\mathcal{O}(v^2/\Lambda^2)}$. Those on the second line
are $\langle g_{eff}^{\mathcal{Z},\psi}\rangle_{\mathcal{O}(v^4/\Lambda^4)}$, as indicated.

The structure of the corrections in  Eq.~\eqref{exampleRell} in the $\mathcal{O}(v^2/\Lambda^2)$ expansion
makes a number of points.
The numerical coefficients of the corrections in the first and second line are identical. This is due to both of these terms
coming about due to linear interference with the SM amplitude, and is consistent with a naive expectation
that corrections in the SMEFT follow a numerical pattern of the form $n \left(x + x^2 + \cdots \right)$,
with $n$ a numerical coefficient, and $x$ a power counting expansion. The first three lines
follow the pattern expected from chiral symmetry, as we are neglecting light fermion masses.
The full
$\mathcal{O}(v^4/\Lambda^4)$ result includes the last four lines.
Note the fact that $g_{eff}^{\psi_L} \times g_{eff}^{\psi'_R}$ interference terms are present in the full result
and this is not inconsistent with chiral symmetry at second order in the SMEFT expansion
even though we are neglecting light quark masses. Here ratios of observables
are considered.

The SMEFT is useful so long as there is at least one small power counting expansion parameter. For EWPD observables, this condition is $ v/\Lambda \ll 1$.
All of the $\mathcal{O}(v^4/\Lambda^4)$ corrections, beyond those in the first line of Eq.~\eqref{exampleRell} are further suppressed.
This numerical factor is absorbed into the presentation of the results via the notation
$\tilde{C}^{(d)}_i \equiv C^{(d)}_i \bar{v}_T^{d-4}/\Lambda^{d-4}$.
Typically, due to the constraints of direct searches, a working hypothesis is $ v/\Lambda \lesssim 0.1$.
Then the second to sixth lines are suppressed, and expected to be percent level corrections to the leading perturbation.
Even so, accidental numerical enhancements occur. This occurs in this observable, note the coefficient
of $56$ for $\langle g_{eff}^{\mathcal{Z},\ell_L}\rangle \,  \langle g_{eff}^{\mathcal{Z},\ell_R}\rangle$.
Although we have shown explicit results for the $\{\hat{m}_W, \hat{m}_Z,\hat{G}_F\}$ scheme, these
points all hold for the $\{\hat{\alpha}_{ew}, \hat{m}_Z,\hat{G}_F\}$ scheme as well.

Fundamentally, scheme dependence is very significant in the SMEFT.
Expanding the effective couplings in terms of the individual Wilson coefficients
and $\delta G_F^{(6)}$ can be done using Tables~\ref{effectivecouplingtoWC},~\ref{effectivecoupling2},~and~\ref{effectivecoupling3}.
The resulting expressions for EWPD are not transparent in interpretation and are lengthy.
The dimension-eight terms are expected to be suppressed by $\lesssim 10^{-2}$ compared to
the dimension-six terms due to the power counting expansion.
On the other hand, the calculable numerical coefficients of dimension-eight terms
compared to dimension-six terms are $ \sim 10^2$ in some cases.
In $\bar{R}_\ell/\hat{R}^{\textrm{SM}}_\ell$, this is reflective of the numerical accident in the enhanced coefficient of
$\langle g_{eff}^{\mathcal{Z},\ell_L}\rangle \,  \langle g_{eff}^{\mathcal{Z},\ell_R}\rangle$.
It does not follow that EWPD offers no constraint on the SMEFT parameter space. Such numerical accidents
in one observable are also expected to be less relevant once multiple measurements are combined in the SMEFT.
This observation does encourage reasonable caution on over-interpreting LEP constraints on $\mathcal{L}^{(6)}$
Wilson coefficients in naive leading-order analyses of LEP data.

\section{SMEFT bottom up}\label{bottomup}

To visually illustrate the $\mathcal O(v^4/\Lambda^4)$ effects, we need to
assign numbers for the unknown Wilson coefficients. Such a numerical
output requires a scheme for numerical inputs. This is true if constraints on a UV model are
studied through its matching to the SMEFT in a global analysis, or if the
SMEFT is studied bottom up as a model-independent EFT. In the latter case, a rough estimate
of the impact of these effects is developed
in this section.

\begin{figure}
\includegraphics[height=0.24\textheight,width=0.45\textwidth]{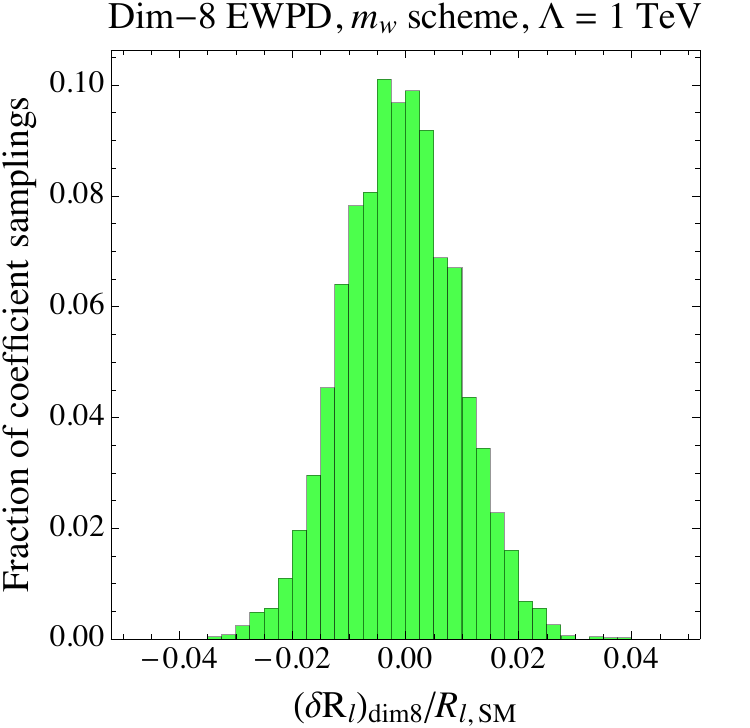}
\includegraphics[height=0.24\textheight,width=0.45\textwidth]{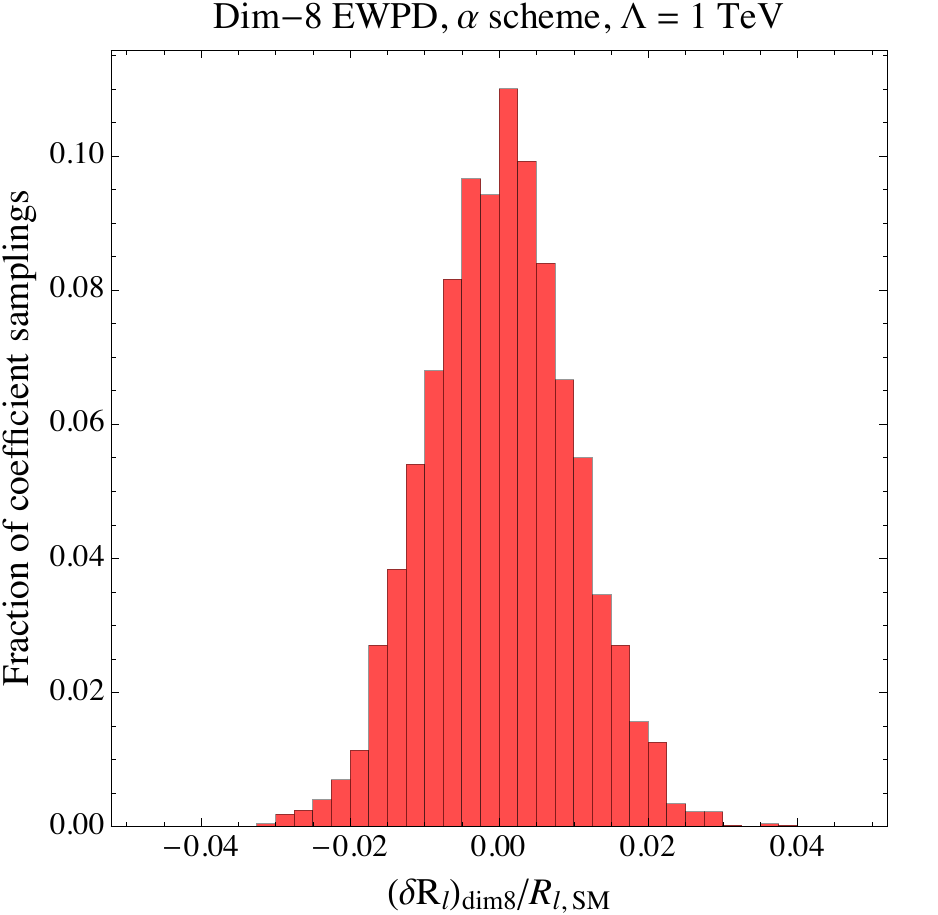}
\includegraphics[height=0.24\textheight,width=0.45\textwidth]{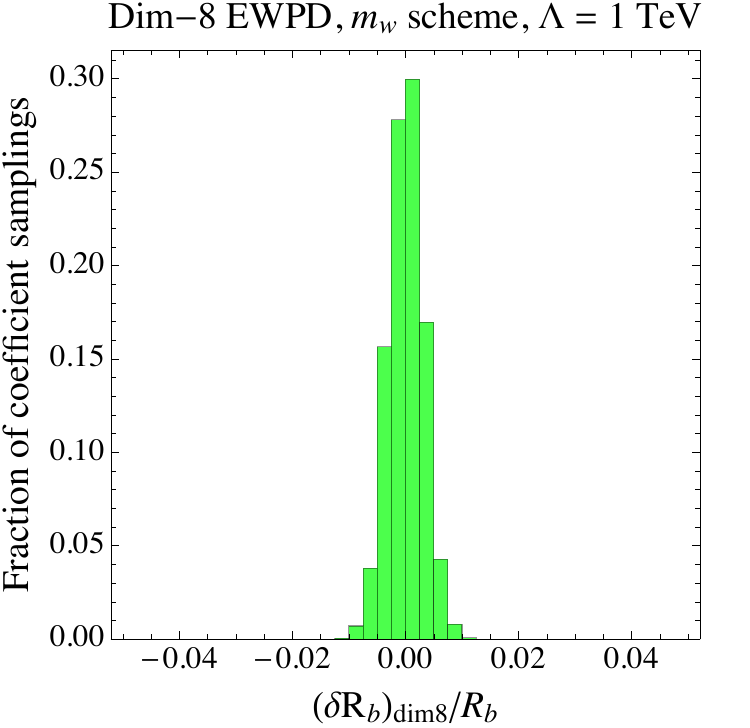}
\includegraphics[height=0.24\textheight,width=0.45\textwidth]{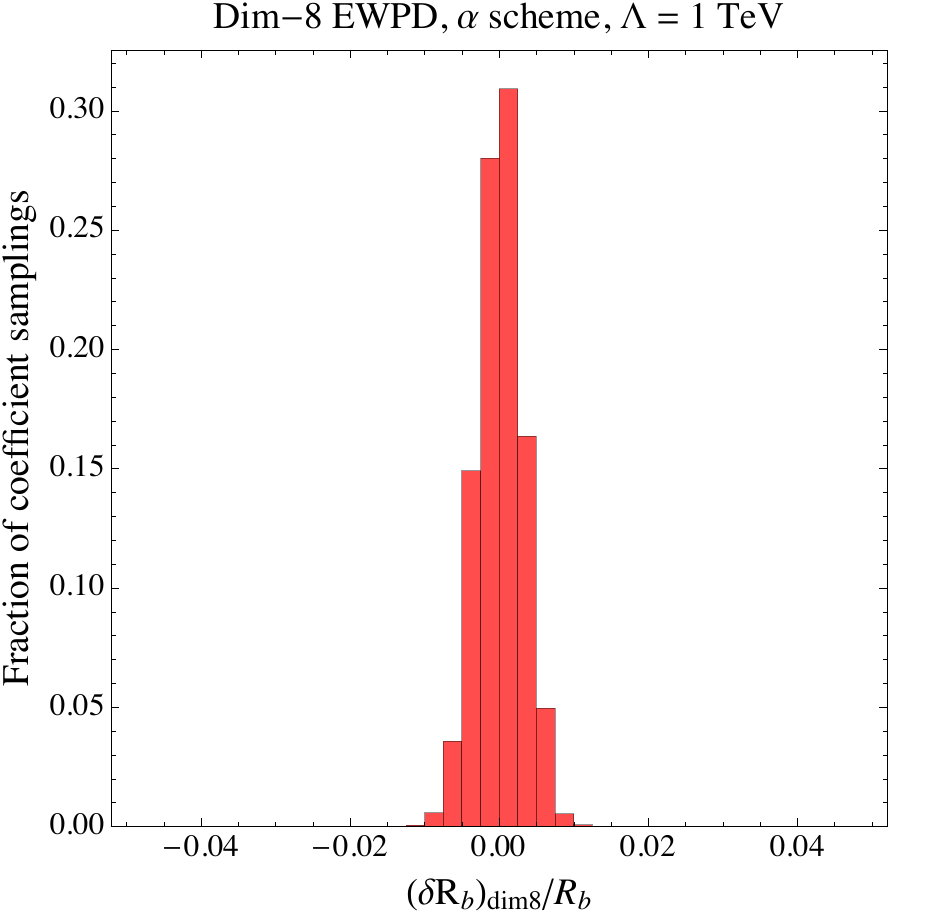}
\includegraphics[height=0.24\textheight,width=0.45\textwidth]{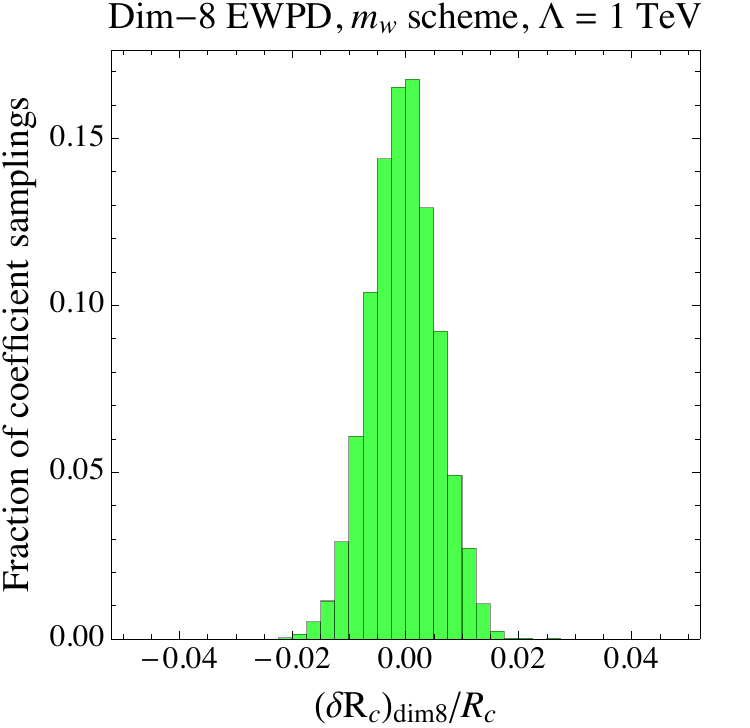}
\includegraphics[height=0.24\textheight,width=0.45\textwidth]{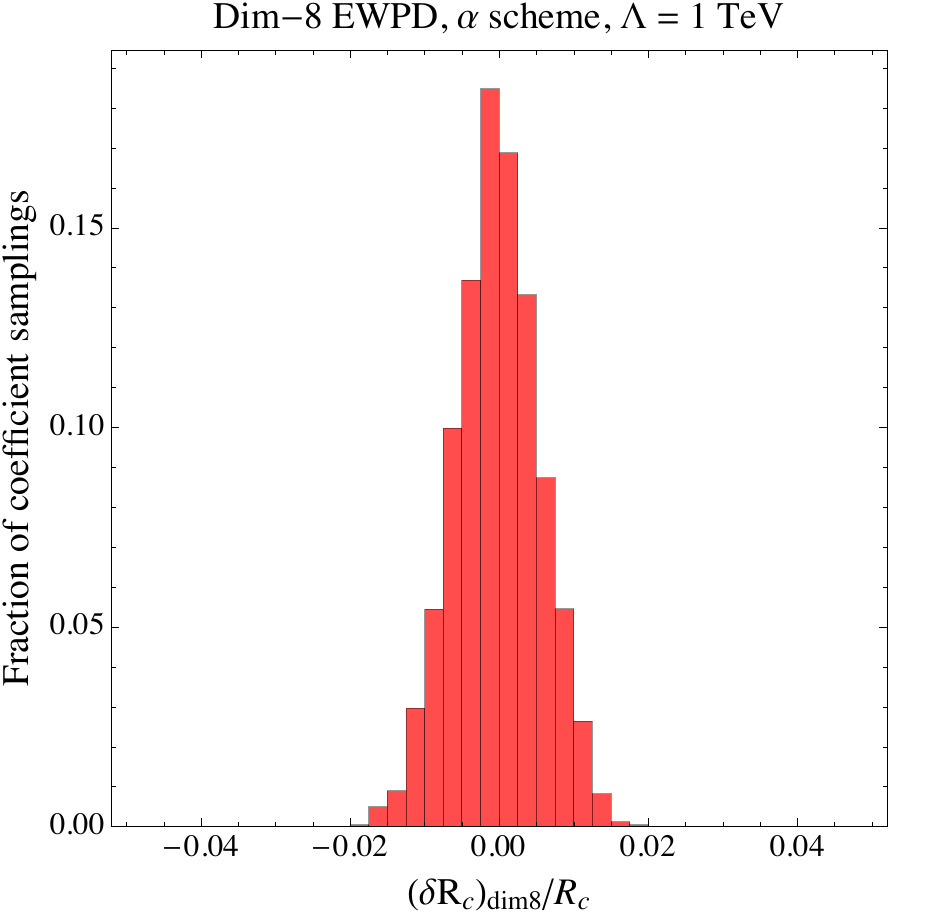}
\caption{Contributions to the $R_\ell, R_b$ and $R_c$ EWPD from $\mathcal{L}^{(8)}$ operators relative to the SM value. Here $\Lambda = 1\, $TeV. The histograms are formed by selecting random values for the coefficients 5000 times following the scheme described in the text. }
\label{fig:bottomupfig1}
\end{figure}

As a first example, we calculate the $\mathcal{L}^{(8)}$
contributions to each EWPD observable relative to the SM EWPD values, $\delta \mathcal O_{i, \rm{dim-8}}/\mathcal O_{i,\textrm{SM}}$.
The $\mathcal{L}^{(8)}$ EWPD contributions are a function of $\bar v_T/\Lambda$ and the $\tilde{C}_i^{(8)}$.
After choosing a $\Lambda$, we select values for the coefficients using the same scheme as in Ref.~\cite{Hays:2020scx}. Specifically, we draw random
coefficient values according to gaussian distributions with zero mean and root mean square equal to 1
for `tree-level' Wilson coefficients and 0.01 for `loop-level' Wilson coefficients
as classified by Refs.~\cite{Arzt:1994gp, Jenkins:2013fya, Craig:2019wmo}. Selecting $\Lambda = 1$ TeV,
the results for the partial width ratios $R_\ell ,R_b, R_c$ for both input schemes and 5000
random coefficient selections are shown in Fig.~\ref{fig:bottomupfig1}; the asymmetries $A_{FB}^\ell$
and $A_{FB}^c$ are show in Fig.~\ref{fig:bottomupfig2}, and the remaining EWPD observables
are shown in Appendix~\ref{app:moreplots}.

\begin{figure}
\includegraphics[height=0.24\textheight,width=0.45\textwidth]{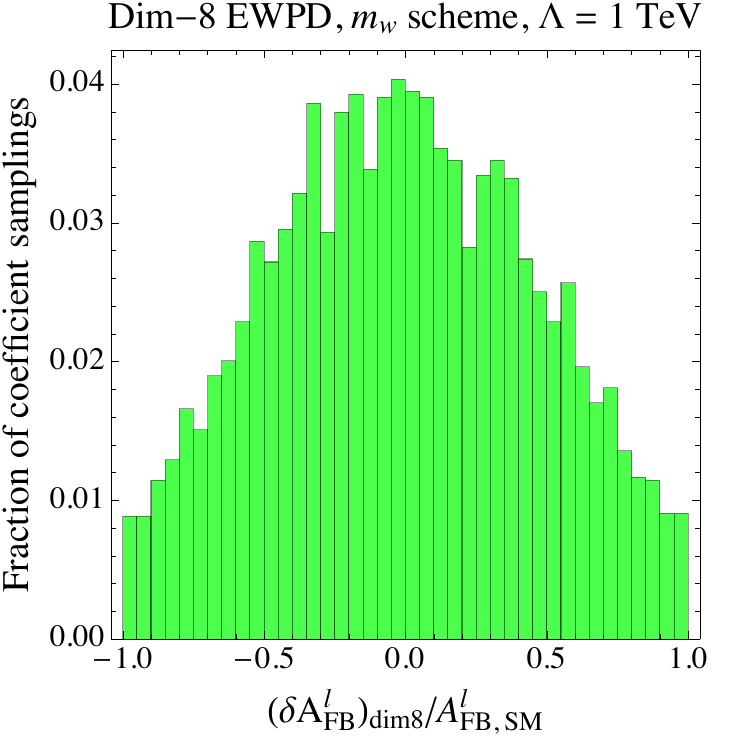}
\includegraphics[height=0.24\textheight,width=0.45\textwidth]{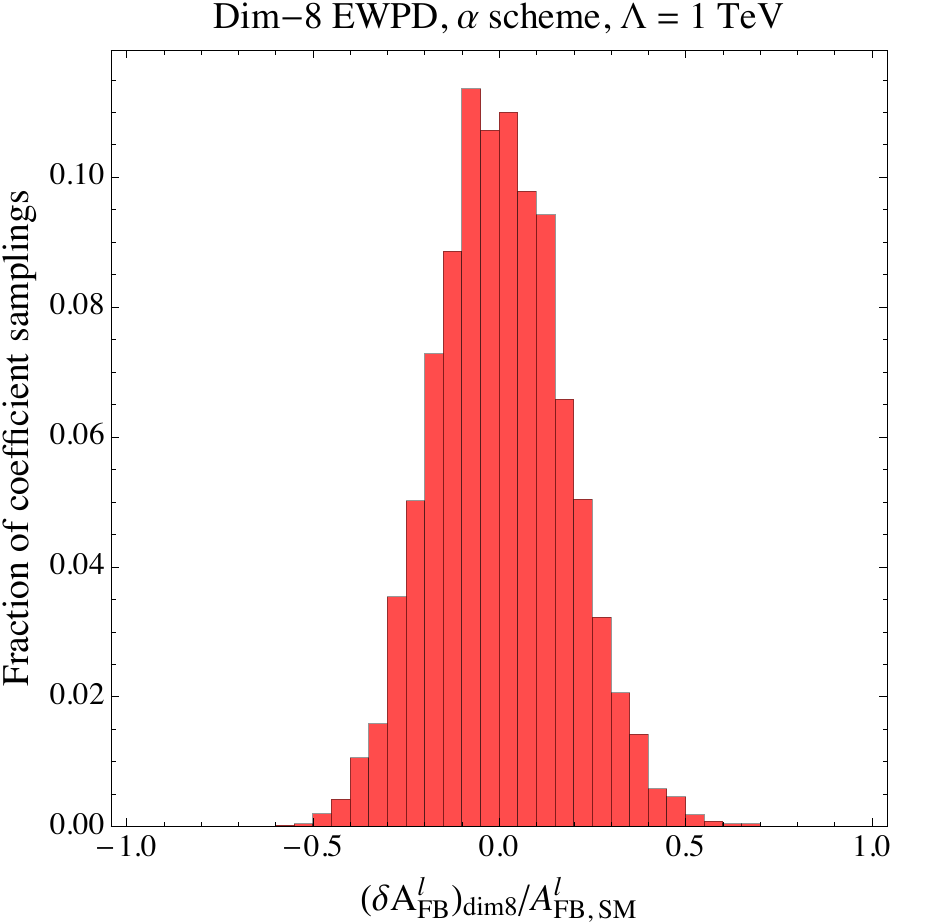}
\includegraphics[height=0.24\textheight,width=0.45\textwidth]{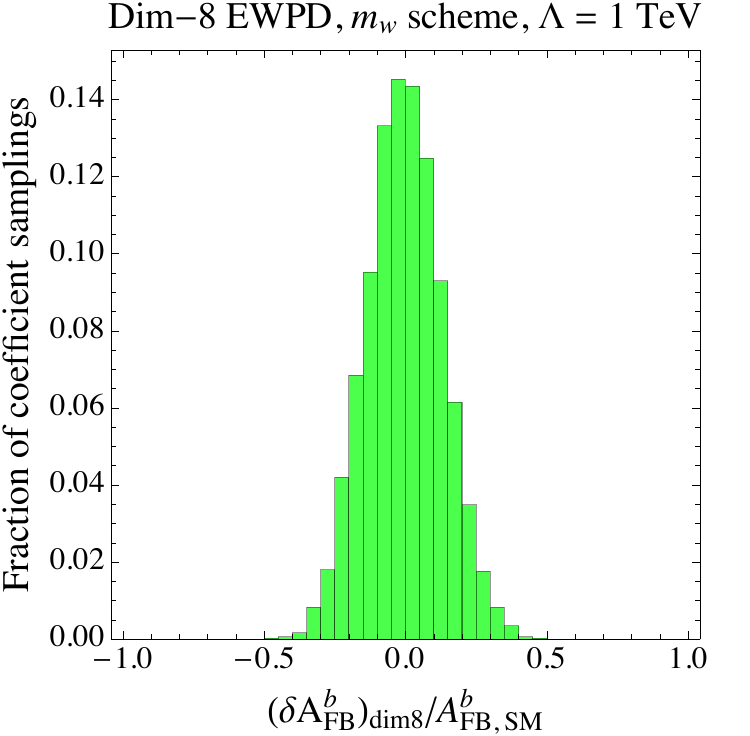}
\includegraphics[height=0.24\textheight,width=0.45\textwidth]{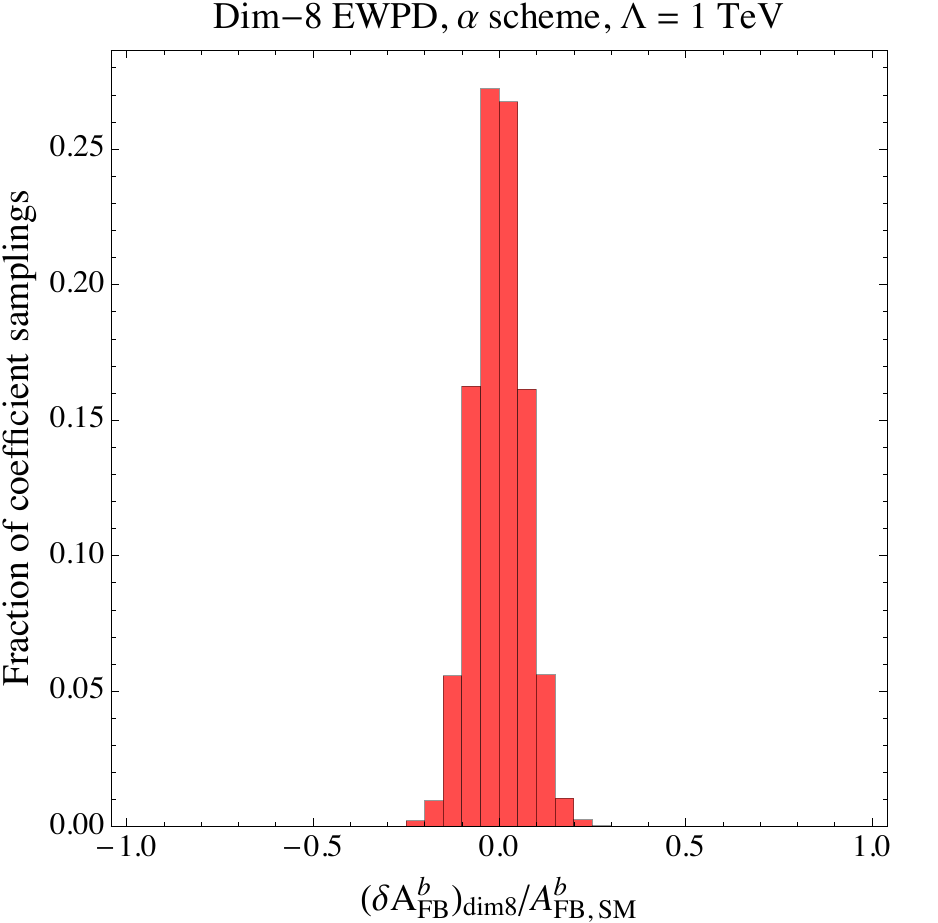}
\includegraphics[height=0.24\textheight,width=0.45\textwidth]{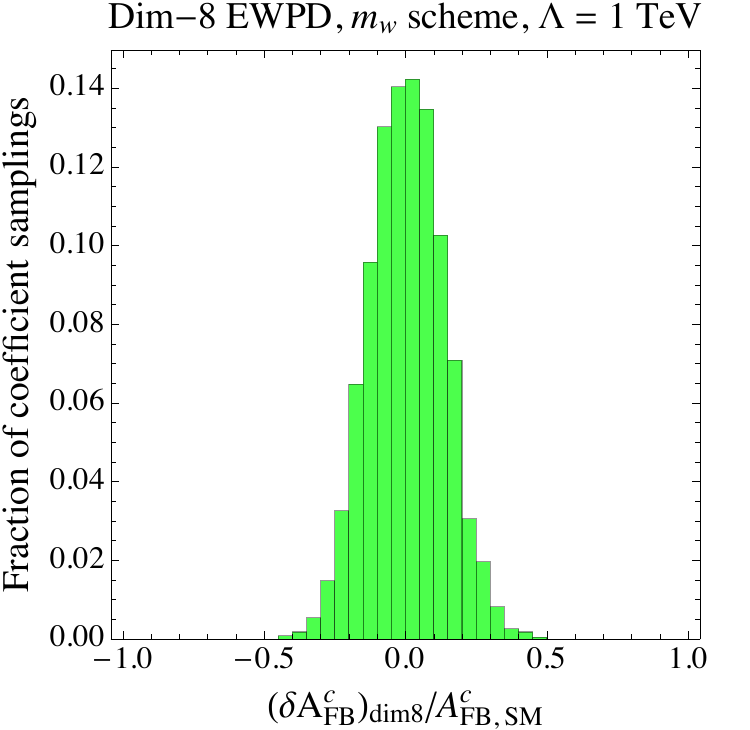}
\includegraphics[height=0.24\textheight,width=0.45\textwidth]{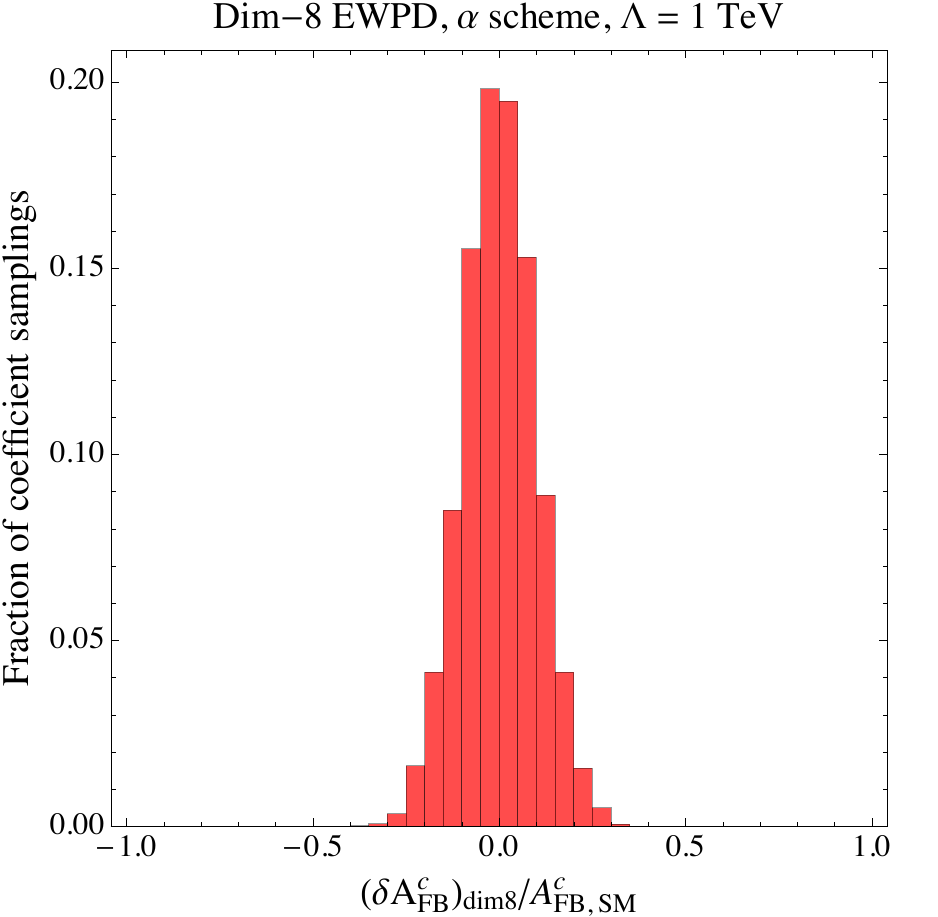}
\caption{Contributions to the $A_{FB}^\ell$ and $A_{FB}^c$ EWPD from $\mathcal{L}^{(8)}$ operators relative to the SM value.  Here $\Lambda = 1\, $TeV. The histograms are formed by selecting random values for the coefficients 5000 times following the scheme described in the text. }
\label{fig:bottomupfig2}
\end{figure}

The resulting distributions have widths of roughly $0.01$ for the partial width ratios and $0.2$ for the asymmetries, with the leptonic versions of both variables being slightly wider. The shapes are not artifacts of our gaussian sampling procedure, as sampling with a flat prior yields qualitatively similar results. For higher choices of $\Lambda$, the distributions narrow, given that the $\mathcal{L}^{(8)}$ contribution scales as $\bar v^4_T/\Lambda^4$.\footnote{The same narrowing/widening would occur if we fixed $\Lambda$ and drew the dimension eight8 coefficients from a thinner/fatter distribution, given that the combination $C^{(8)}_i/\Lambda^4$ is what appears in all observables. By this logic, the distributions for $C^{(8)} \sim 1/0.01$ and $\Lambda = 1$ TeV are the same as $C^{(8)} \sim 0.06/6\times 10^{-4}, \Lambda = 0.5$ TeV or $\sim 16/0.16, \Lambda = 2$ TeV. } While the numerical results shown are specific to how we chose dimension-eight coefficients, other coefficient choices can easily be tested using the formulae in Sec.~\ref{ewpdresults}. Finally, it is important to remember that the $\mathcal{L}^{(8)}$ terms are only a portion of the $\mathcal O(v^4/\Lambda^4)$ contribution. However, if we repeat the simple calculation above with both $\mathcal{L}^{(6)}$ and $\mathcal{L}^{(8)}$ operators, the $\mathcal O(v^4/\Lambda^4)$ effects are correlated with the $\mathcal O(v^2/\Lambda^2)$ effects. In order to make sure we are not biased by scenarios with $\mathcal O(v^2/\Lambda^2)$ effects that are experimentally excluded, a more careful calculation is necessary.

As a second example, we will zero all $\mathcal{L}^{(6)}$ coefficients except for two, then calculate the $\chi^2$ bounds on that 2-d coefficient space with and without $\mathcal O(v^4/\Lambda^4)$ effects. Said more plainly, we want to see how the S-T
analysis \cite{Kennedy:1988sn,Altarelli:1990zd,Golden:1990ig,Holdom:1990tc,Peskin:1990zt,Peskin:1991sw,Maksymyk:1993zm,Burgess:1993mg,Burgess:1993vc,Bamert:1996px} (and other, less famous 2-d slices of coefficient space) fare at $\mathcal O(v^4/\Lambda^4)$.

Using the expressions in Sec.~\ref{ewpdresults}, we form the full $\chi^2$ following the procedure laid out in Refs.~\cite{Berthier:2015oma, Berthier:2015gja} for $\mathcal O(v^2/\Lambda^2)$ SMEFT; we take the experimental correlation matrix from Ref.~\cite{Z-Pole} and assume theoretical uncertainties are completely uncorrelated. We also calculate the $\chi^2$ using EWPD observables calculated to $\mathcal O(v^2/\Lambda^2)$ only.

Next, we zero all $\mathcal{L}^{(6)}$ Wilson coefficients except two: our first choice
is to zero all coefficients but $C_{HWB}$ and $C_{HD}$ -- S and T up to normalization factors,
while for our second choice we zero all but $C^{(6)}_{H\ell}$ and $C_{HD}$. $C^{(6)}_{H\ell}$ affects
the coupling of $\mathcal{Z}$ to leptons and is therefore assumed to be tightly constrained by LEP, hence it is an interesting candidate to study including $\mathcal O(v^4/\Lambda^4)$ effects. At this point, $\chi^2_{\mathcal O(v^2/\Lambda^2)}$ is a function of the two nonzero coefficients
and the scale $\Lambda$, while $\chi^2_{\mathcal O(v^4/\Lambda^4)}$ also depends
on the $\mathcal{L}^{(8)}$ Wilson coefficients. Rather than rely on random coefficients,
we adopt a simpler approach here -- setting all tree-level $\mathcal{L}^{(8)}$ Wilson coefficients
to 1 and all loop-level to 0.01.

\begin{figure}
\includegraphics[height=0.24\textheight,width=0.45\textwidth]{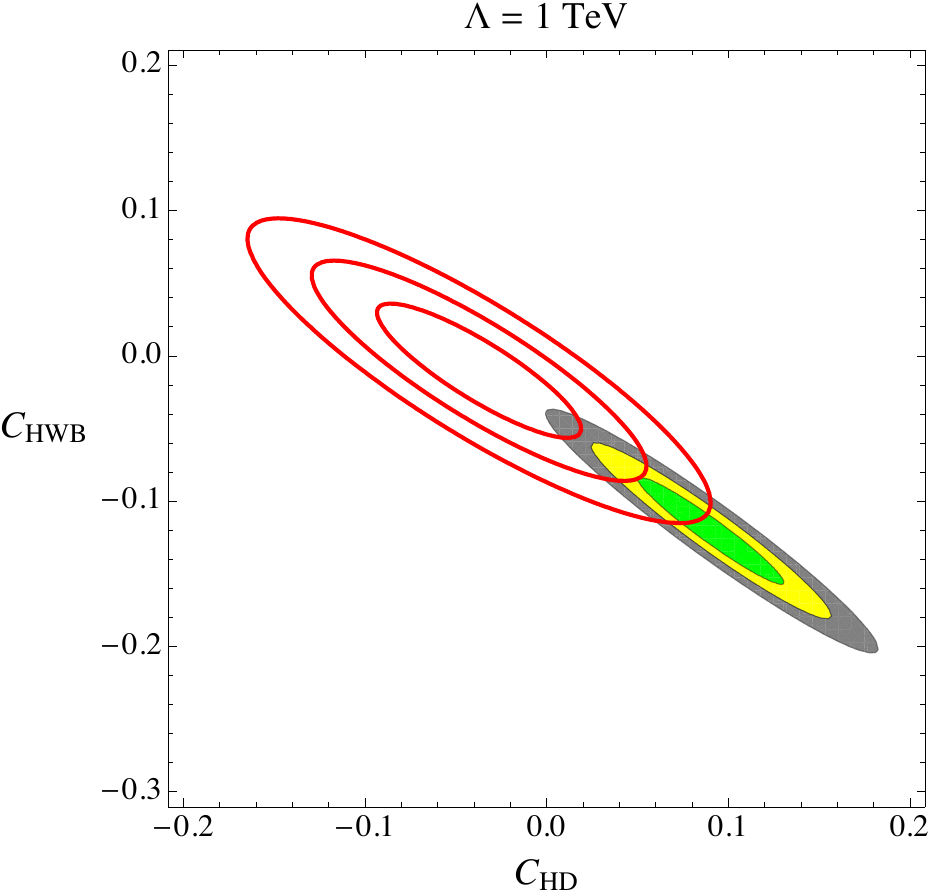}
\includegraphics[height=0.24\textheight,width=0.45\textwidth]{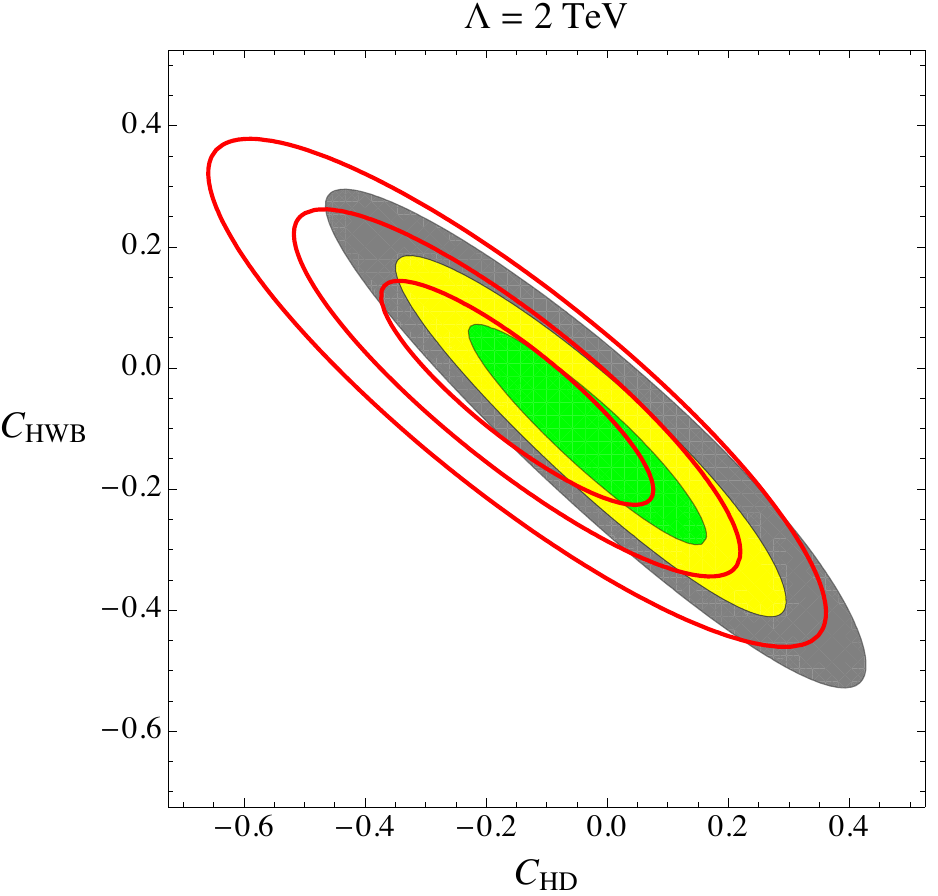}
\includegraphics[height=0.24\textheight,width=0.45\textwidth]{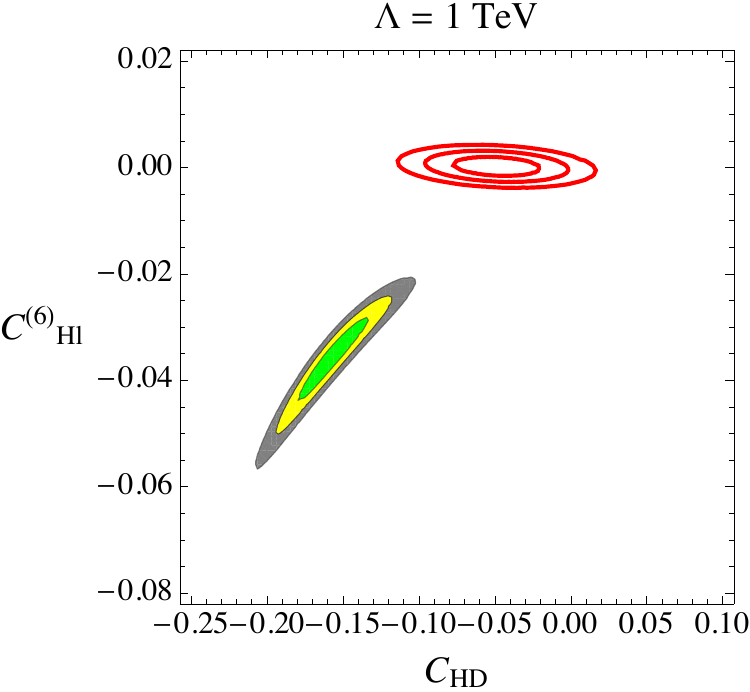}
\includegraphics[height=0.24\textheight,width=0.45\textwidth]{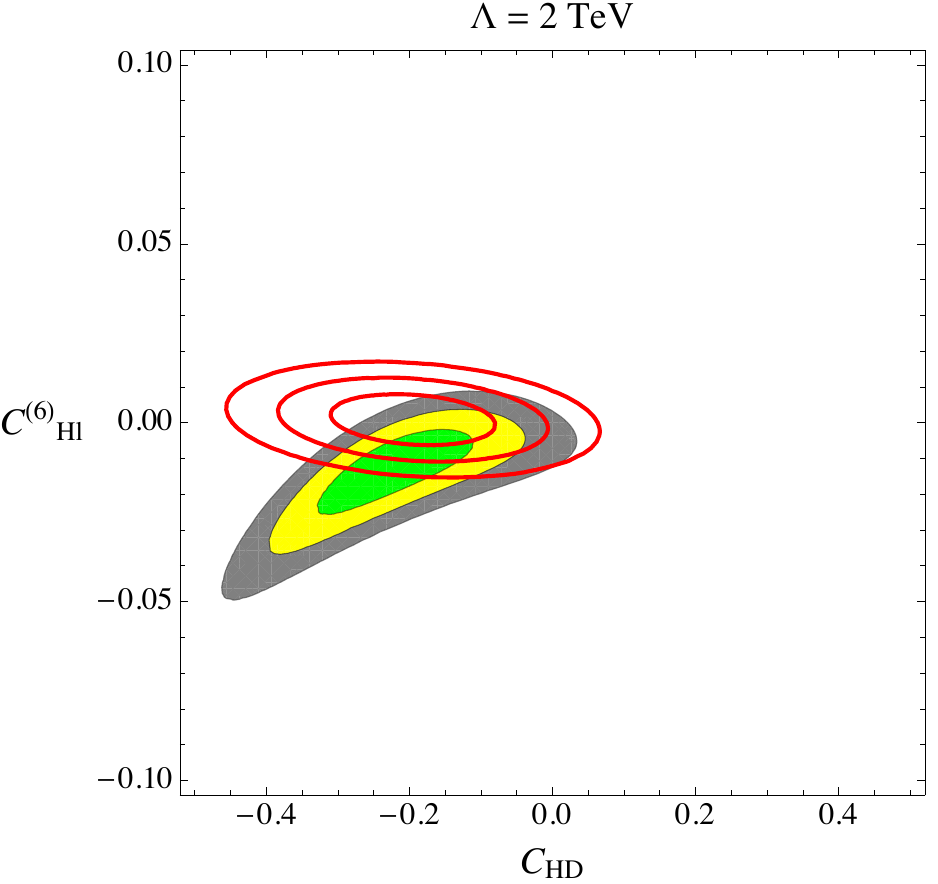}
\caption{The green/yellow/gray contours correspond to the $68\%/95\%/99.9\%$ CL two parameter fit determined by $\Delta \chi^2_{\mathcal O(v^4/\Lambda^4)}$, while the red rings correspond to the same CL determined using $\Delta \chi^2_{\mathcal O(v^2/\Lambda^2)}$. In the top panels the free parameters are $C_{HD}$ and $C_{HWB}$, while in the bottom panels the free parameters are $C_{HD}$ and $C^{(6)}_{H\ell}$. Note that the axes ranges vary from panel to panel. In the left panels, we have taken the scale $\Lambda = 1$ TeV, while in the right panels $\Lambda = 2$ TeV. All calculations use the $\hat m_W$ scheme. }
\label{fig:deltachisq}
\end{figure}

For a given $\Lambda$, we determine the minima of $\chi^2_{\mathcal O(v^2/\Lambda^2)}$ and
$\chi^2_{\mathcal O(v^4/\Lambda^4)}$ and plot the $\Delta \chi^2$  contours in Fig.~\ref{fig:deltachisq}. The green, yellow, grey regions correspond to the $68 \%, 95\%$ and $99.9 \%$ CL
regions for a two parameter fit around the minimum
of $\chi^2_{\mathcal O(v^4/\Lambda^4)}$.
The regions correspond to $\chi^2 = \chi^2_{min}+ \Delta \chi^2$
with $\Delta \chi^2 =  2.30$ ($1 \sigma$, green), $6.18$ ($2 \sigma$,yellow), $11.83$ ($3 \sigma$, grey)
defined via the Cumulative Distribution function for a two-parameter fit. The same $\Delta \chi^2$ regions are shown in red for the fit using $\chi^2_{\mathcal O(v^2/\Lambda^2)}$ (inner contour is $68\%$ CL, intermediate is $95\%$ and the outer is $99.9\%$ CL).

The difference between the contours shows the effect of going from $\mathcal O(v^2/\Lambda^2)$ to $\mathcal O(v^4/\Lambda^4)$. For $\Lambda = 1$ TeV, the shift is striking, to the extent that the different order contours don't even overlap for the $C_{HD} - C^{(6)}_{H\ell}$ case. The effect is smaller for $\Lambda = 2$ TeV, unsurprising given that the difference between the contours scales as  $v^4/\Lambda^4$, but it is not negligible.  Of course, the details of how the fit shifts depends strongly on our treatment of the dimension-eight operators, so these results should be viewed as qualitative. However, we emphasize that the choice of $1/0.01$ for tree/loop level dimension-eight coefficients was made for simplicity and not to amplify the effect. Repeating the study using random coefficients for the dimension-eight coefficients (following the procedure used earlier in the section), we observe a wide range of shifts, from significantly smaller to significantly larger than what is shown in Fig.~\ref{fig:deltachisq}.

\section{Explicit matching examples}\label{matchingsection}
\begin{table}
  \begin{center}
	  \caption{$\mathcal{L}^{(6)}$ matching coefficients in the U(1) model \cite{Hays:2020scx},
    here $b_1 = k^2 - 2\lambda \, (k^2-k^4) \, \frac{\bar{v}_T^2}{m_K^2}$. Our notation is such that $\hyp_i$ is the hypercharge of
    field $i$.}
    \label{tab:dim6Matching}
\begin{tabular}{l|c}
      \multicolumn{2}{c}{$H^2\psi^2 D$}   \\
      \toprule
      $C_{H\ell}^{1,(6)}$ & $-\frac{\hyp_\ell g_1^2}{2m_K^2}b_1$ \\
      \midrule
      $C_{He}^{(6)}$ & $-\frac{\hyp_e g_1^2}{2m_K^2}b_1$ \\
      \midrule
      $C_{Hq}^{1,(6)}$ & $-\frac{\hyp_q g_1^2}{2m_K^2}b_1$ \\
      \midrule
      $C_{Hu}^{(6)}$ & $-\frac{\hyp_u g_1^2}{2m_K^2}b_1$ \\
      \midrule
      $C_{Hd}^{(6)}$ & $-\frac{\hyp_d g_1^2}{2m_K^2}b_1$ \\
      \bottomrule
    \end{tabular}
    \quad
	  \begin{tabular}{l|c}
      \multicolumn{2}{c}{$H^4 D^2$}   \\
      \toprule
      $C_{H\Box}^{(6)}$ & $-\frac{g_1^2 k^2}{8m_K^2}$ \\
      \midrule
      $C_{HD}^{(6)}$ &  $-\frac{g_1^2 k^2}{2m_K^2}$ \\
      \bottomrule
    \multicolumn{2}{c}{}   \\
    \multicolumn{2}{c}{$\psi^4: (\bar{L}L)(\bar{L}L)$}   \\
      \toprule
      $C_{\ell\ell}^{(6)}$ & $-\frac{1}{8}\frac{g_1^2 k^2}{m_K^2}$ \\
      \midrule
      $C_{qq}^{1,(6)}$ & $-\frac{1}{72}\frac{g_1^2 k^2}{m_K^2}$ \\
      \midrule
      $C_{\ell q}^{1,(6)}$ &  $\frac{1}{12}\frac{g_1^2 k^2}{m_K^2}$ \\
      \bottomrule
    \end{tabular}
   	\quad
    \begin{tabular}{l|c}
      \multicolumn{2}{c}{$ \psi^4: (\bar{R}R)(\bar{R}R)$}   \\
      \toprule
      $C_{ee}^{(6)}$ & $-\frac{1}{2}\frac{g_1^2 k^2}{m_K^2}$ \\
      \midrule
      $C_{uu}^{(6)}$ &  $-\frac{2}{9}\frac{g_1^2 k^2}{m_K^2}$ \\
          \midrule
      $C_{dd}^{(6)}$ &  $-\frac{1}{18}\frac{g_1^2 k^2}{m_K^2}$ \\
      \midrule
      $C_{eu}^{(6)}$ &  $\frac{2}{3}\frac{g_1^2 k^2}{m_K^2}$ \\
      \midrule
      $C_{ed}^{(6)}$ &  $-\frac{1}{3}\frac{g_1^2 k^2}{m_K^2}$ \\
      \midrule
      $C_{ud}^{1,(6)}$ &  $\frac{2}{9}\frac{g_1^2 k^2}{m_K^2}$ \\
      \bottomrule
    \end{tabular}
\quad
    \begin{tabular}{l|c}
      \multicolumn{2}{c}{$ \psi^4: (\bar{L}L)(\bar{R}R)$}   \\
      \toprule
      $C_{\ell e}^{(6)}$ & $-\frac{1}{2}\frac{g_1^2 k^2}{m_K^2}$ \\
      \midrule
      $C_{\ell u}^{(6)}$ &  $\frac{1}{3}\frac{g_1^2 k^2}{m_K^2}$ \\
          \midrule
      $C_{\ell d}^{(6)}$ &  $-\frac{1}{6}\frac{g_1^2 k^2}{m_K^2}$ \\
      \midrule
      $C_{qe}^{(6)}$ &  $\frac{1}{6}\frac{g_1^2 k^2}{m_K^2}$ \\
      \midrule
      $C_{qu}^{1,(6)}$ &  $-\frac{1}{9}\frac{g_1^2 k^2}{m_K^2}$ \\
      \midrule
      $C_{qd}^{1,(6)}$ &  $\frac{1}{18}\frac{g_1^2 k^2}{m_K^2}$ \\
      \bottomrule
    \end{tabular}
  \end{center}
\end{table}
The expressions for EWPD to dimension eight in the SMEFT are lengthy. While a bottom up
analysis in the SMEFT is reported in Section \ref{bottomup}, it is also useful to examine
some cases where models are matched to dimension eight, and EWPD constraints
are studied. Restricting to UV models with few parameters
allows the results to be visually represented. In the following sections we explore two such UV models -- the $\rm U(1)$ model developed to dimension eight in matching
in Ref.~\cite{Hays:2020scx}, and a model containing a scalar triplet. The details of matching the triplet model to dimension eight can be found in Appendix~\ref{app:matching}.

\subsection{$\rm U(1)$ kinetic mixing}\label{sec:UVmodel}
In this model, a heavy U(1) gauge boson $K_\mu$ with Stueckelberg mass \cite{Stueckelberg:1900zz} $m_K$ kinetically mixes with $B_\mu$,
the U(1)$_{\rm Y}$ gauge boson in the SM.
The SM Lagrangian is extended with the UV Lagrangian
\begin{align}
	\Delta \Lagr = -\frac{1}{4} K_{\mu\nu} K^{\mu\nu} + \frac{1}{2}m_K^2 K_\mu K^\mu
	- \frac{k}{2} B^{\mu\nu}K_{\mu\nu},
\end{align}
where the field strength is $K_{\mu\nu} = \partial_\mu K_\nu - \partial_\nu K_\mu$.
Integrating out the heavy $K^{\mu}$ field, the matching pattern in the SMEFT, with geoSMEFT
operator form conventions, is given in Table \ref{tab:dim6Matching}
and Table \ref{tab:dim8Matching}. This weakly coupled, renormalizable model has one scale and one coupling,
but its matching pattern does not follow the pattern claimed to follow from a UV of this form in some literature.
The matching pattern is consistent with the results of Ref.~\cite{Arzt:1994gp,Jenkins:2013fya,Craig:2019wmo,Hays:2020scx}.
\begin{table}[ht!]
  \begin{center}
	  \caption{Matching coefficients onto operators in $\mathcal{L}^{(8)}$ \cite{Hays:2020scx}. In the U(1) model,
    in addition to these matching contributions, there are four-fermion operators and four-point contributions.}
    \label{tab:dim8Matching}
\begin{tabular}{l|c}
      \multicolumn{2}{c}{$H^4\psi^2 D$}   \\
      \toprule
      $C_{H\ell}^{1,(8)}$ & $\frac{\hyp_\ell g_1^4}{4 \, m_K^4} \, k^4
			-\frac{g_1^2 \, \hyp_\ell}{m_K^4} (k^2-k^4) (2 \lambda + \frac{g_1^2+ g_2^2}{4}) $ \\
      \midrule
      $C_{He}^{1,(8)}$ & $\frac{\hyp_e g_1^4}{4 \, m_K^4} \, k^4
			-\frac{g_1^2 \, \hyp_e}{m_K^4}  (k^2-k^4) (2 \lambda + \frac{g_1^2+ g_2^2}{4}) $ \\
      \midrule
      $C_{Hq}^{1,(8)}$ & $\frac{\hyp_q g_1^4}{4 \, m_K^4} \, k^4
			-\frac{g_1^2 \, \hyp_q}{m_K^4} (k^2-k^4) (2 \lambda + \frac{g_1^2+ g_2^2}{4}) $ \\
      \midrule
      $C_{Hu}^{1,(8)}$ & 	$\frac{\hyp_u g_1^4}{4 \, m_K^4} \, k^4
			-\frac{g_1^2 \, \hyp_u}{m_K^4}  (k^2-k^4) (2 \lambda + \frac{g_1^2+ g_2^2}{4}) $ \\
      \midrule
      $C_{Hd}^{1,(8)}$ & $\frac{\hyp_d g_1^4}{4 \, m_K^4} \, k^4
			-\frac{g_1^2 \, \hyp_d}{m_K^4}  (k^2-k^4) (2 \lambda + \frac{g_1^2+ g_2^2}{4}) $ \\
			\midrule
      $C_{H\ell}^{2,(8)}$ & $-\frac{g_1^2 \, g_2^2}{16 \, m_K^4} (k^2-k^4) $ \\
			\midrule
      $C_{Hq}^{2,(8)}$ & $-\frac{g_1^2 \, g_2^2}{16 \, m_K^4} (k^2-k^4) $ \\
			\midrule
      $C_{H\ell}^{3,(8)}$ & $-\frac{g_1^2 \, g_2^2}{16 \, m_K^4} (k^2-k^4) $ \\
			\midrule
      $C_{Hq}^{3,(8)}$ & $-\frac{g_1^2 \, g_2^2}{16 \, m_K^4} (k^2-k^4) $ \\
      \bottomrule
    \end{tabular}
    \quad
	  \begin{tabular}{l|c}
      \multicolumn{2}{c}{$H^6 D^2$}   \\
      \toprule
      $C_{H,D2}^{(8)}$ & $\frac{g_1^4 \, k^4}{8 \, m_K^4} - \frac{g_1^2 \, g_2^2}{2 \, m_K^4} (k^2-k^4)$ \\
      \midrule
      $C_{HD}^{(8)}$ &  $\frac{3 \, g_1^4 \, k^4}{16 \, m_K^4} - \frac{g_1^2 \, g_2^2}{2 \, m_K^4} (k^2-k^4)$ \\
      \bottomrule
    \multicolumn{2}{c}{}   \\
    \multicolumn{2}{c}{$X^2 H^4$}   \\
      \toprule
        $C_{HB}^{(8)}$ & $- \frac{g_1^4}{16 \, m_K^4}(k^2-k^4)$ \\
				\midrule
				$C_{HW}^{(8)}$ & $ \frac{g_1^2 \,g_2^2}{16 \, m_K^4}(k^2-k^4)$ \\
      \bottomrule
    \end{tabular}
   	\end{center}
\end{table}

In the SM we have the leading-order effective couplings given in Ref.~\cite{Hays:2020scx}.
Using the results of Ref.~\cite{Hays:2020scx} and this work, the $\mathcal O(v^2/\Lambda^2)$ corrections are:
\begin{align}
\langle g_{\rm eff, pp}^{\mathcal{Z},u_R}\rangle^{[\hat{m}_{W}/\hat{\alpha}_{ew}]}_{\mathcal{O}(v^2/\Lambda^2)} &=-
[0.0017/0.019] k^2 \frac{\bar{v}_T^2}{m_K^2}
+ [0.030/0.031] \lambda \, (k^2- k^4)\frac{v^2 \, \bar{v}_T^2}{m_K^4}, \nn
\langle g_{\rm eff, pp}^{\mathcal{Z},d_R}\rangle^{[\hat{m}_{W}/\hat{\alpha}_{ew}]}_{\mathcal{O}(v^2/\Lambda^2)} &=
[0.00084/0.0095] k^2 \frac{\bar{v}_T^2}{m_K^2}
- [0.015/0.016] \lambda \,  (k^2- k^4)\frac{v^2 \, \bar{v}_T^2}{m_K^4}, \nn
\langle g_{\rm eff, pp}^{\mathcal{Z},\ell_R}\rangle^{[\hat{m}_{W}/\hat{\alpha}_{ew}]}_{\mathcal{O}(v^2/\Lambda^2)} &=
[0.0025/0.029] k^2 \frac{\bar{v}_T^2}{m_K^2}
- [0.045/0.047] \lambda \,  (k^2- k^4)\frac{v^2 \, \bar{v}_T^2}{m_K^4},
\label{eq:U1cpl1}
\end{align}
\begin{align}
\langle g_{\rm eff, pp}^{\mathcal{Z},u_L}\rangle^{[\hat{m}_{W}/\hat{\alpha}_{ew}]}_{\mathcal{O}(v^2/\Lambda^2)} &=
[0.0040/-0.013] k^2 \frac{\bar{v}_T^2}{m_K^2}
+ [0.0075/0.0078] \lambda \,  (k^2- k^4)\frac{v^2 \, \bar{v}_T^2}{m_K^4}, \nn
\langle g_{\rm eff, pp}^{\mathcal{Z},d_L}\rangle^{[\hat{m}_{W}/\hat{\alpha}_{ew}]}_{\mathcal{O}(v^2/\Lambda^2)} &=
[-0.0048/+0.0036] k^2 \frac{\bar{v}_T^2}{m_K^2}
+ [0.0075/0.0078] \lambda \,  (k^2- k^4)\frac{v^2 \, \bar{v}_T^2}{m_K^4}, \nn
\langle g_{\rm eff, pp}^{\mathcal{Z},\ell_L}\rangle^{[\hat{m}_{W}/\hat{\alpha}_{ew}]}_{\mathcal{O}(v^2/\Lambda^2)} &=
[-0.0031/0.023] k^2 \frac{\bar{v}_T^2}{m_K^2}
- [0.023/0.023] \lambda \,  (k^2- k^4)\frac{v^2 \, \bar{v}_T^2}{m_K^4}, \nn
\langle g_{\rm eff, pp}^{\mathcal{Z},\nu_L}\rangle^{[\hat{m}_{W}/\hat{\alpha}_{ew}]}_{\mathcal{O}(v^2/\Lambda^2)} &=
[0.0057/0.0059] k^2 \frac{\bar{v}_T^2}{m_K^2}
- [0.023/0.023] \lambda \,  (k^2- k^4)\frac{v^2 \, \bar{v}_T^2}{m_K^4}, \nn,
\label{eq:U1cpl2}
\end{align}
where the first value in square brackets is the value in the $\hat m_W$ scheme and the second is the $\hat \alpha_{ew}$ scheme value. The $\lambda$ that appears is the Higgs quartic coupling, which arises because the Higgs EOM is needed to massage the operators one gets from integrating out $K_\mu$ into the geoSMEFT basis.  The $\mathcal{O}(v^4/\Lambda^4)$ corrections are:
\begin{align}
\langle g_{\rm eff, pp}^{\mathcal{Z},u_R}\rangle^{[\hat{m}_{W}/\hat{\alpha}_{ew}]}_{\mathcal{O}(v^4/\Lambda^4)} &=
[3.8/-3.9] 10^{-3} k^2 \frac{\bar{v}_T^4}{m_K^4} + [-3.8/4.3] 10^{-3} k^4 \frac{\bar{v}_T^4}{m_K^4}
- [3.0/3.1] 10^{-2} \lambda \, (k^2- k^4)\frac{v^2 \, \bar{v}_T^2}{m_K^4},\nn
\langle g_{\rm eff, pp}^{\mathcal{Z},d_R}\rangle^{[\hat{m}_{W}/\hat{\alpha}_{ew}]}_{\mathcal{O}(v^4/\Lambda^4)} &=
[-1.9/2.0] 10^{-3} k^2\frac{\bar{v}_T^4}{m_K^4} + [1.9/-2.2] 10^{-3}k^4 \frac{\bar{v}_T^4}{m_K^4}
+ [1.5/1.6] 10^{-2}\lambda \,  (k^2- k^4)\frac{v^2 \, \bar{v}_T^2}{m_K^4}, \nn
\langle g_{\rm eff, pp}^{\mathcal{Z},\ell_R}\rangle^{[\hat{m}_{W}/\hat{\alpha}_{ew}]}_{\mathcal{O}(v^4/\Lambda^4)} &=
[-5.7/5.9] 10^{-3}k^2 \frac{\bar{v}_T^4}{m_K^4} + [5.7/-6.5] 10^{-3}k^4 \frac{\bar{v}_T^4}{m_K^4}
+ [4.5/4.7] 10^{-2}\lambda \,  (k^2- k^4)\frac{v^2 \, \bar{v}_T^2}{m_K^4},
\label{eq:U1cpl3}
\end{align}
\begin{align}
\langle g_{\rm eff, pp}^{\mathcal{Z},u_L}\rangle^{[\hat{m}_{W}/\hat{\alpha}_{ew}]}_{\mathcal{O}(v^4/\Lambda^4)} &=
[3.6/-4.2] 10^{-3} k^2 \frac{\bar{v}_T^4}{m_K^4} - [3.7/-4.4] 10^{-3} k^4 \frac{\bar{v}_T^4}{m_K^4}
- [7.5/7.8] 10^{-3}\lambda \,  (k^2- k^4)\frac{v^2 \, \bar{v}_T^2}{m_K^4}, \nn
\langle g_{\rm eff, pp}^{\mathcal{Z},d_L}\rangle^{[\hat{m}_{W}/\hat{\alpha}_{ew}]}_{\mathcal{O}(v^4/\Lambda^4)} &=
[-1.7/2.2] 10^{-3}k^2 \frac{\bar{v}_T^4}{m_K^4} +[1.8/-2.3] 10^{-3} k^4 \frac{\bar{v}_T^4}{m_K^4}
-[7.5/7.8] 10^{-3}\lambda \,  (k^2- k^4)\frac{v^2 \, \bar{v}_T^2}{m_K^4}, \nn
\langle g_{\rm eff, pp}^{\mathcal{Z},\ell_L}\rangle^{[\hat{m}_{W}/\hat{\alpha}_{ew}]}_{\mathcal{O}(v^4/\Lambda^4)} &=
[-5.5/6.1] 10^{-3}k^2 \frac{\bar{v}_T^4}{m_K^4} + [5.6/-6.6] 10^{-3}k^4 \frac{\bar{v}_T^4}{m_K^4}
+ [2.3/2.3] 10^{-2}\lambda \,  (k^2- k^4)\frac{v^2 \, \bar{v}_T^2}{m_K^4}, \nn
\langle g_{\rm eff, pp}^{\mathcal{Z},\nu_L}\rangle^{[\hat{m}_{W}/\hat{\alpha}_{ew}]}_{\mathcal{O}(v^4/\Lambda^4)} &=
-[2.6/2.5] 10^{-4} k^2 \frac{\bar{v}_T^4}{m_K^4}+[1.3/1.1]10^{-4} k^4 \frac{\bar{v}_T^4}{m_K^4}
+ [2.3/2.3] 10^{-2}\lambda \,  (k^2- k^4)\frac{v^2 \, \bar{v}_T^2}{m_K^4}, \nn
\label{eq:U1cpl4}
\end{align}


Note that the $\lambda$ dependence, which is a basis dependent artifact
in this matching,  cancels exactly {\it between} the $\mathcal{O}(v^2/\Lambda^2)$
and $\mathcal{O}(v^4/\Lambda^4)$ contributions to the effective couplings (e.g. when one adds Eq.~\eqref{eq:U1cpl1},~\eqref{eq:U1cpl2} to Eq.~\eqref{eq:U1cpl3},~\eqref{eq:U1cpl4}).
These terms come about
due to correlated matching at $\mathcal{L}^{(6)}$, and $\mathcal{L}^{(8)}$ in the SMEFT
in both models. This occurs quite generally,
due to the presence of the classical dimensionful parameter, SM Higgs vev $v$ in the EFT.
Matching contributions that are naively assumed restricted to $\mathcal{L}^{(8)}$ corrections
{\it descend down in mass dimension} to give matching contributions to $\mathcal{L}^{(6)}$ Wilson coefficients.
These matching contributions can be overlooked until matching results are developed to $\mathcal{L}^{(8)}$
and are an example of the intrinsic ambiguity in a $\mathcal{L}^{(6)}$ SMEFT treatment
related to higher-order terms in the power counting expansion.

This cancelation in $\lambda$ dependence in observable quantities has an important implication.
Quantities such as the $\mathcal{Z}$ effective couplings, and subsequently the amplitudes they define
are not exact in the SMEFT, but are only defined order by order systematically
in $1/\Lambda$. In this case
an ambiguity is present of order ${v}^2/\Lambda^4$ in the matching
to $\mathcal{L}^{(6)}$. This leads to an intrinsic ambiguity
in an amplitude that depends on $C_i^{(6)}$ parameters of order $1/\Lambda^4$.
The square of the SM perturbed with such a correction is then ambiguous and not
precisely defined at order $1/\Lambda^4$.
Due to the classical presence of a
parameter in the theory carrying mass dimension, the SM Higgs vev,
all contributions
to observables at each order in  $1/\Lambda$ are required to obtain basis independent and well-defined
results in an observable. Although we have discussed this point considering
$\mathcal{L}^{(8)}$ corrections leading to matching ambiguities in $\mathcal{L}^{(6)}$
operators of order $1/\Lambda^4$, the same effect is present
for all higher order $\mathcal{L}^{(6+ 2n)}$ matching corrections
with $n>1$.

$\Gamma_\mathcal{Z}$ was reported in Ref.~\cite{Hays:2020scx} in
each input parameter scheme. These numerical results differ in the last significant digit
compared to Ref.~\cite{Hays:2020scx}. This is due
to the use of updated SM predictions produced and reported in Table \ref{schemeresults} and
numerical approximations differing in this work. The total width is
\bea
 \frac{\sum_{\psi} \bar{\Gamma}^{{\rm SMEFT},\hat{\alpha}_{ew}}_{\mathcal{Z} \rightarrow \bar{\psi}_p \psi_p}}{\sum_{\psi} \hat{\Gamma}^{{\rm SM}, \hat{\alpha}_{ew}}_{\mathcal{Z} \rightarrow \bar{\psi}_p \psi_p}}
&=& 1 + 4.8 \times 10^{-3}  \, \frac{\bar{v}_T^2 \, k^2}{m_K^2} - 5.6 \times 10^{-3}\, (k^4 - 1.70 k^2) \, \frac{\bar{v}_T^4}{m_K^4},
\eea
\bea
 \frac{\sum_{\psi} \bar{\Gamma}^{{\rm SMEFT}, \hat{m}_{W}}_{\mathcal{Z} \rightarrow \bar{\psi}_p \psi_p}}{\sum_{\psi} \hat{\Gamma}^{{\rm SM},\hat{m}_{W}}_{\mathcal{Z} \rightarrow \bar{\psi}_p \psi_p}}
&=& 1 -3.0 \times 10^{-2} \, \frac{\bar{v}_T^2 \, k^2}{m_K^2} + 7.8 \times 10^{-3} \, (k^4 - 0.88 k^2) \, \frac{\bar{v}_T^4}{m_K^4}.
\eea
\begin{figure}
\includegraphics[height=0.24\textheight,width=0.45\textwidth]{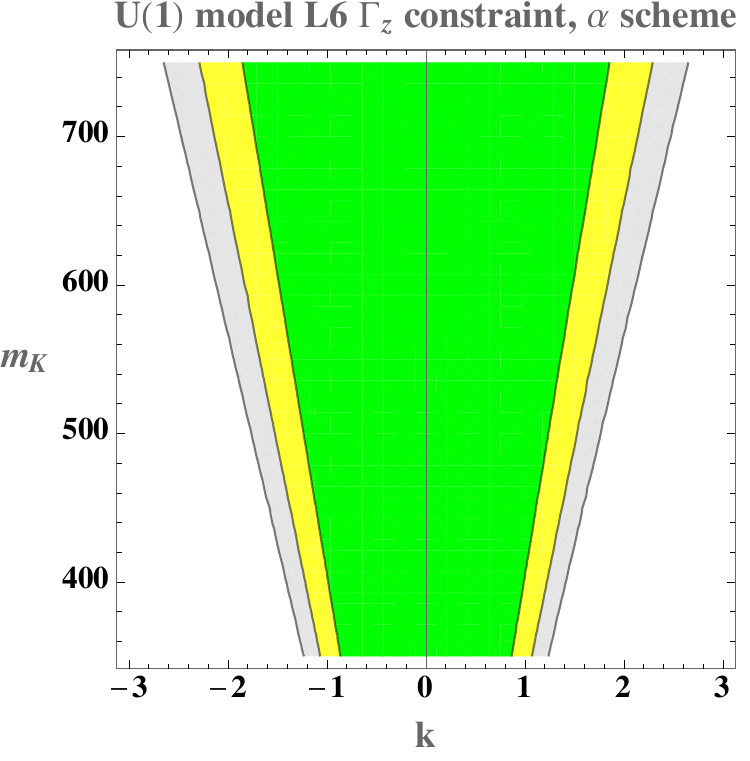}
\includegraphics[height=0.24\textheight,width=0.45\textwidth]{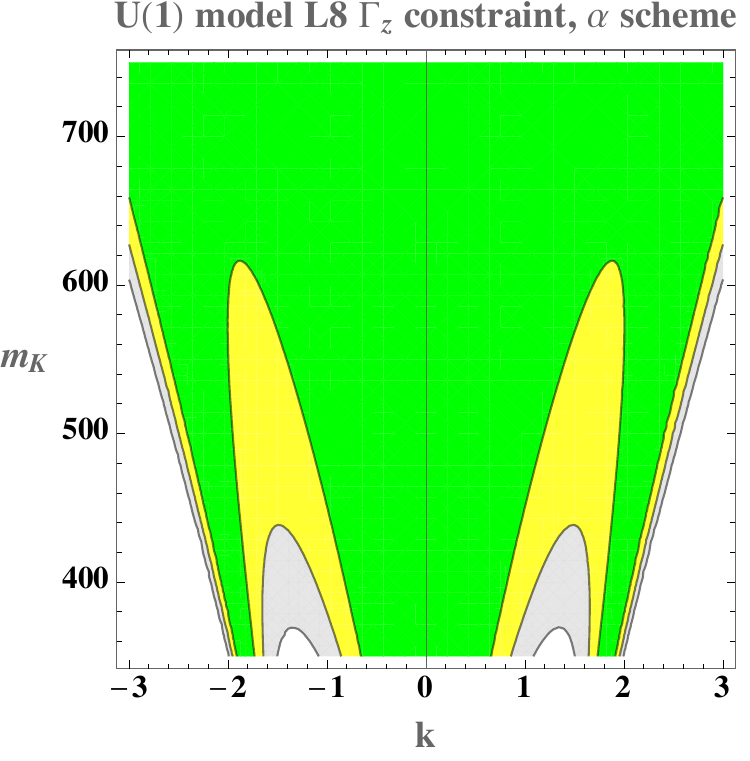}
\includegraphics[height=0.24\textheight,width=0.45\textwidth]{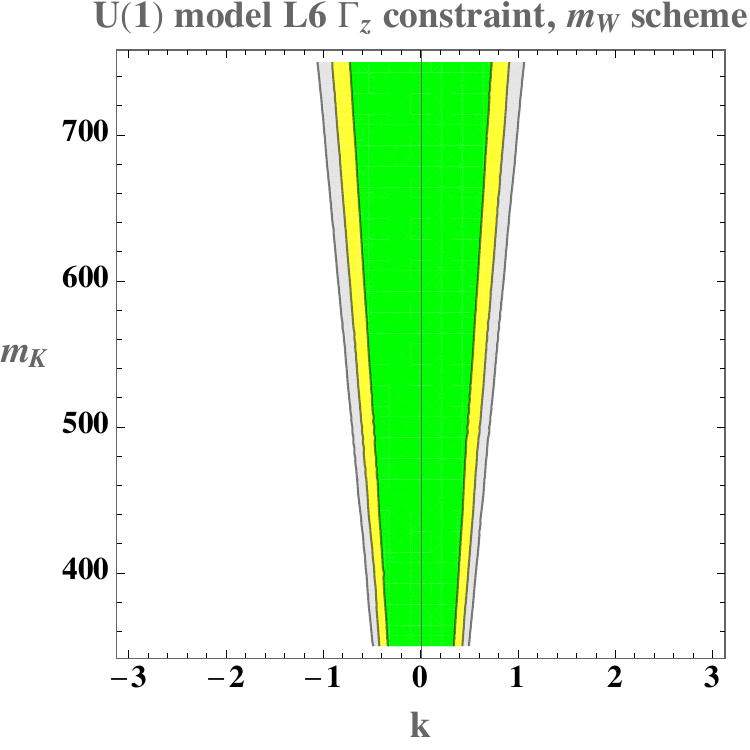}
\includegraphics[height=0.24\textheight,width=0.45\textwidth]{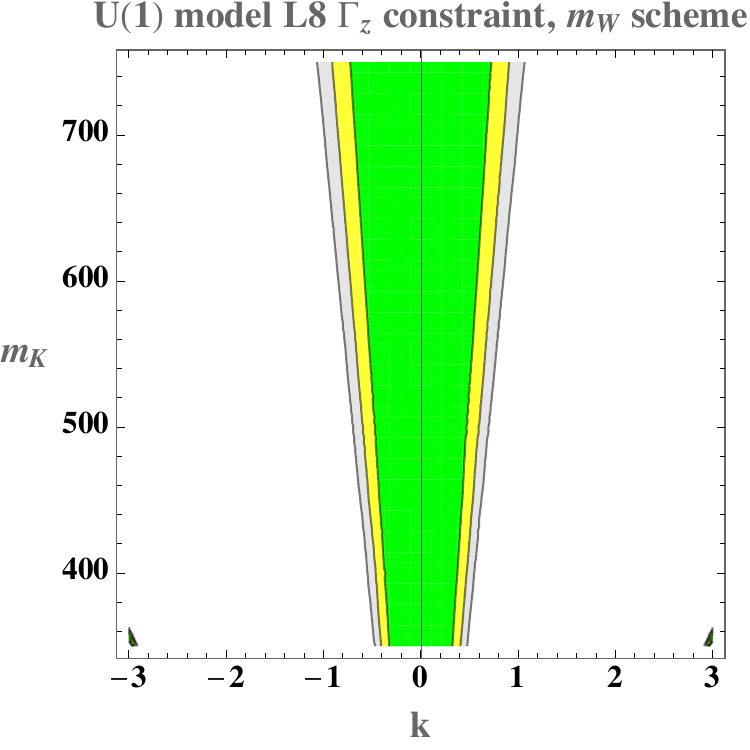}
\includegraphics[height=0.24\textheight,width=0.45\textwidth]{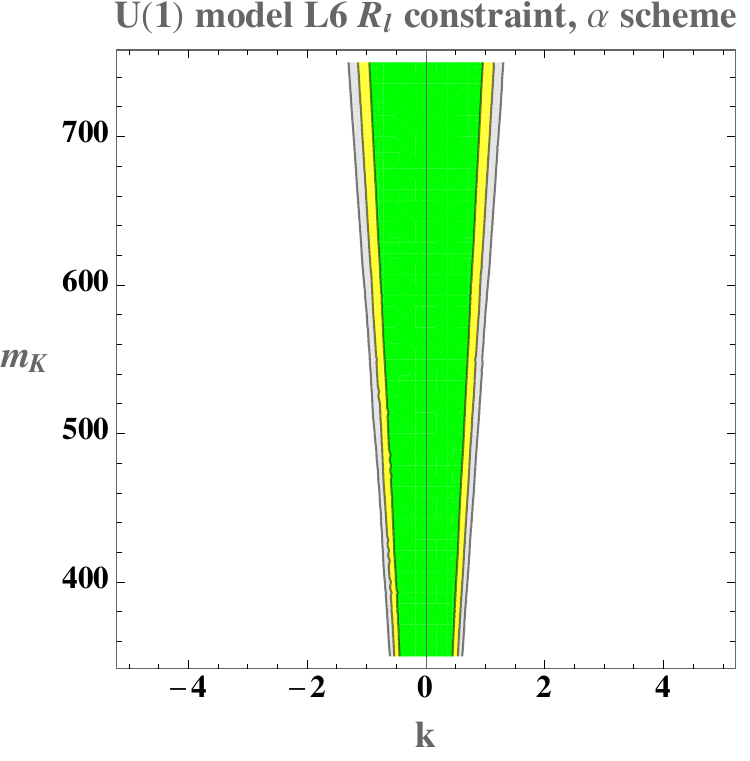}
\includegraphics[height=0.24\textheight,width=0.45\textwidth]{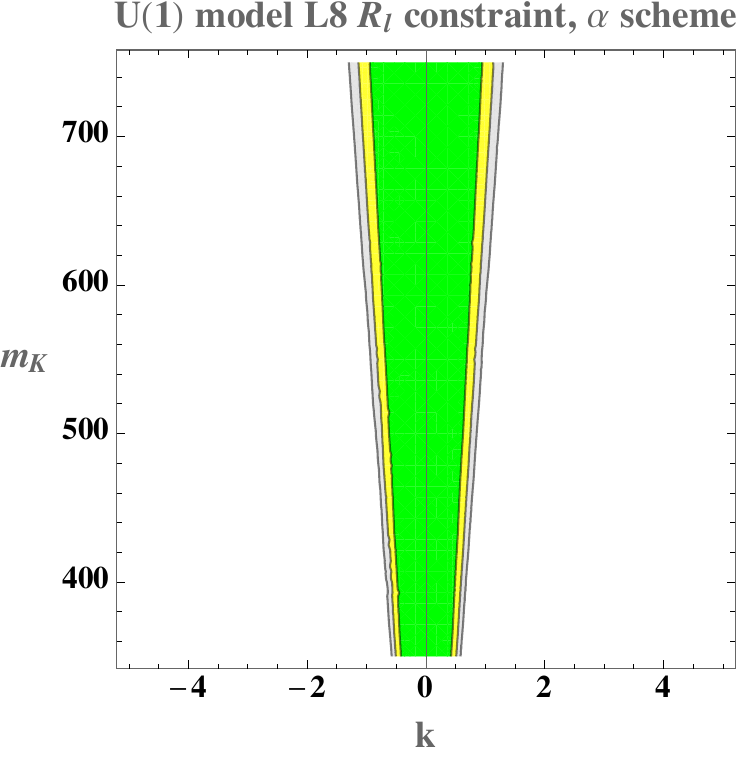}
\includegraphics[height=0.24\textheight,width=0.45\textwidth]{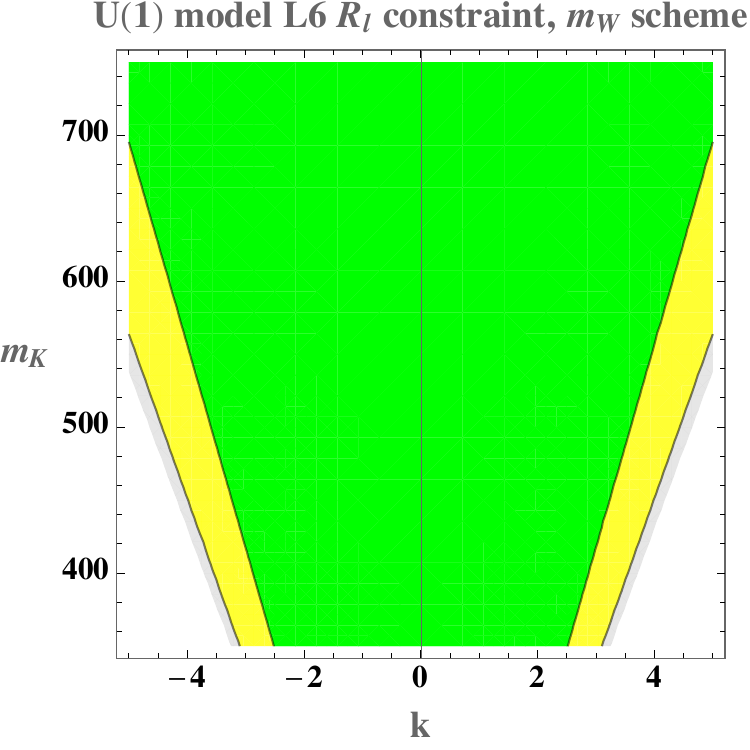}
\includegraphics[height=0.24\textheight,width=0.45\textwidth]{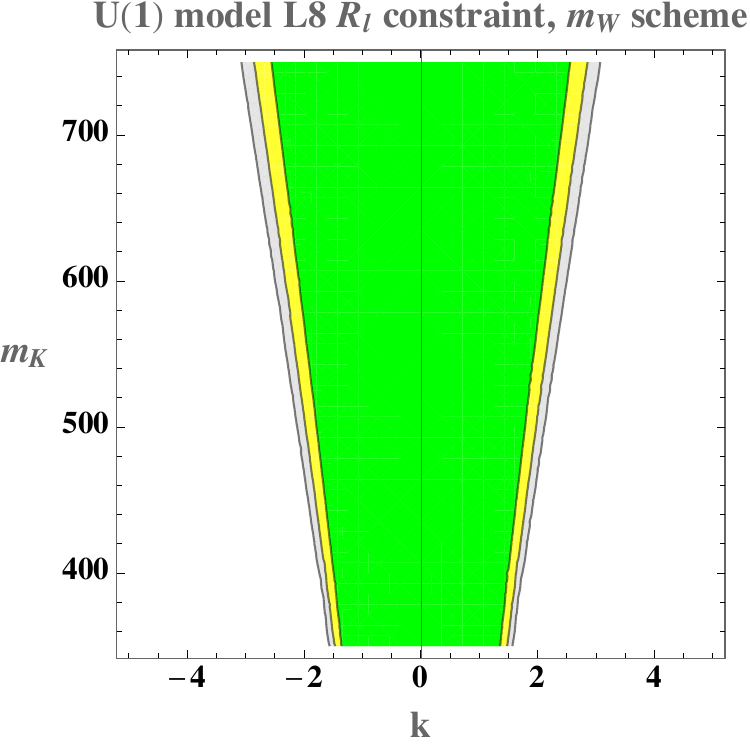}
\caption{Constraints from EWPD observables in the $\rm U(1)$ mixing model.
The results are organised so that increasing the precision of the theoretical predication
from $\mathcal{O}(v^2/\Lambda^2)$ to $\mathcal{O}(v^4/\Lambda^4)$ from left to right.
Both the $\alpha$ and $m_W$ schemes results are shown, and individual observables
carry a significant scheme dependence. Shown are the constraints on the model
space from the $\Gamma_Z$ and $R_\ell$ observables.\label{U1plotfig1}}
\end{figure}

\begin{figure}
\includegraphics[height=0.24\textheight,width=0.45\textwidth]{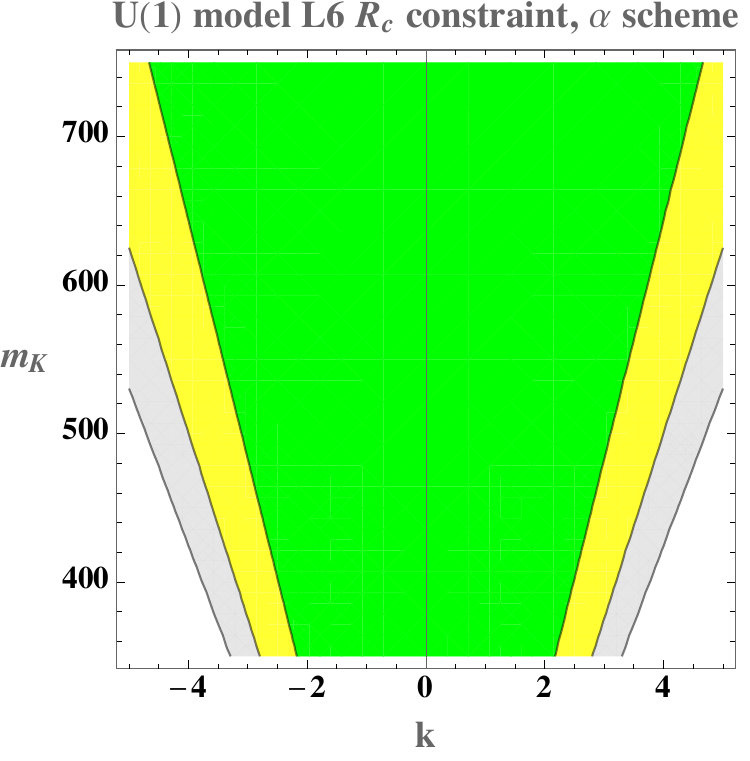}
\includegraphics[height=0.24\textheight,width=0.45\textwidth]{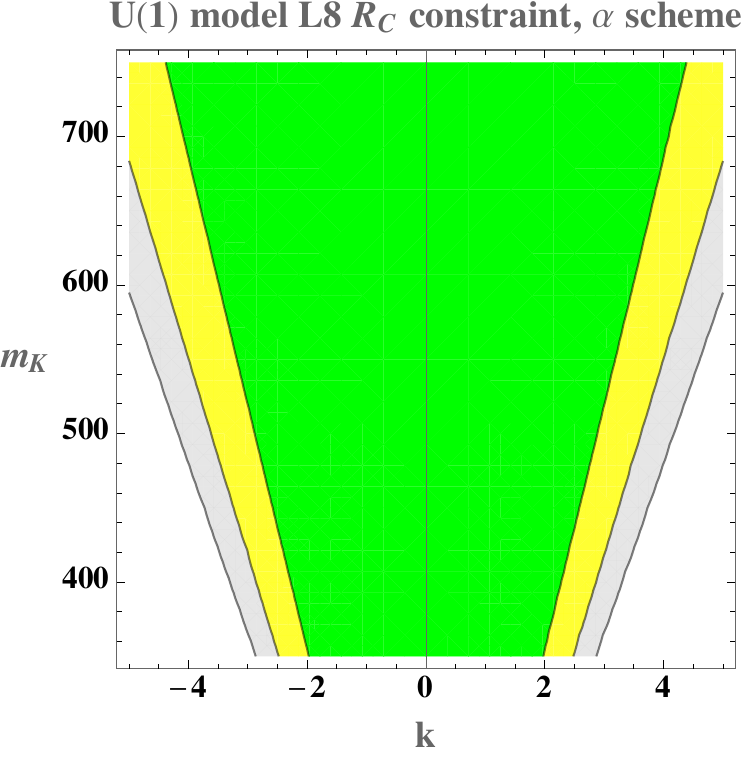}
\includegraphics[height=0.24\textheight,width=0.45\textwidth]{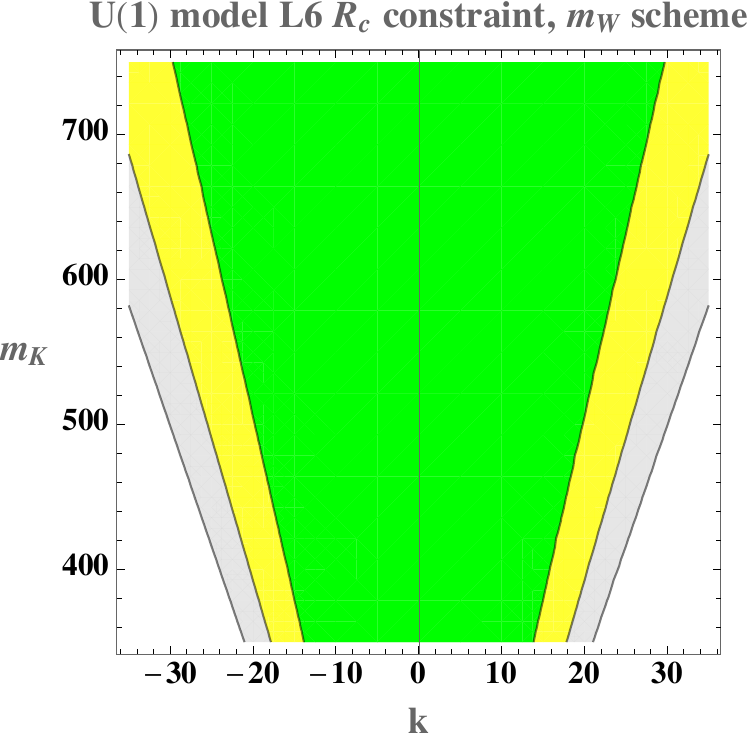}
\includegraphics[height=0.24\textheight,width=0.45\textwidth]{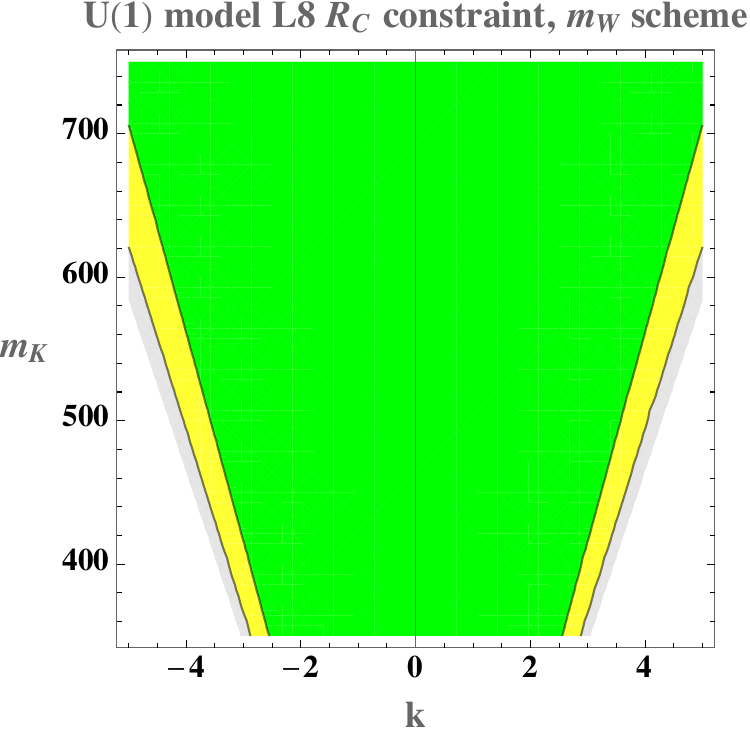}
\includegraphics[height=0.24\textheight,width=0.45\textwidth]{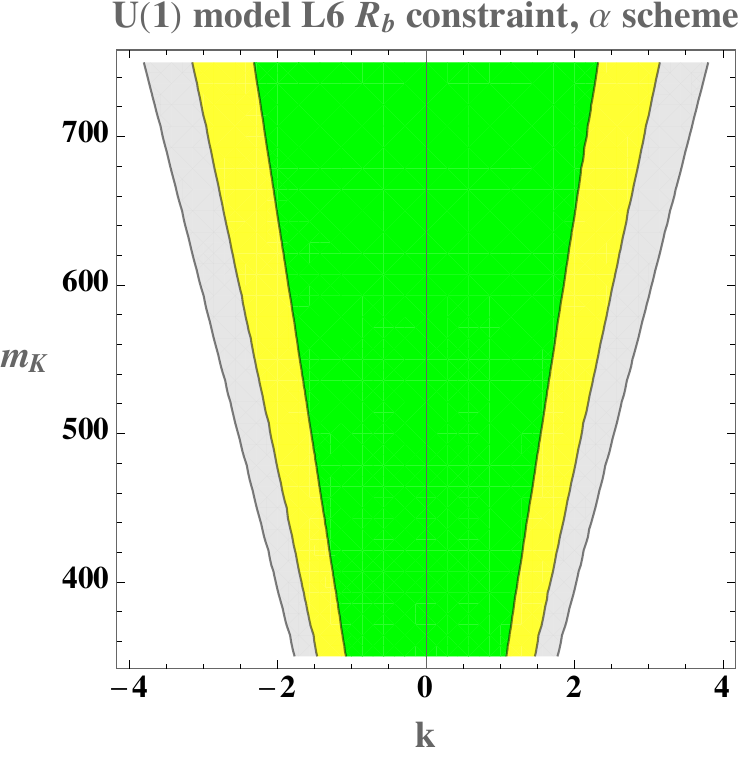}
\includegraphics[height=0.24\textheight,width=0.45\textwidth]{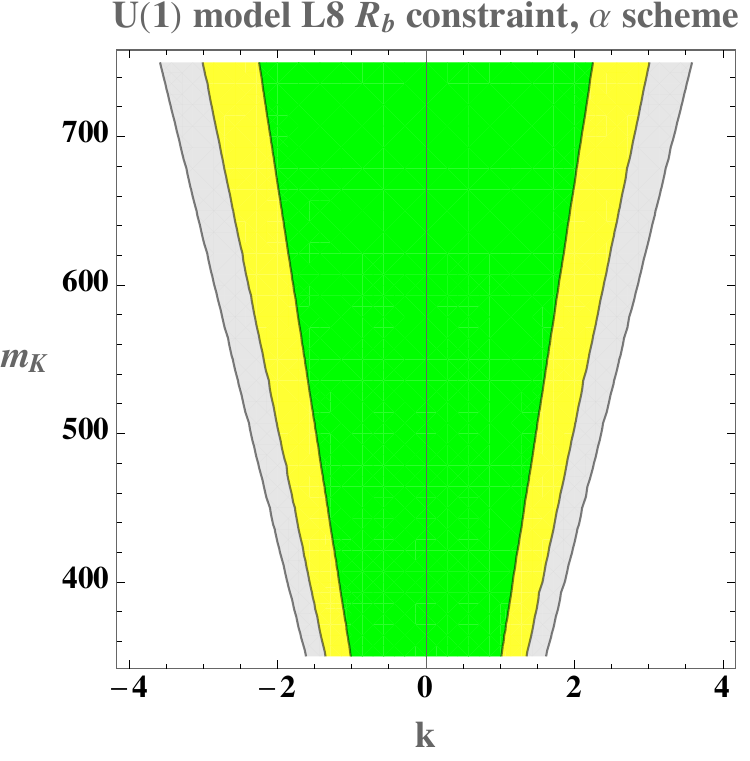}
\includegraphics[height=0.24\textheight,width=0.45\textwidth]{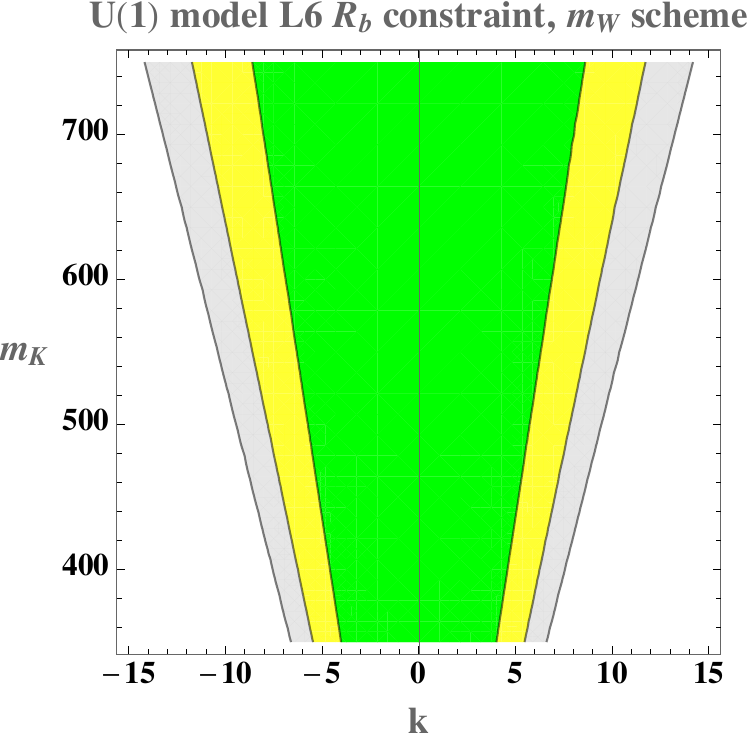}
\includegraphics[height=0.24\textheight,width=0.45\textwidth]{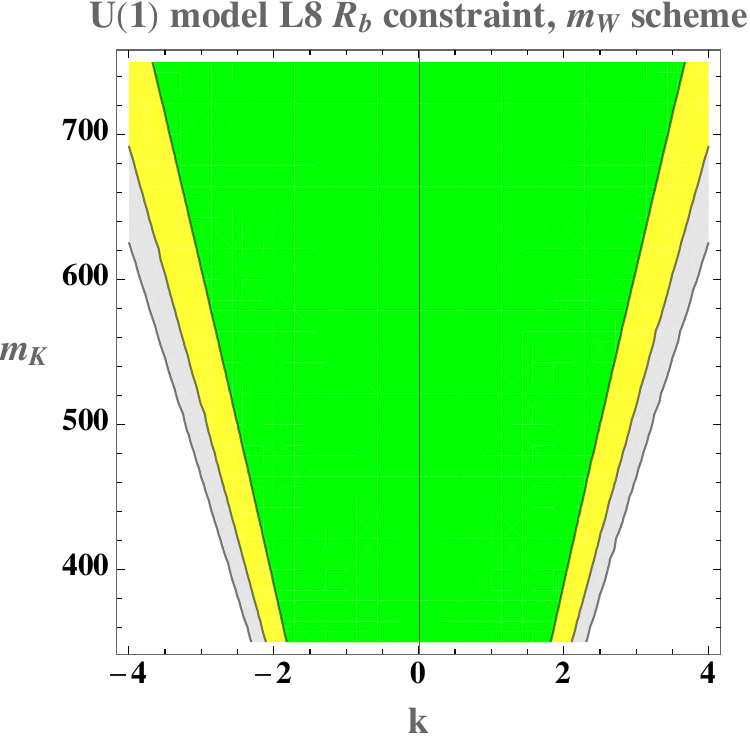}
\caption{Shown are the constraints on the model
space from the $R_c$ and $R_b$ EWPD observables.\label{U1plotfig2}}
\end{figure}

\begin{figure}
\includegraphics[height=0.24\textheight,width=0.45\textwidth]{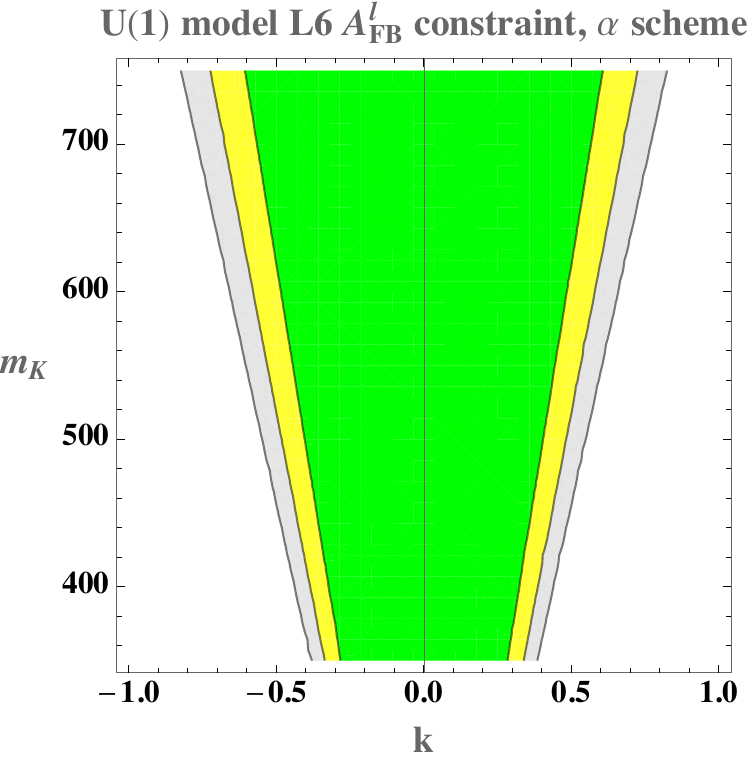}
\includegraphics[height=0.24\textheight,width=0.45\textwidth]{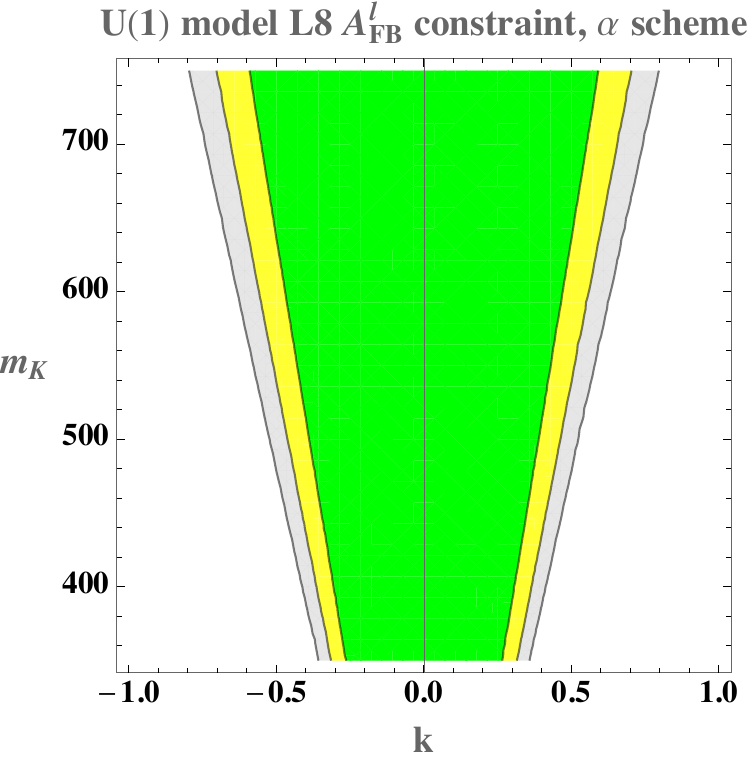}
\includegraphics[height=0.24\textheight,width=0.45\textwidth]{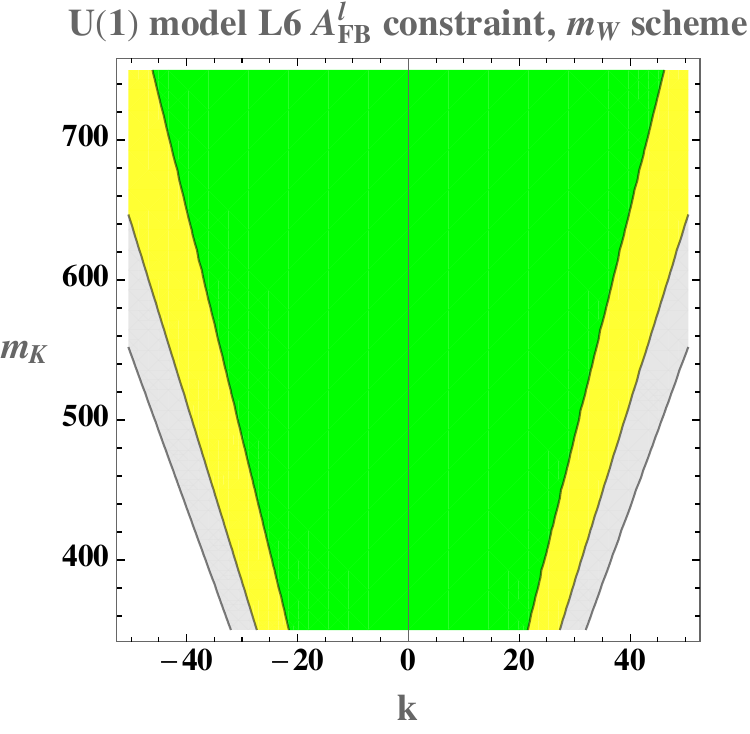}
\includegraphics[height=0.24\textheight,width=0.45\textwidth]{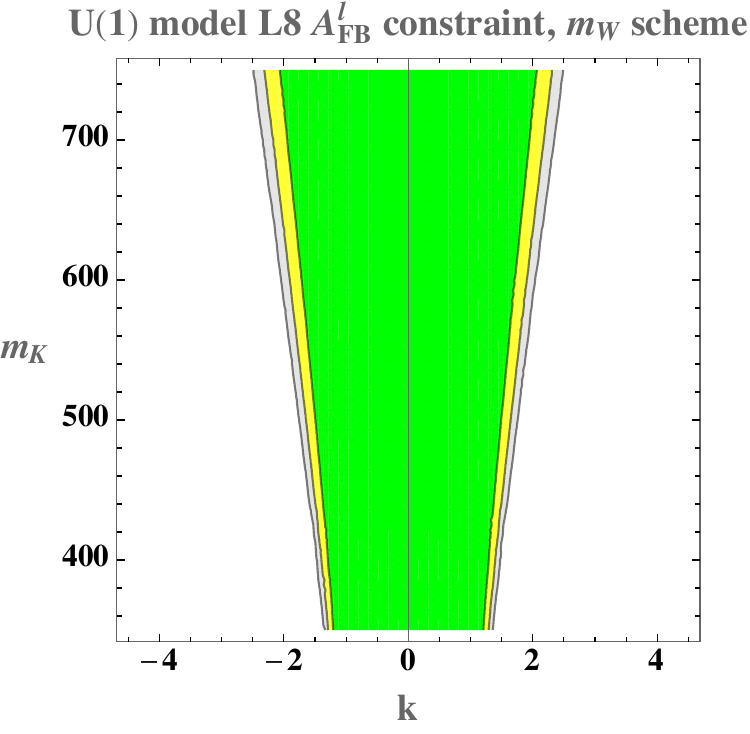}
\includegraphics[height=0.24\textheight,width=0.45\textwidth]{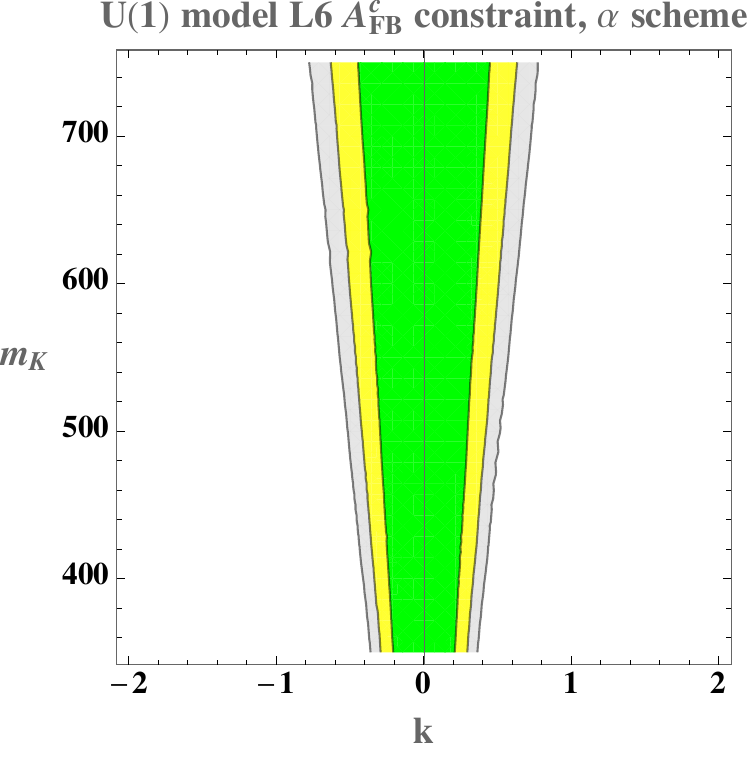}
\includegraphics[height=0.24\textheight,width=0.45\textwidth]{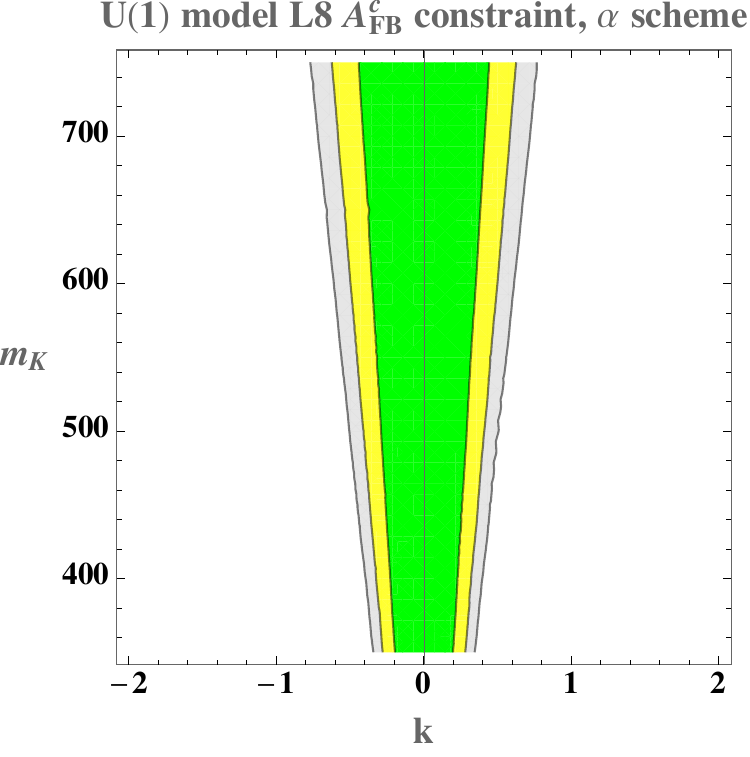}
\includegraphics[height=0.24\textheight,width=0.45\textwidth]{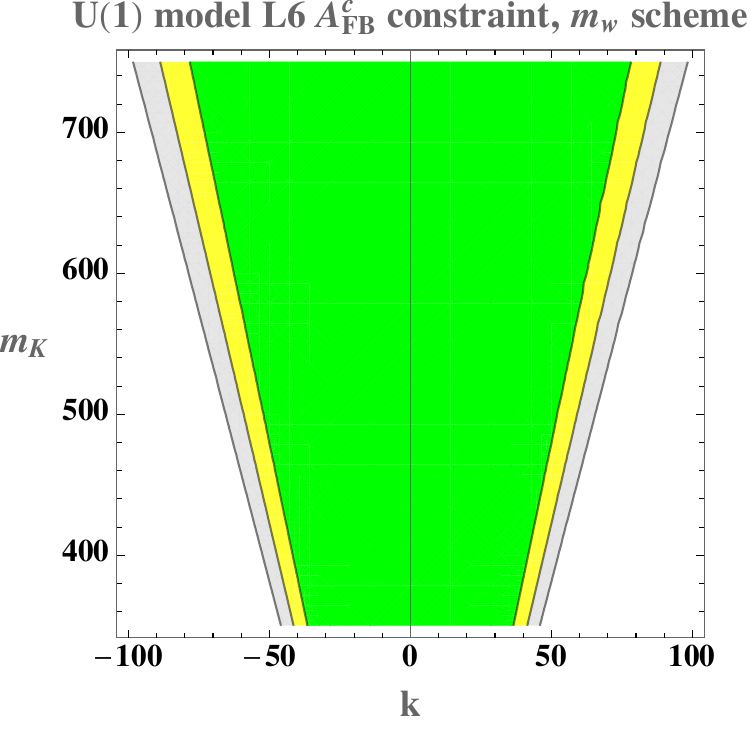}
\includegraphics[height=0.24\textheight,width=0.45\textwidth]{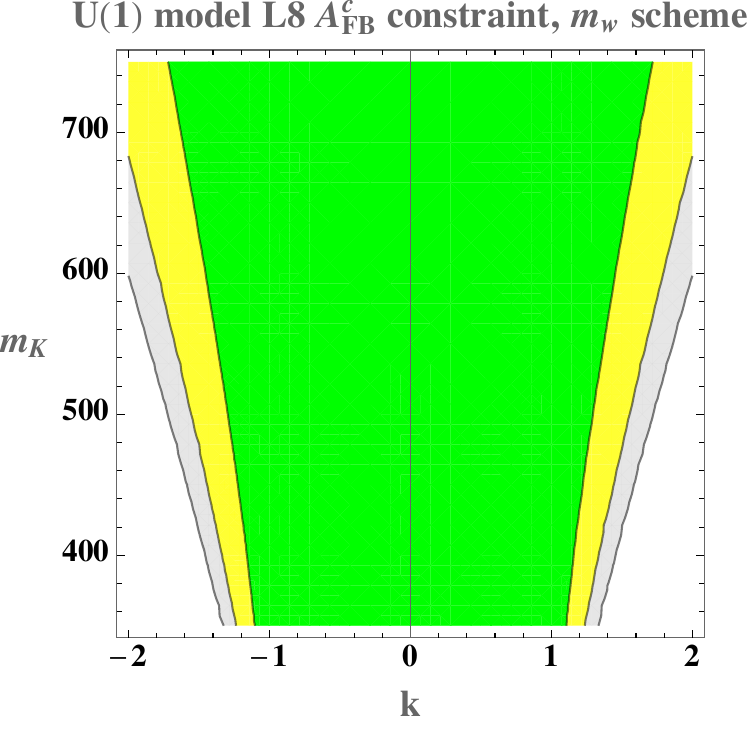}
\caption{Shown are the constraints on the model
space from the $A_{FB}^\ell$ and $A_{FB}^c$ EWPD observables.\label{U1plotfig3}}
\end{figure}

\begin{figure}
\includegraphics[height=0.24\textheight,width=0.45\textwidth]{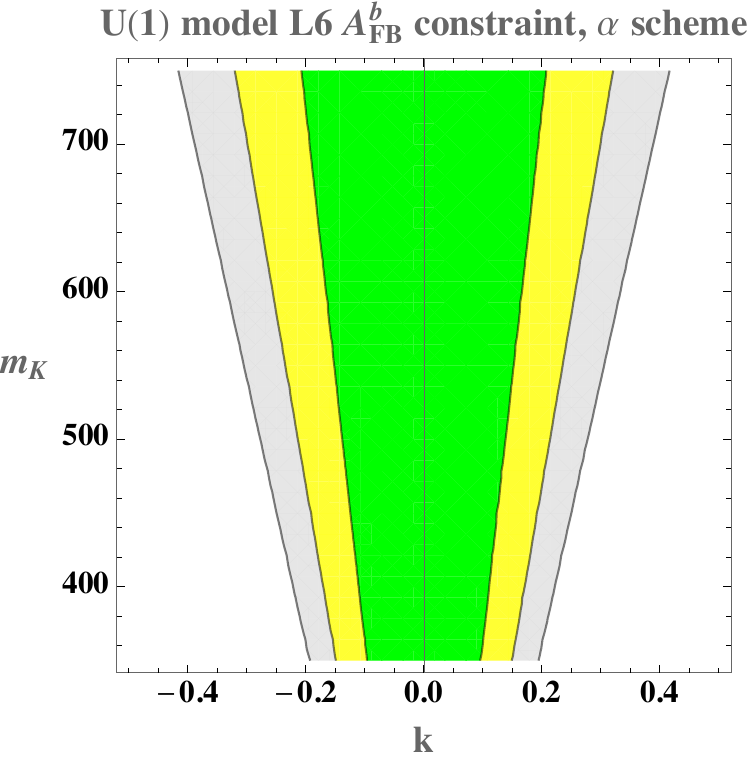}
\includegraphics[height=0.24\textheight,width=0.45\textwidth]{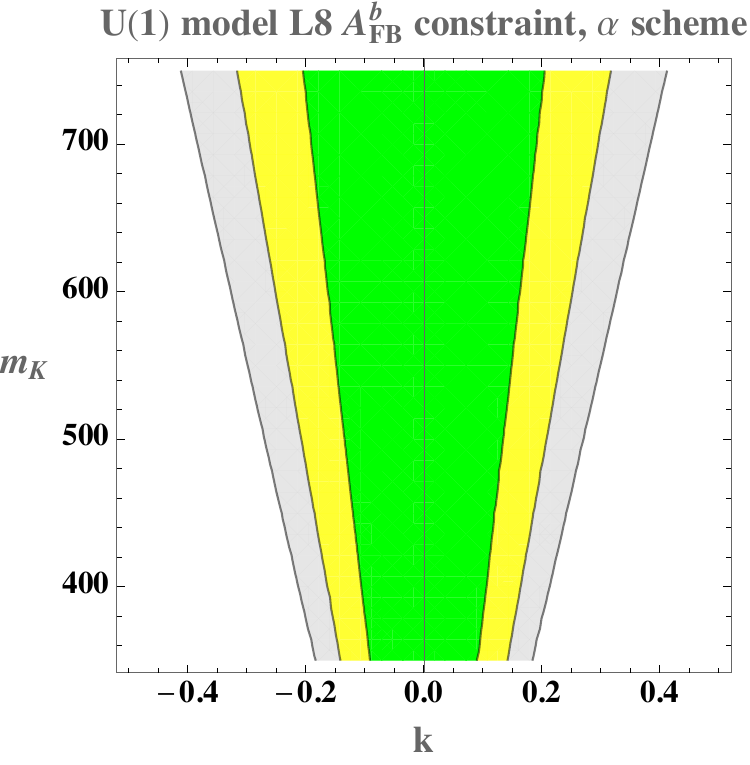}
\includegraphics[height=0.24\textheight,width=0.45\textwidth]{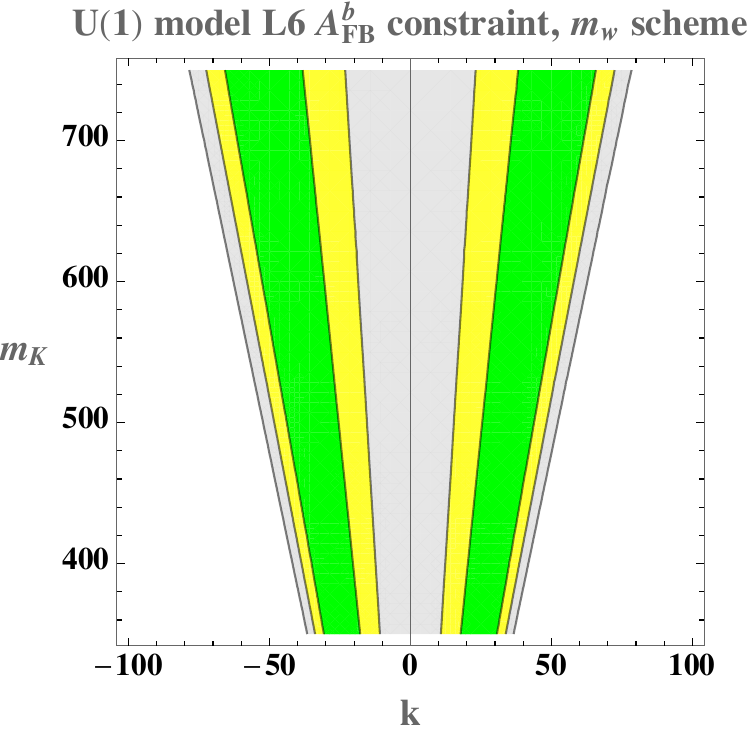}
\includegraphics[height=0.24\textheight,width=0.45\textwidth]{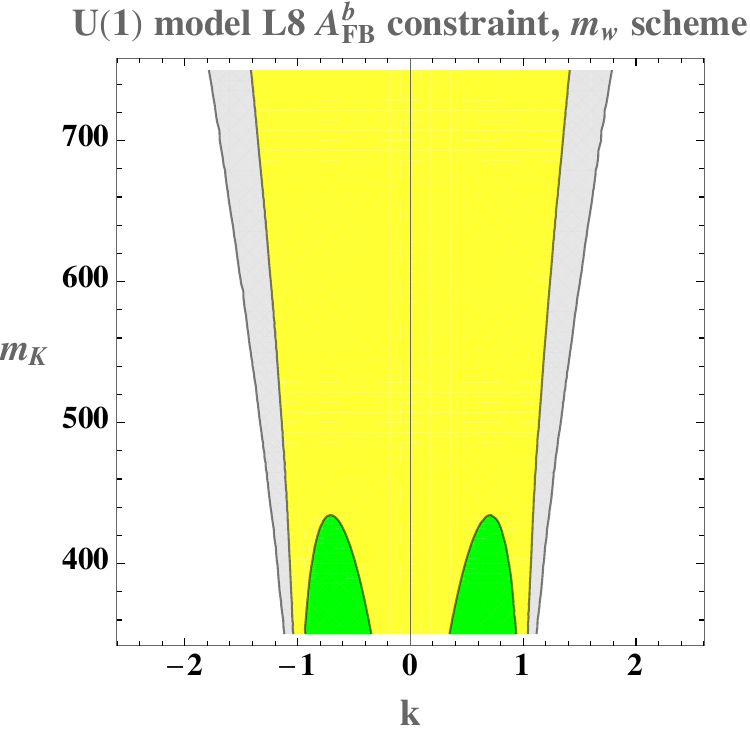}
\includegraphics[height=0.24\textheight,width=0.45\textwidth]{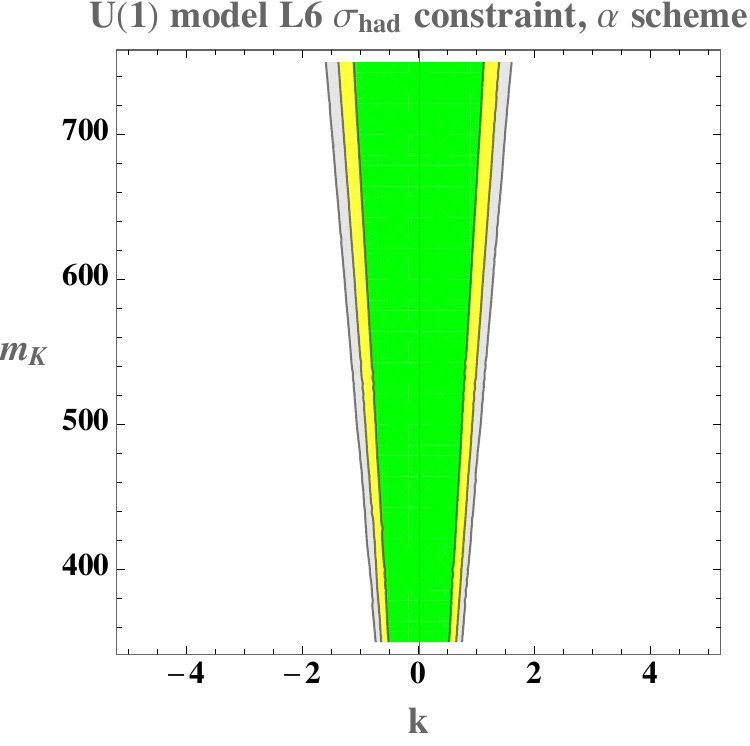}
\includegraphics[height=0.24\textheight,width=0.45\textwidth]{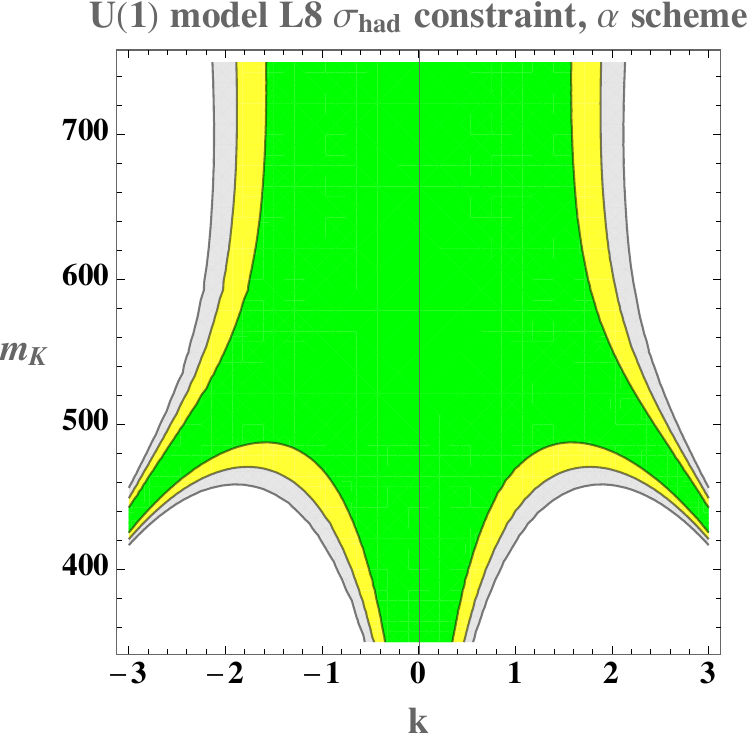}
\includegraphics[height=0.24\textheight,width=0.45\textwidth]{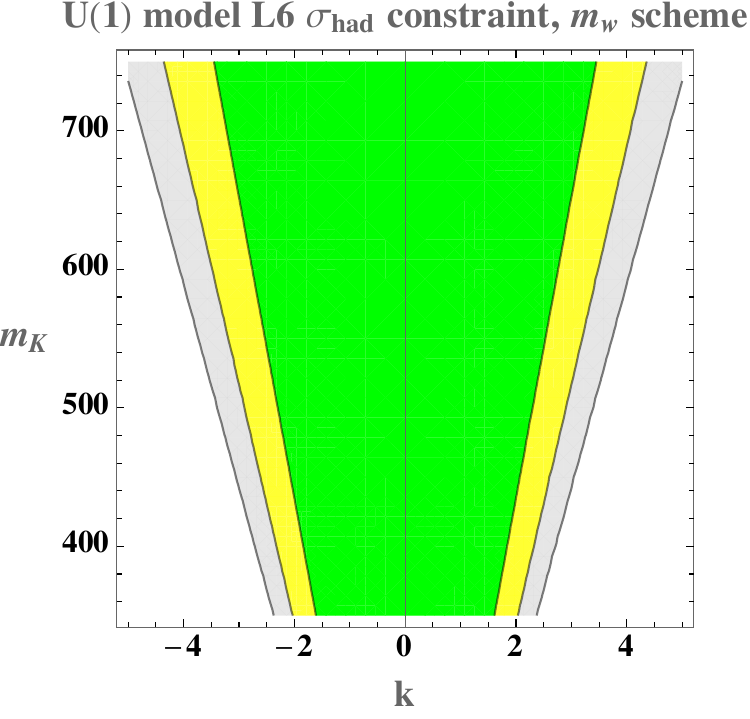}
\includegraphics[height=0.24\textheight,width=0.45\textwidth]{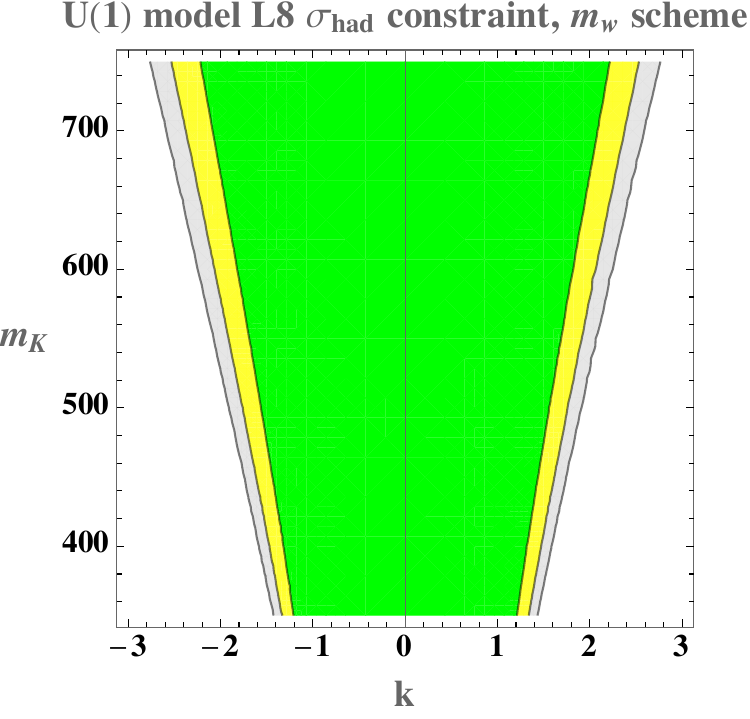}
\caption{Shown are the constraints on the model
space from the $A_{FB}^b$ and $\sigma_{had}^0$ EWPD observables. \label{U1plotfig4}}
\end{figure}

The remaining EWPD observables, with numerical values ordered in the $[\hat{m}_W,\alpha_{EW}]$ schemes, are
\bea
\frac{\bar{R}_{c}^{\textrm{SMEFT}}}{\hat{R}_{c}^{\textrm{SM}}} =
[2.7/110] \times 10^{-4} \,  \frac{k^2 \, \bar{v}_T^2}{m_K^2}
+ [0.0028/0.00056]  \,  \frac{k^4 \, \bar{v}_T^4}{m_K^4}
+ [-0.0028/0.0026] \, \frac{k^2 \, \bar{v}_T^4}{m_K^4}, \nn
\eea

\bea
\frac{\bar{R}_{b}^{\textrm{SMEFT}}}{\hat{R}_{b}^{\textrm{SM}}} =
-[4.2/59] \times 10^{-4}  \,  \frac{k^2 \, \bar{v}_T^2}{m_K^2}
- [1.4/0.44] \times 10^{-3} \,  \frac{k^4 \, \bar{v}_T^4}{m_K^4}
+ [1.4/-1.2] \times 10^{-3}\, \frac{k^2 \, \bar{v}_T^4}{m_K^4}, \nn
\eea

\bea
\frac{\bar{R}_{\ell}^{\textrm{SMEFT}}}{\hat{R}_{\ell}^{\textrm{SM}}} =
[7.5/280] \times 10^{-4} \,  \frac{k^2 \, \bar{v}_T^2}{m_K^2}
+ [4.3/-2.3] \times 10^{-3} \,  \frac{k^4 \, \bar{v}_T^4}{m_K^4}
+[-4.3/6.2] \times 10^{-3} \, \frac{k^2 \, \bar{v}_T^4}{m_K^4}, \nn
\eea

\bea
\frac{(\bar{\sigma}_{had}^{0})^{\textrm{SMEFT}}}{(\hat{\sigma}_{had}^{0})^{\textrm{SM}}} =
-[2.6/28] \times 10^{-4}  \frac{k^2 \, \bar{v}_T^2}{m_K^2}
- [8.7/6.0]\times 10^{-4} \,  \frac{k^2 \, \bar{v}_T^4}{m_K^4}
+ [8.7/140]\times 10^{-4} \,  \frac{k^4 \, \bar{v}_T^4}{m_K^4}, \nn
\eea

\bea
\frac{(\bar{A}_{FB}^{0,c})^{\textrm{SMEFT}}}{(\hat{A}_{FB}^{0,c})^{\textrm{SM}}} =
[-2.3/2.0] \times 10^{-4} \frac{k^2 \, \bar{v}_T^2}{m_K^2}
+ [0.31/-0.22] \, k^4 \frac{\bar{v}_T^4}{m_K^4} + [-0.31/0.46] \, k^2 \frac{\bar{v}_T^4}{m_K^4}, \nn
\eea

\bea
\frac{(\bar{A}_{FB}^{0,b})^{\textrm{SMEFT}}}{(\hat{A}_{FB}^{0,b})^{\textrm{SM}}} =
[-2.1/1.8]\times 10^{-4} \,  \frac{k^2 \, \bar{v}_T^2}{m_K^2}
+ [0.28/-0.45] \, k^4  \frac{\bar{v}_T^4}{m_K^4} + [-0.28/0.43] \, k^2  \frac{\bar{v}_T^4}{m_K^4}, \nn
\eea

\bea
\frac{(\bar{A}_{FB}^{0,\ell})^{\textrm{SMEFT}}}{(\hat{A}_{FB}^{0,\ell})^{\textrm{SM}}} =
[-4.3/3.6] \times 10^{-4} \,  \frac{k^2 \, \bar{v}_T^2}{m_K^2}
+ [0.55/2.3] \, k^4\, \frac{\bar{v}_T^4}{m_K^4} + [-0.55/0.84] \, k^2\, \frac{\bar{v}_T^4}{m_K^4}. \nn
\eea

\begin{figure}
\includegraphics[height=0.24\textheight,width=0.45\textwidth]{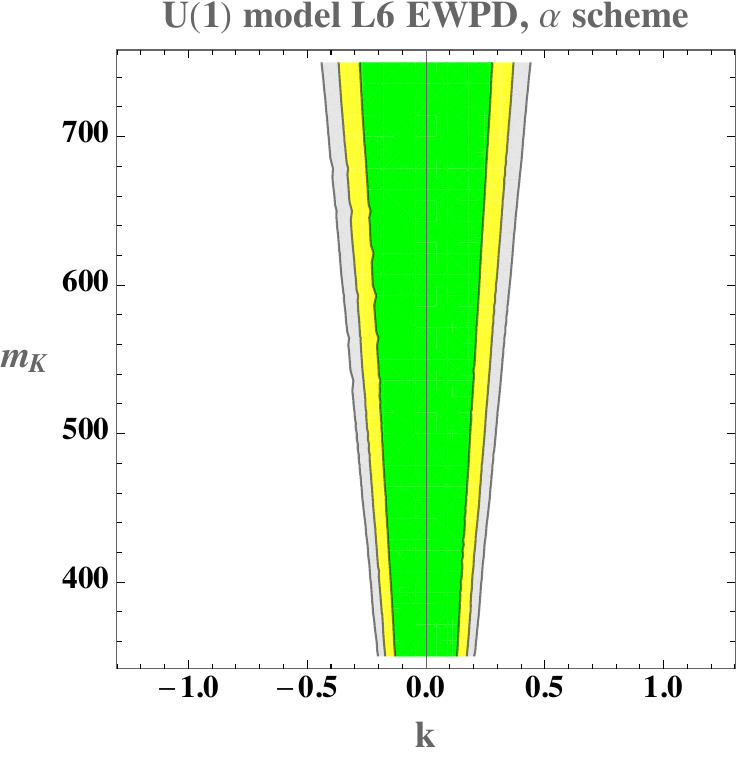}
\includegraphics[height=0.24\textheight,width=0.45\textwidth]{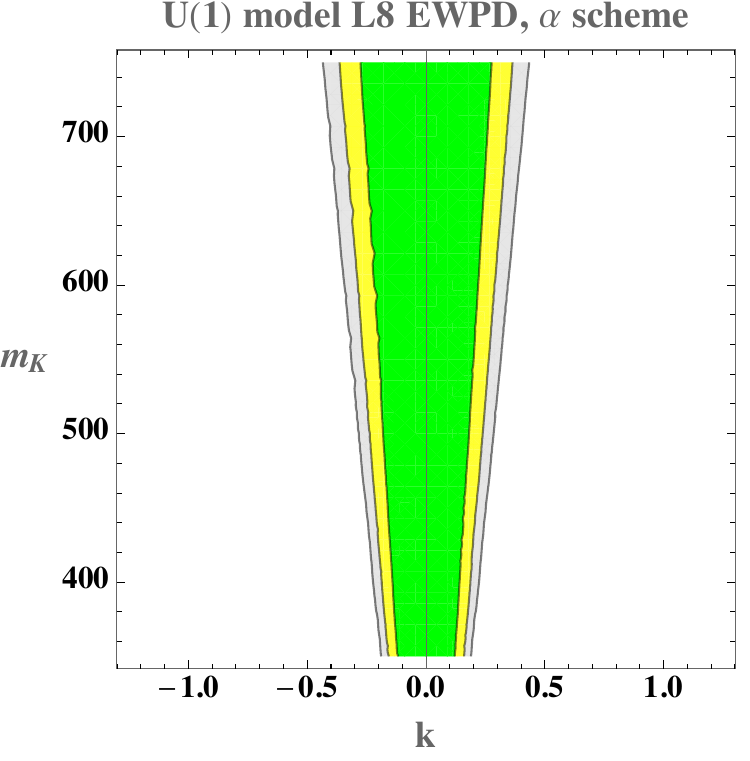}
\includegraphics[height=0.24\textheight,width=0.45\textwidth]{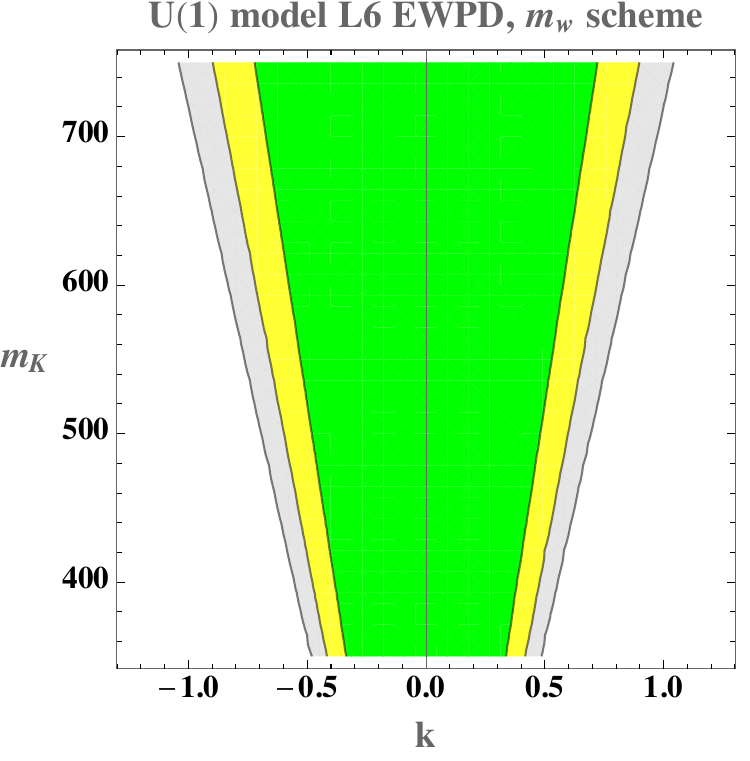}
\includegraphics[height=0.24\textheight,width=0.45\textwidth]{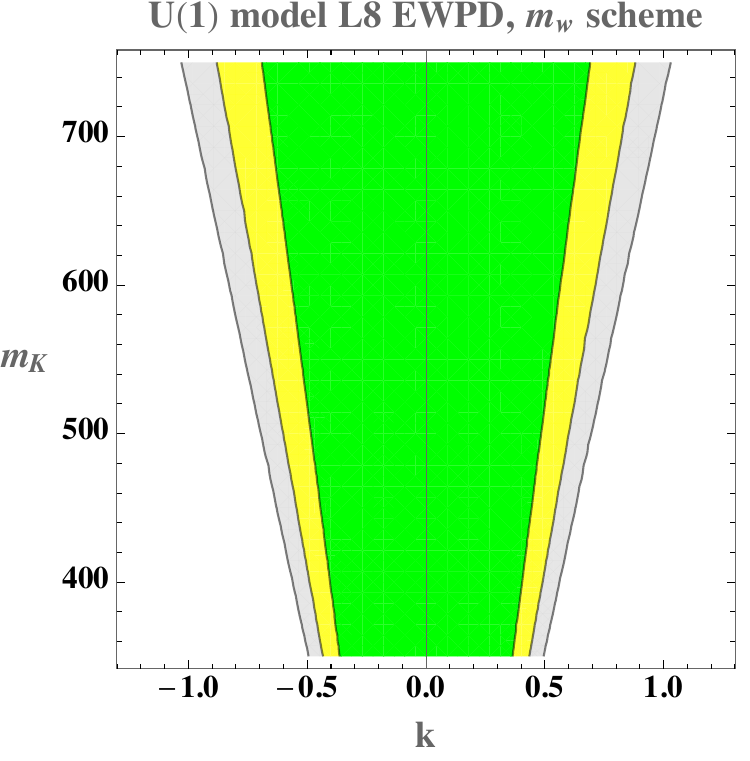}
\caption{Combined constraints from the full set of EWPD LEP observables in the $\rm U(1)$
mixing model.\label{U1plotfig5}}
\end{figure}

The above expressions are the perturbations to the SM, as such there is an implied $1+$ in each equation.
There are several generic expectations on how SMEFT constraints projected onto a specific UV model
come about in a global analysis. Input parameter scheme dependence, and the effects of $\mathcal{L}^{(8)}$
corrections are both expected to be significant in some observables due to the decoupling theorem and numerical
accidents. Both effects are expected to be reduced as more observables are consistently combined in
a global SMEFT fit, and this expectation is born out in the models we study in this section.
In the case of the $\rm U(1)$ kinetic mixing model, the results for each observable are shown in
Figs.~\ref{U1plotfig1},\ref{U1plotfig2},\ref{U1plotfig3},\ref{U1plotfig4},\ref{U1plotfig5}.
The plots show $m_K$ in the range $[350,750] \, {\rm GeV}$, chosen to illustrate some of the structure of the allowed parameter
space that shows the largest effects at low mass scales. We make these same choices
in the scalar triplet model. Larger suppression scales are considered in the previous section.\footnote{The lower bound on $m_K > \bar{v}_T$ is imposed in the global minimum found to define the best fit region as
a prior assumption.}

Note the large x axis in the case of the $A_{FB}^{c,b}$ constraints
at $\mathcal{L}^{(6)}$ in the $m_W$ scheme. This is due to an accidental numerical
suppression of the $\mathcal{L}^{(6)}$ perturbations in these observables.
For such large coupling values, perturbation theory is breaking down, so the axis was
expanded simply to illustrate the allowed parameter space.

The effect of the $\mathcal{L}^{(8)}$ corrections in the $\rm U(1)$
model damps in the allowed parameter space when all LEP observables are combined, as expected.
Input parameter scheme dependence is reduced as observables are combined,
but remains even in the global fit combination.
This is driven by the accidental numerical scheme dependence in the observables ($A_{FB}^{c,b}$) that
introduce the largest pulls in the EWPD fit.

\subsection{Electroweak triplet scalar}

In this model, we introduce a scalar $\Phi$ which is an
electroweak triplet with zero hyperchange.
The SM Lagrangian is then augmented by the Lagrangian for the new
heavy scalar,
\begin{align}
	\label{eq:tripletLagr}
	\mathcal{L}_{\Phi} =& \frac{1}{2} \left( D_{\mu} \Phi^{a} \right)^{2}
	- \frac{1}{2} m^{2}_{\Phi} \Phi^{a} \Phi^{a}
	+ 2 \kappa H^{\dagger} \tau^{a} H \Phi^{a}
	- \eta ( H^{\dagger} H ) \Phi^{a} \Phi^{a}
	- \frac{1}{4} \lambda_{\Phi} ( \Phi^{a} \Phi^{a} )^{2} .
\end{align}
We now integrate out the heavy triplet scalar, with the details given in Appendix
\ref{app:matching}. The resulting matching pattern for the $\mathcal{L}^{(6)}$
and $\mathcal{L}^{(8)}$ Wilson coefficients in the SMEFT is given in Table \ref{tab:matchingTriplet}.
For the effective couplings we find:
\begin{table}
  \begin{center}
	  \caption{Matching for the electroweak triplet scalar up to $\mathcal{L}^{(8)}$. These matching results
    are consistent with geoSMEFT conventions on operator forms and are sufficient to examine observables
    constructed via two and three point contributions.}
    \label{tab:matchingTriplet}
\begin{tabular}{l|c}
	\multicolumn{2}{c}{$H^{2n}$}   \\
      \toprule
      $C_{H}^{(4)}$ & $ \frac{\kappa^{2}}{2m^{2}_{\Phi}} \left( 1 - \frac{4v^{2}\lambda}{m^{2}_{\Phi}} \right) + \frac{6v^{4}\lambda^{2} \kappa^{2}}{m^{6}_{\Phi}} $ \\
      $C_{H}^{(6)}$ & $ \frac{(4\lambda - \eta) \kappa^{2}}{m^{4}_{\Phi}}
      - \frac{2\kappa^{2} (\kappa^{2} + 2 v^{2}\lambda (4\lambda - \eta))}{m^{6}_{\Phi}}
      + \frac{11 v^{2}\lambda \kappa^{4}}{m^{8}_{\Phi}} $ \\
      $C_{H}^{(8)}$ & $ \frac{2 (2\lambda - \eta)^{2} \kappa^{2}}{m^{6}_{\Phi}}
      + \frac{(40 \eta - 72 \lambda - \lambda_{\Phi})\kappa^{4}}{4m^{8}_{\Phi}}
      + \frac{7 \kappa^{6}}{m^{10}_{\Phi}}
      $ \\
      \hline
      \\
      \multicolumn{2}{c}{$H^{2n+1} \psi^{2}$}   \\
    \toprule
      $C_{\psi H}^{(6)}$ & $ y_{\psi} \frac{\kappa^{2}}{m^{4}_{\Phi}}\left( 1 - \frac{4v^{2}\lambda}{m^{2}_{\Phi}} \right)$ \\
      $C_{\psi H}^{(8)}$ & $ y_{\psi} \frac{\kappa^{2}}{m^{6}_{\Phi}} \left( 4\lambda - 2\eta - \frac{3\kappa^{2}}{m^{2}_{\Phi}} \right) $ \\
        \end{tabular}
    \quad
	  \begin{tabular}{l|c}
      \multicolumn{2}{c}{$H^4 D^2$}   \\
      \toprule
      $C_{H\Box}^{(6)}$ & $ \frac{\kappa^{2}}{2m^{4}_{\Phi}} \left( 1 - \frac{4 v^{2}\lambda}{m^{2}_{\Phi}} \right) $ \\
      $C_{HD}^{(6)}$ &  $ - \frac{2\kappa^{2}}{m^{4}_{\Phi}} \left( 1 - \frac{4 v^{2} \lambda }{m^{2}_{\Phi}} \right) $ \\
      \multicolumn{2}{c}{}   \\
      \multicolumn{2}{c}{$H^6 D^2$}   \\
      \toprule
      $C_{HD}^{(8)}$ & $ - \frac{\kappa^{4}}{m^{8}_{\Phi}} $ \\
      $C_{HD2}^{(8)}$ &  $  \frac{4(\eta - 2\lambda )\kappa^{2}}{m^{6}_{\Phi}} + \frac{8\kappa^{4}}{m^{8}_{\Phi}} $ \\
    \end{tabular}
  \end{center}
\end{table}

\begin{align}
\langle g_{\rm eff, pp}^{\mathcal{Z},u_R}\rangle^{[\hat{m}_{W}/\hat{\alpha}_{ew}]}_{\mathcal{O}(v^2/\Lambda^2)} &=[0.44/-0.11]\frac{\kappa^2\, \bar v^2_T}{m^4_\Phi} + [-1.8/0.42]\lambda \frac{\kappa^2 \, v^2 \, \bar v^2_T}{m^6_\Phi}, \nn
\langle g_{\rm eff, pp}^{\mathcal{Z},d_R}\rangle^{[\hat{m}_{W}/\hat{\alpha}_{ew}]}_{\mathcal{O}(v^2/\Lambda^2)} &=[-0.22/0.53]\frac{\kappa^2\, \bar v^2_T}{m^4_\Phi} + [0.88/-0.21]\lambda \frac{\kappa^2  v^2 \, \bar v^2_T}{m^6_\Phi}, \nn
\langle g_{\rm eff, pp}^{\mathcal{Z},\ell_R}\rangle^{[\hat{m}_{W}/\hat{\alpha}_{ew}]}_{\mathcal{O}(v^2/\Lambda^2)} &=[-0.66/0.16]\frac{\kappa^2\, \bar v^2_T}{m^4_\Phi} + [2.6/-0.64]\lambda \frac{\kappa^2  v^2 \, \bar v^2_T}{m^6_\Phi},
\end{align}

\begin{align}
\langle g_{\rm eff, pp}^{\mathcal{Z},u_L}\rangle^{[\hat{m}_{W}/\hat{\alpha}_{ew}]}_{\mathcal{O}(v^2/\Lambda^2)} &= [0.25/-0.29]\frac{\kappa^2\, \bar v^2_T}{m^4_\Phi} + [-1.0/1.2]\lambda \frac{\kappa^2 \, v^2 \, \bar v^2_T}{m^6_\Phi}, \nn
\langle g_{\rm eff, pp}^{\mathcal{Z},d_L}\rangle^{[\hat{m}_{W}/\hat{\alpha}_{ew}]}_{\mathcal{O}(v^2/\Lambda^2)} &= [-0.034/0.24]\frac{\kappa^2\, \bar v^2_T}{m^4_\Phi} + [0.14/-0.95]\lambda \frac{\kappa^2 \, v^2 \, \bar v^2_T}{m^6_\Phi}, \nn
\langle g_{\rm eff, pp}^{\mathcal{Z},\ell_L}\rangle^{[\hat{m}_{W}/\hat{\alpha}_{ew}]}_{\mathcal{O}(v^2/\Lambda^2)} &= [-0.47/0.34]\frac{\kappa^2\, \bar v^2_T}{m^4_\Phi} + [1.9/-1.4]\lambda \frac{\kappa^2 \, v^2 \, \bar v^2_T}{m^6_\Phi}, \nn
\langle g_{\rm eff, pp}^{\mathcal{Z},\nu_L}\rangle^{[\hat{m}_{W}/\hat{\alpha}_{ew}]}_{\mathcal{O}(v^2/\Lambda^2)} &= [-0.19/-0.19]\frac{\kappa^2\, \bar v^2_T}{m^4_\Phi} + [0.74/0.74]\lambda \frac{\kappa^2 \, v^2 \, \bar v^2_T}{m^6_\Phi},
\end{align}
with $\mathcal{O}(v^4/\Lambda^4)$ corrections:
\begin{align}
\langle g_{\rm eff, pp}^{\mathcal{Z},u_R}\rangle^{[\hat{m}_{W}/\hat{\alpha}_{ew}]}_{\mathcal{O}(v^4/\Lambda^4)} &=[-1.5/0.49]\frac{\kappa^4\, \bar v^4_T}{m^8_\Phi} + [-0.88/0.21]\eta\frac{\kappa^2\, \bar v^4_T}{m^6_\Phi} + [1.8/-0.42]\lambda\frac{\kappa^2\, v^2 \, \bar{v}_T^2}{m^6_\Phi},\nn
\langle g_{\rm eff, pp}^{\mathcal{Z},d_R}\rangle^{[\hat{m}_{W}/\hat{\alpha}_{ew}]}_{\mathcal{O}(v^4/\Lambda^4)} &=[0.76/-0.24]\frac{\kappa^4\, \bar v^4_T}{m^8_\Phi} + [0.44/-0.11]\eta\frac{\kappa^2\, \bar v^4_T}{m^6_\Phi} + [-0.88/0.21]\lambda\frac{\kappa^2\, v^2 \, \bar{v}_T^2}{m^6_\Phi}, \nn
\langle g_{\rm eff, pp}^{\mathcal{Z},\ell_R}\rangle^{[\hat{m}_{W}/\hat{\alpha}_{ew}]}_{\mathcal{O}(v^4/\Lambda^4)} &=[2.3/-0.73]\frac{\kappa^4\, \bar v^4_T}{m^8_\Phi} + [1.3/-0.32]\eta\frac{\kappa^2\, \bar v^4_T}{m^6_\Phi} + [-2.6/0.64]\lambda\frac{\kappa^2\,v^2 \, \bar{v}_T^2}{m^6_\Phi},
\end{align}
\begin{align}
\langle g_{\rm eff, pp}^{\mathcal{Z},u_L}\rangle^{[\hat{m}_{W}/\hat{\alpha}_{ew}]}_{\mathcal{O}(v^4/\Lambda^4)} &=[-0.92/1.1]\frac{\kappa^4\, \bar v^4_T}{m^8_\Phi} + [-0.51/0.6]\eta\frac{\kappa^2\, \bar v^4_T}{m^6_\Phi} + [1.0/-1.2]\lambda\frac{\kappa^2\, v^2 \, \bar{v}_T^2}{m^6_\Phi}, \nn
\langle g_{\rm eff, pp}^{\mathcal{Z},d_L}\rangle^{[\hat{m}_{W}/\hat{\alpha}_{ew}]}_{\mathcal{O}(v^4/\Lambda^4)} &=[0.16/-0.84]\frac{\kappa^4\, \bar v^4_T}{m^8_\Phi} + [0.068/-0.48]\eta\frac{\kappa^2\, \bar v^4_T}{m^6_\Phi} + [-0.14/0.95]\lambda\frac{\kappa^2\, v^2 \, \bar{v}_T^2}{m^6_\Phi}, \nn
\langle g_{\rm eff, pp}^{\mathcal{Z},\ell_L}\rangle^{[\hat{m}_{W}/\hat{\alpha}_{ew}]}_{\mathcal{O}(v^4/\Lambda^4)} &=[1.7/-1.3]\frac{\kappa^4\, \bar v^4_T}{m^8_\Phi} + [0.95/-0.69]\eta\frac{\kappa^2\, \bar v^4_T}{m^6_\Phi} + [-1.9/1.4]\lambda\frac{\kappa^2\, v^2 \, \bar{v}_T^2}{m^6_\Phi}, \nn
\langle g_{\rm eff, pp}^{\mathcal{Z},\nu_L}\rangle^{[\hat{m}_{W}/\hat{\alpha}_{ew}]}_{\mathcal{O}(v^4/\Lambda^4)} &= [0.60/0.60]\frac{\kappa^4\, \bar v^4_T}{m^8_\Phi} + [0.37/0.37]\eta\frac{\kappa^2\, \bar v^4_T}{m^6_\Phi} + [-0.74/-0.74]\lambda\frac{\kappa^2\,v^2 \, \bar{v}_T^2}{m^6_\Phi},\nn
\end{align}

Again we note that the $\lambda$ dependence exactly cancels between matching effects at $\mathcal{L}^{(6)}$
and $\mathcal{L}^{(8)}$ in observables. This occurs at the effective coupling level, as the
on shell $\mathcal{Z}$ effective couplings are closely related to observable quantities.
\begin{figure}
\includegraphics[height=0.24\textheight,width=0.45\textwidth]{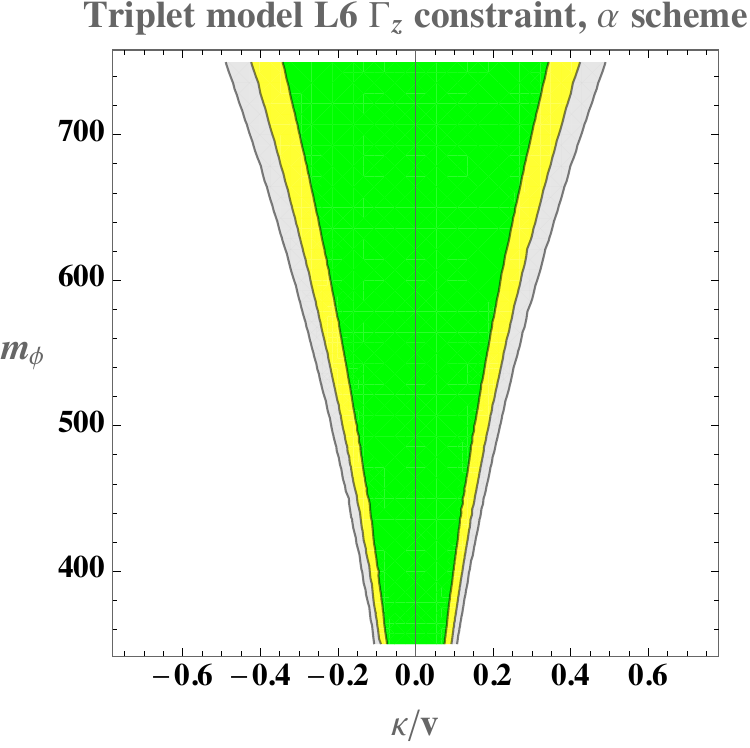}
\includegraphics[height=0.24\textheight,width=0.45\textwidth]{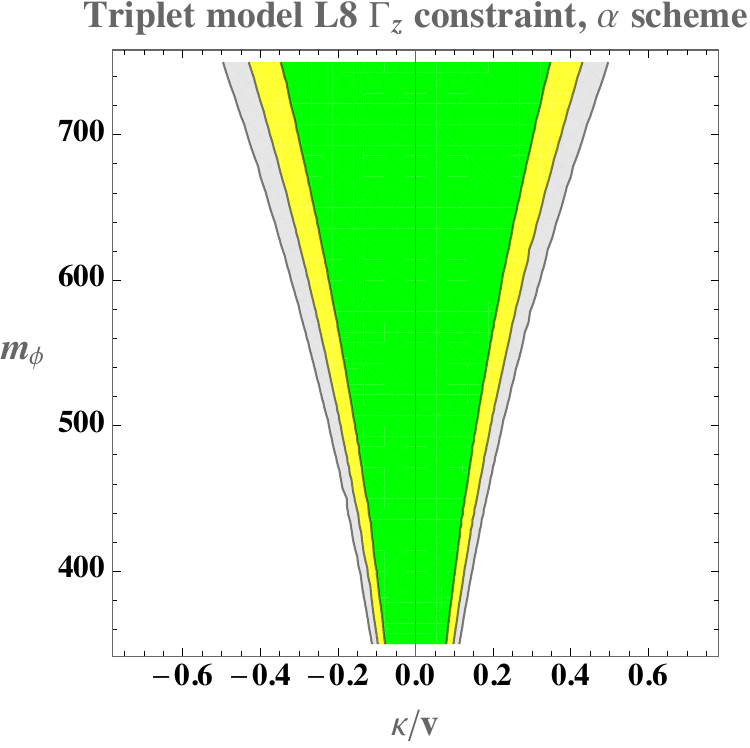}
\includegraphics[height=0.24\textheight,width=0.45\textwidth]{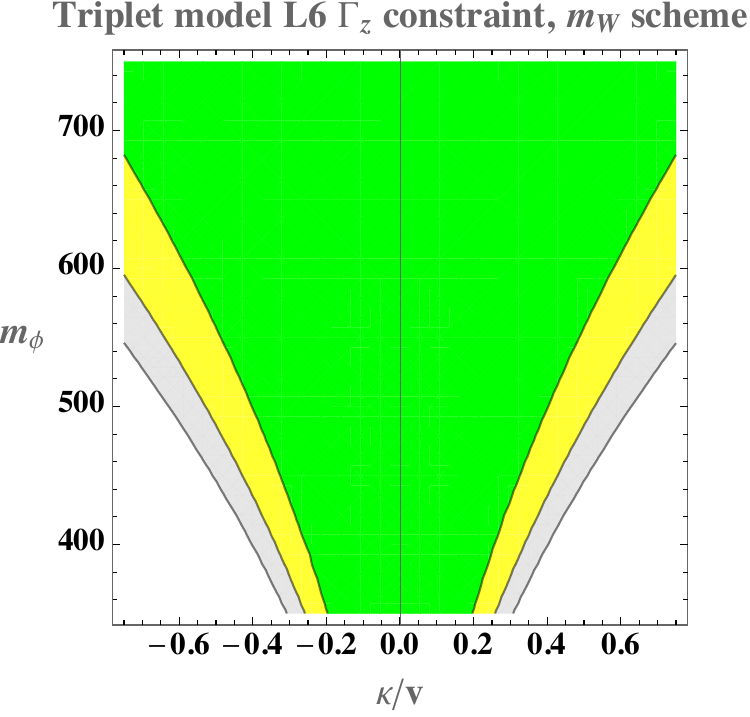}
\includegraphics[height=0.24\textheight,width=0.45\textwidth]{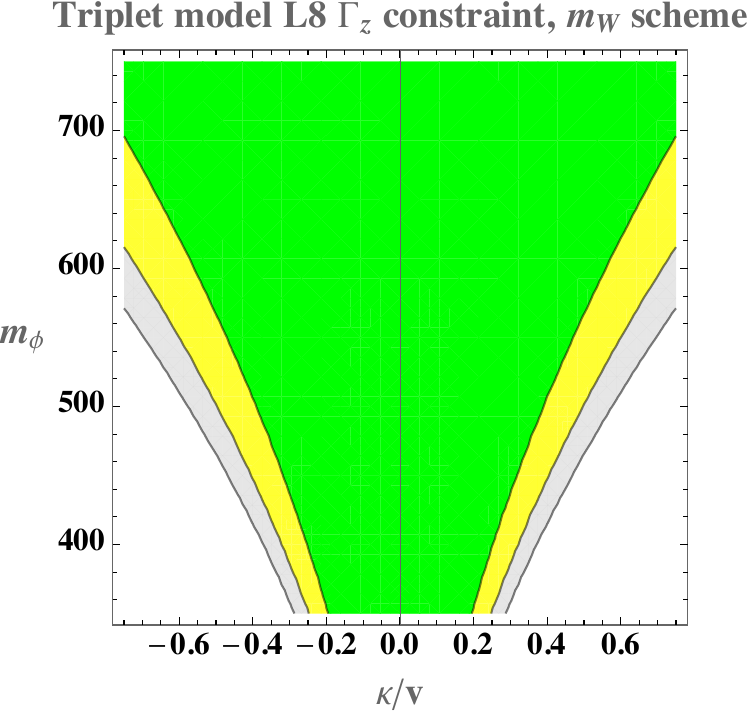}
\includegraphics[height=0.24\textheight,width=0.45\textwidth]{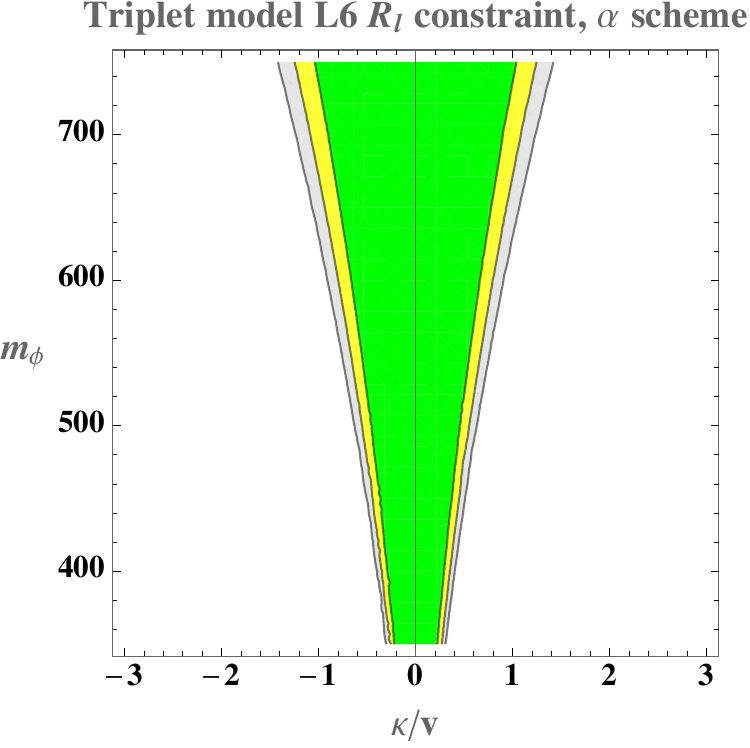}
\includegraphics[height=0.24\textheight,width=0.45\textwidth]{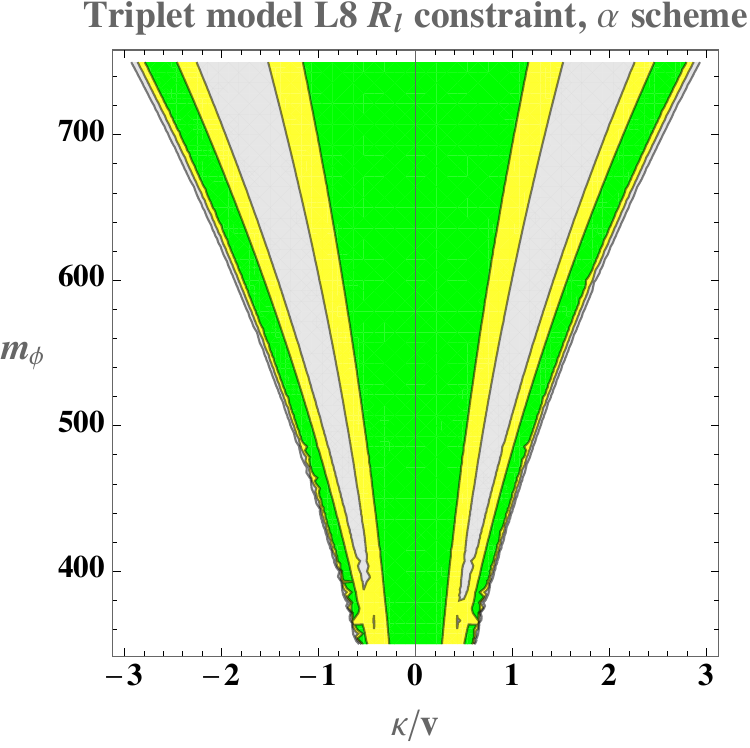}
\includegraphics[height=0.24\textheight,width=0.45\textwidth]{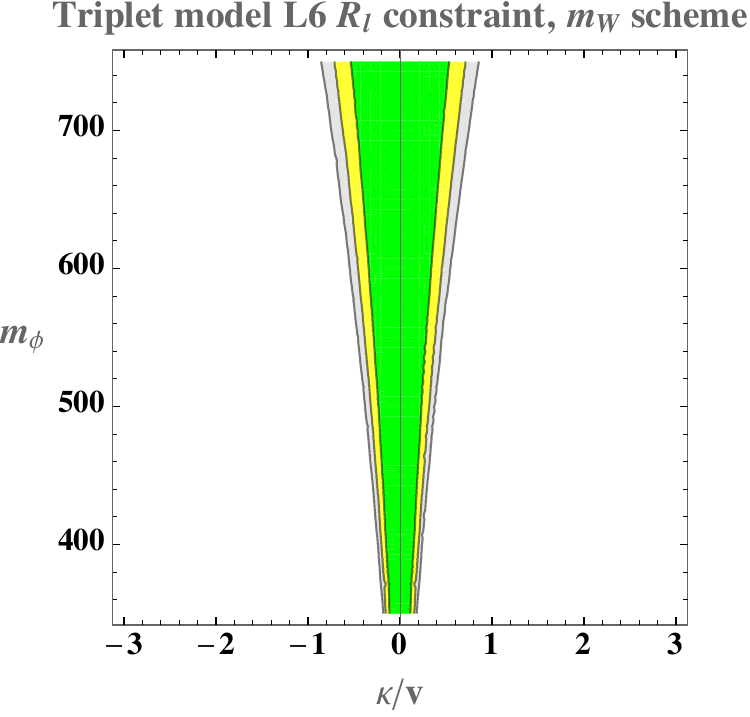}
\includegraphics[height=0.24\textheight,width=0.45\textwidth]{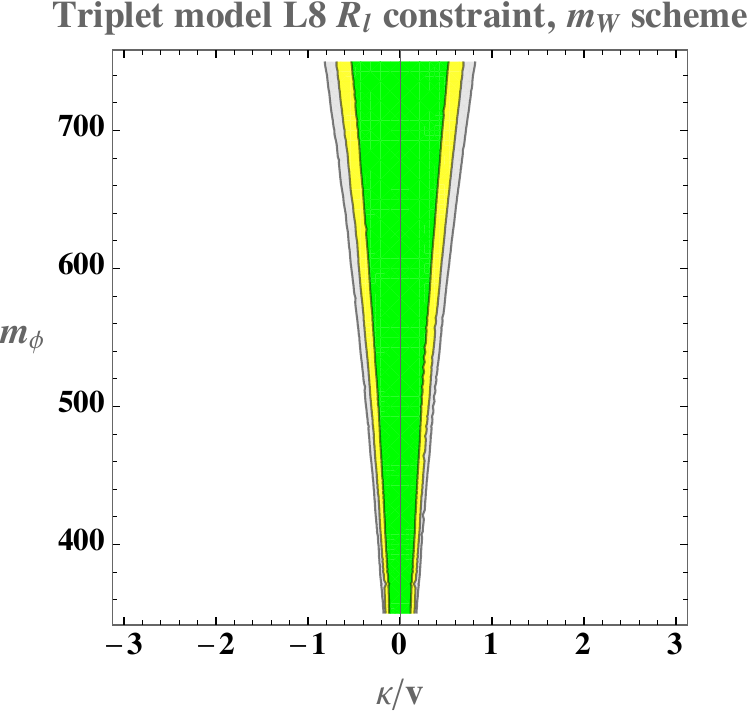}
\caption{Constraints from EWPD observables in the scalar triplet model.
Same conventions as in previous plots.
Shown are the constraints on the model
space from the $\Gamma_Z$ and $R_\ell$ observables.\label{scalarplotfig1}}
\end{figure}

\begin{figure}
\includegraphics[height=0.24\textheight,width=0.45\textwidth]{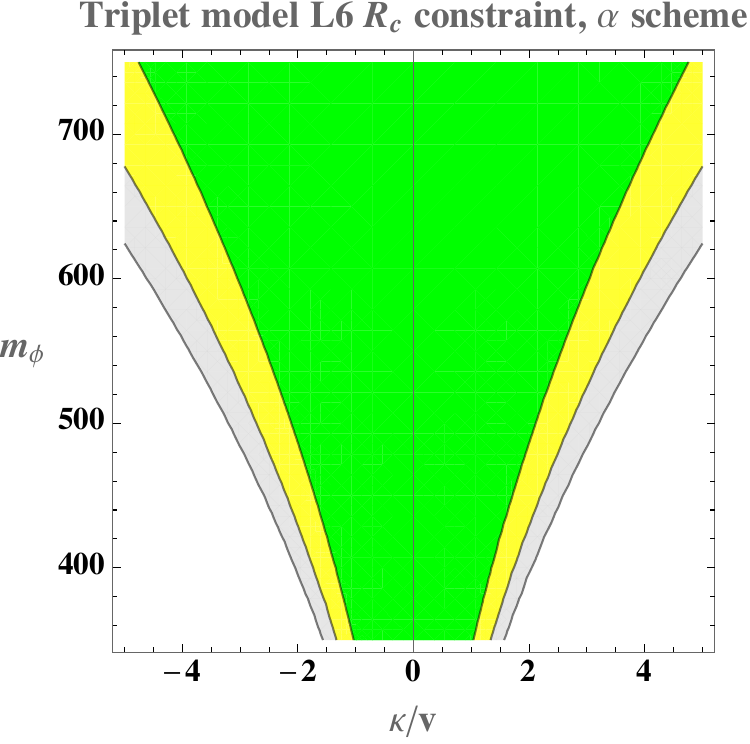}
\includegraphics[height=0.24\textheight,width=0.45\textwidth]{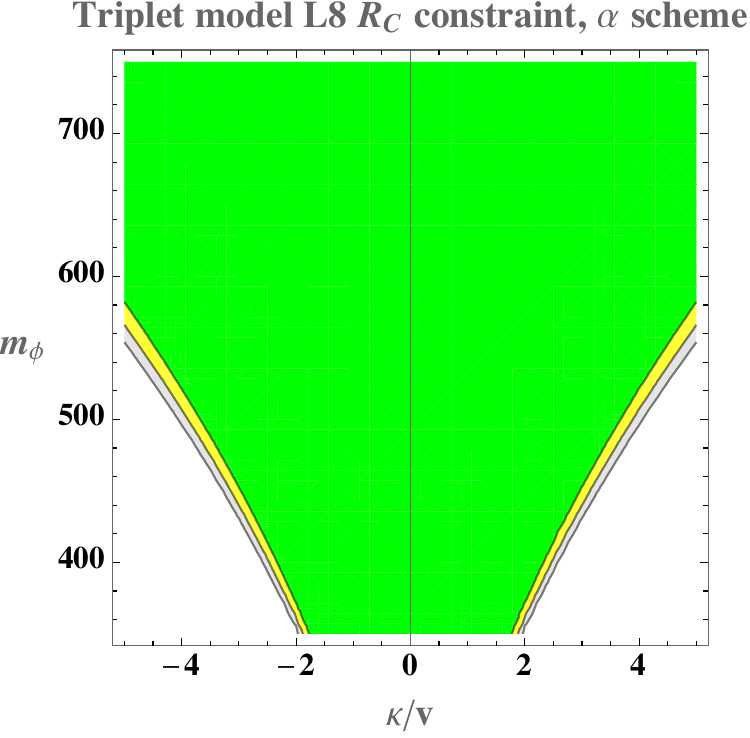}
\includegraphics[height=0.24\textheight,width=0.45\textwidth]{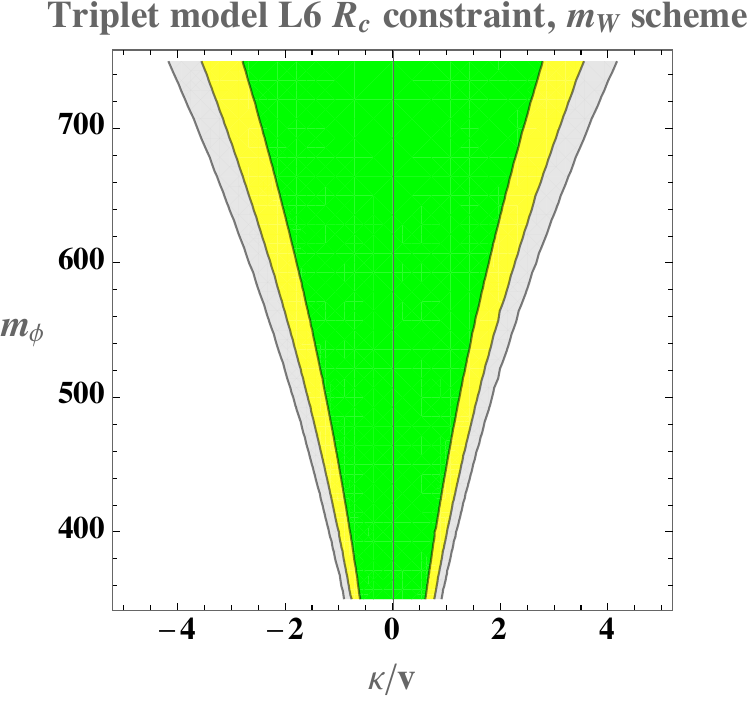}
\includegraphics[height=0.24\textheight,width=0.45\textwidth]{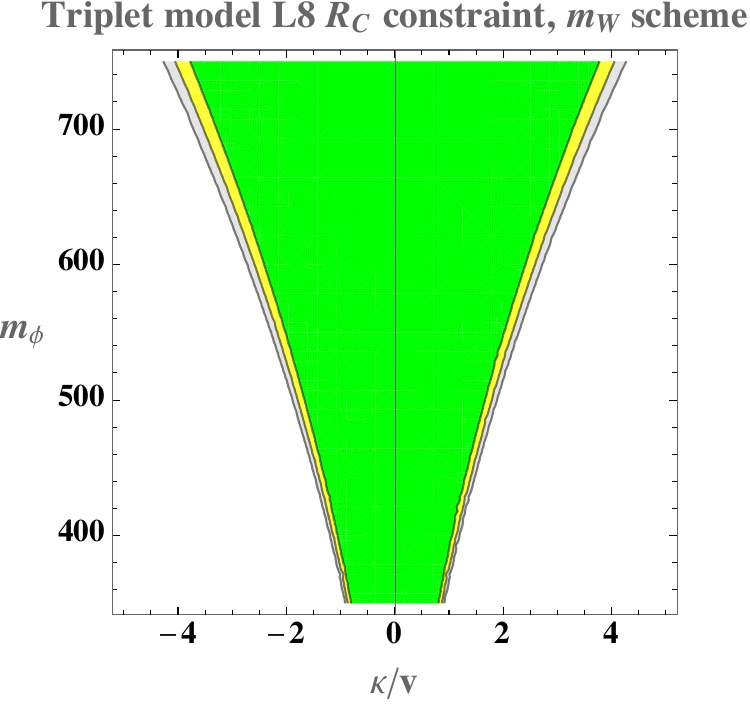}
\includegraphics[height=0.24\textheight,width=0.45\textwidth]{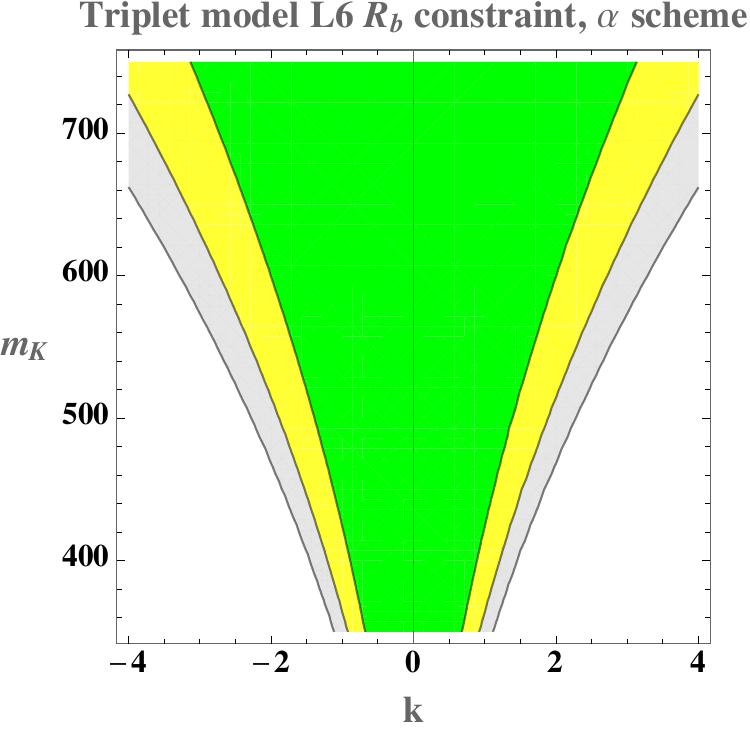}
\includegraphics[height=0.24\textheight,width=0.45\textwidth]{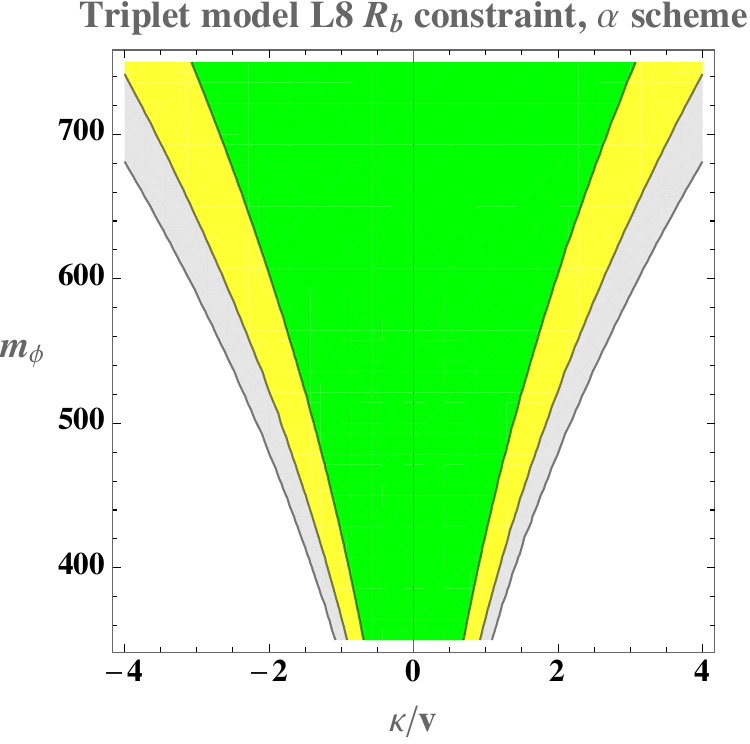}
\includegraphics[height=0.24\textheight,width=0.45\textwidth]{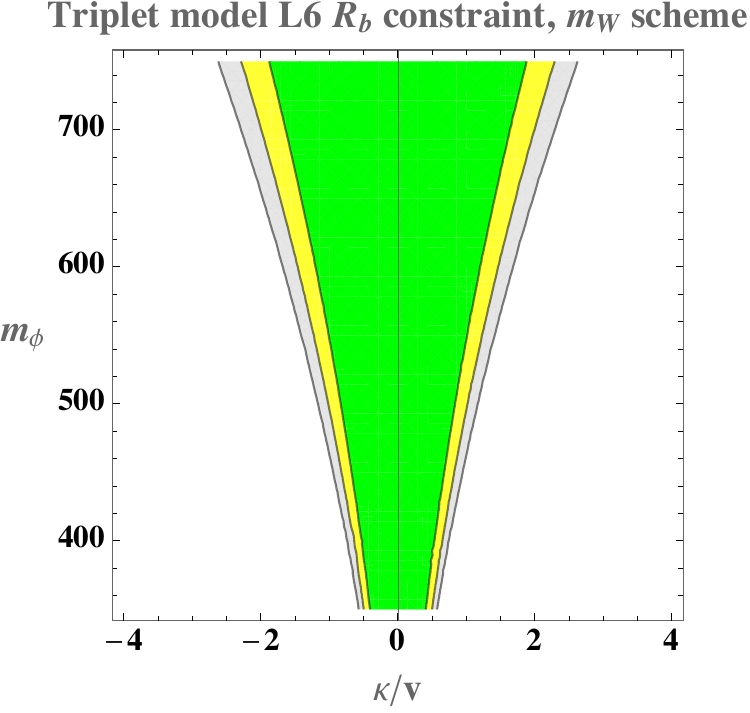}
\includegraphics[height=0.24\textheight,width=0.45\textwidth]{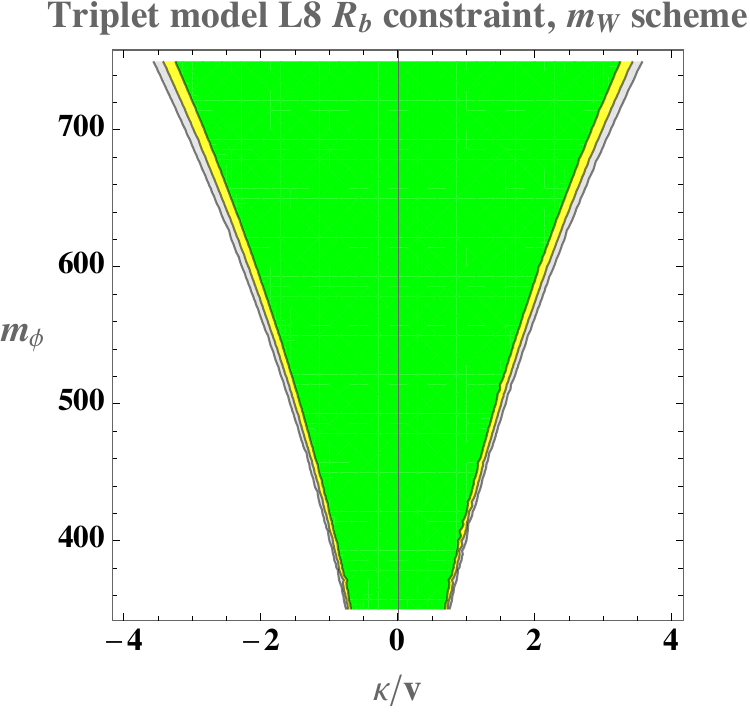}
\caption{Shown are the constraints on the model
space from the $R_c$ and $R_b$ EWPD observables.\label{scalarplotfig2}}
\end{figure}

\begin{figure}
\includegraphics[height=0.24\textheight,width=0.45\textwidth]{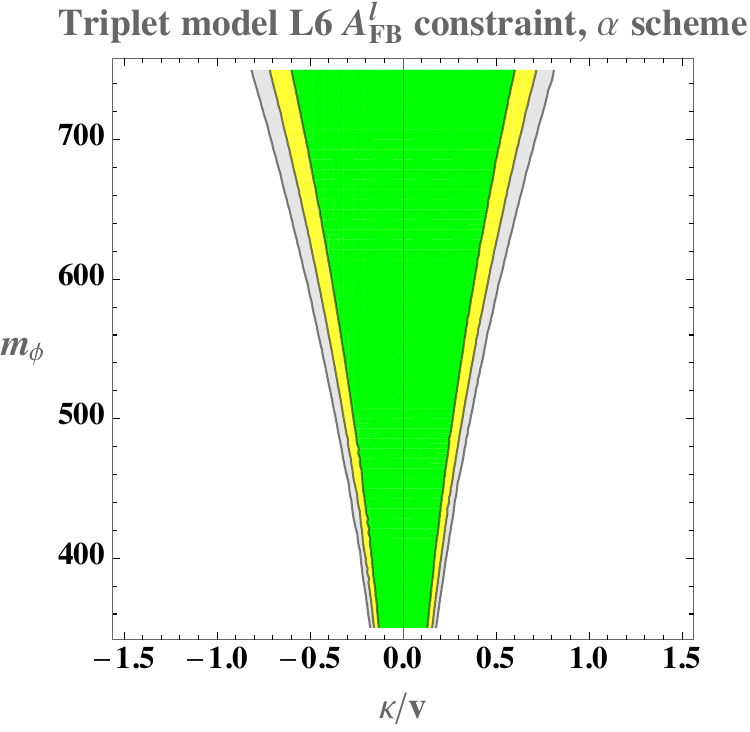}
\includegraphics[height=0.24\textheight,width=0.45\textwidth]{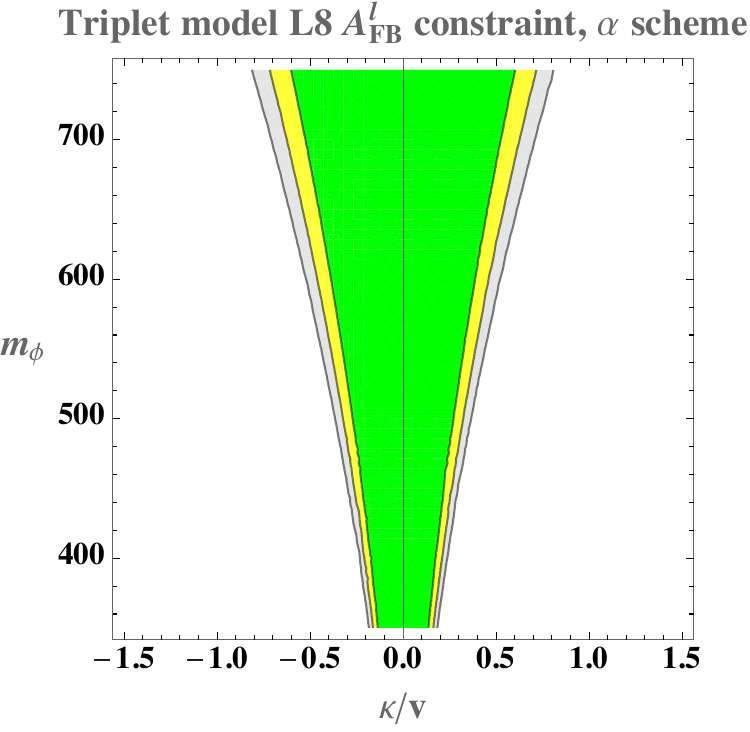}
\includegraphics[height=0.24\textheight,width=0.45\textwidth]{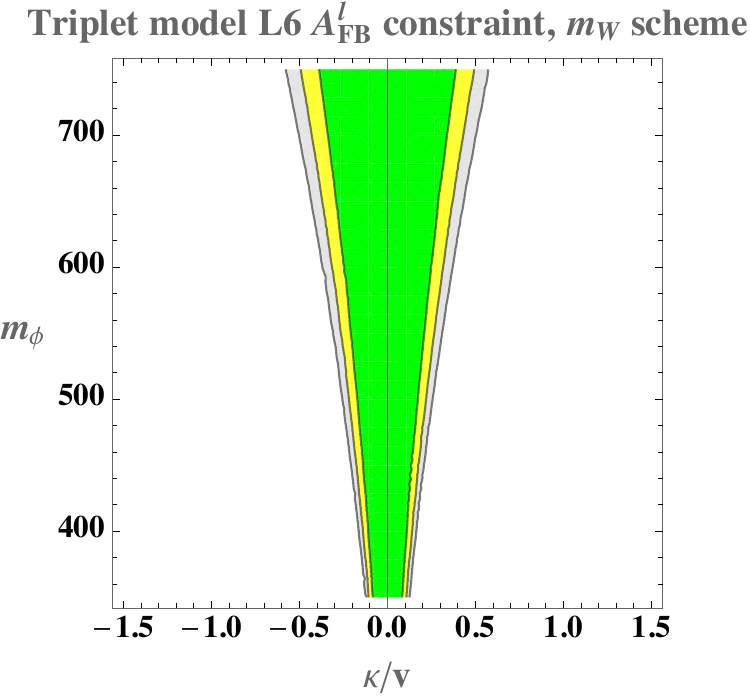}
\includegraphics[height=0.24\textheight,width=0.45\textwidth]{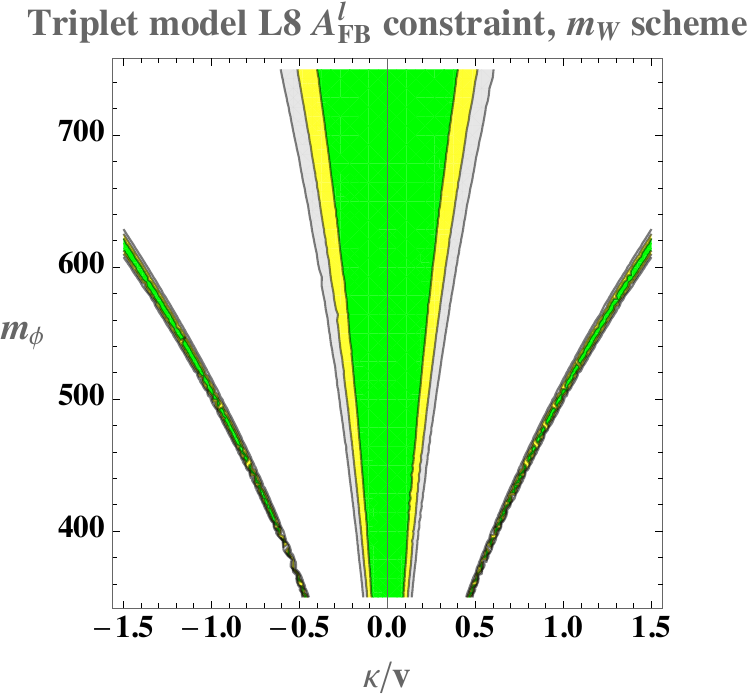}
\includegraphics[height=0.24\textheight,width=0.45\textwidth]{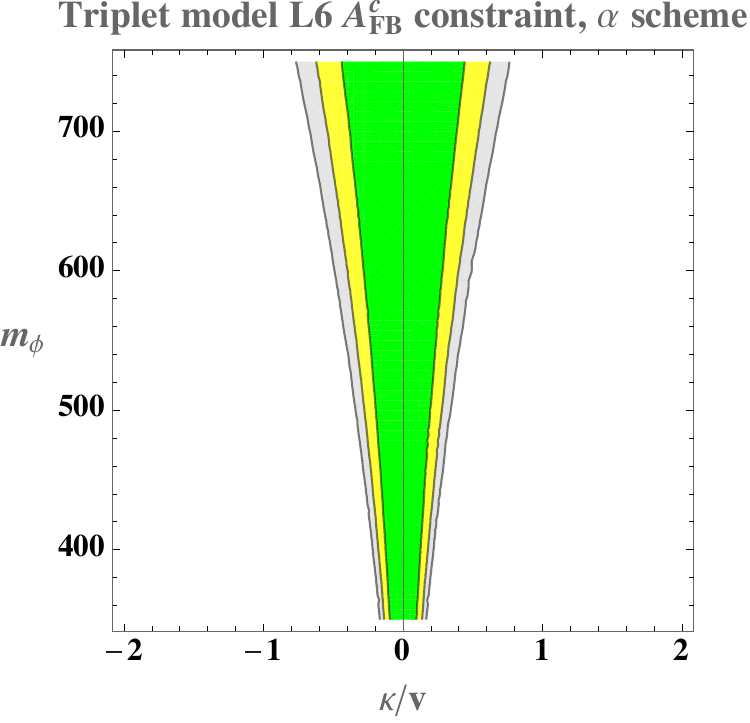}
\includegraphics[height=0.24\textheight,width=0.45\textwidth]{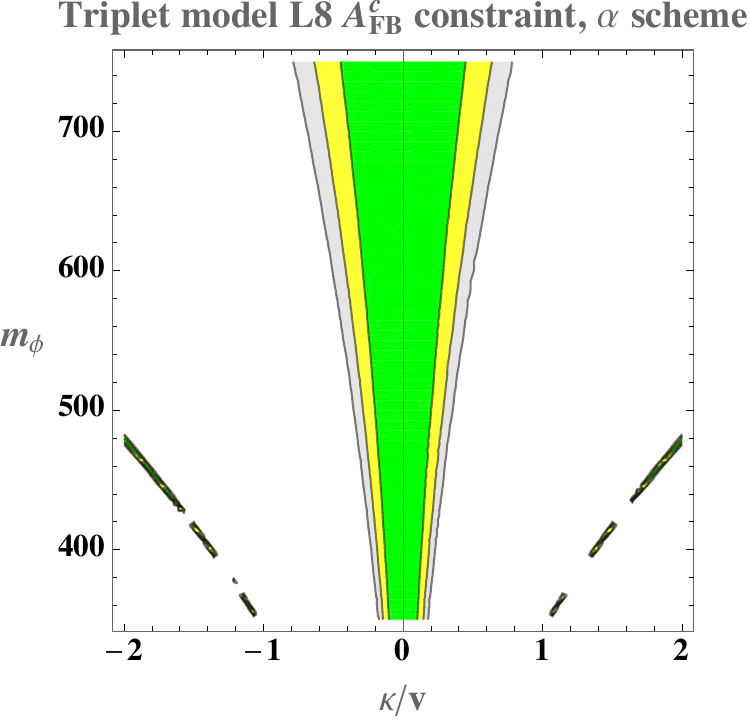}
\includegraphics[height=0.24\textheight,width=0.45\textwidth]{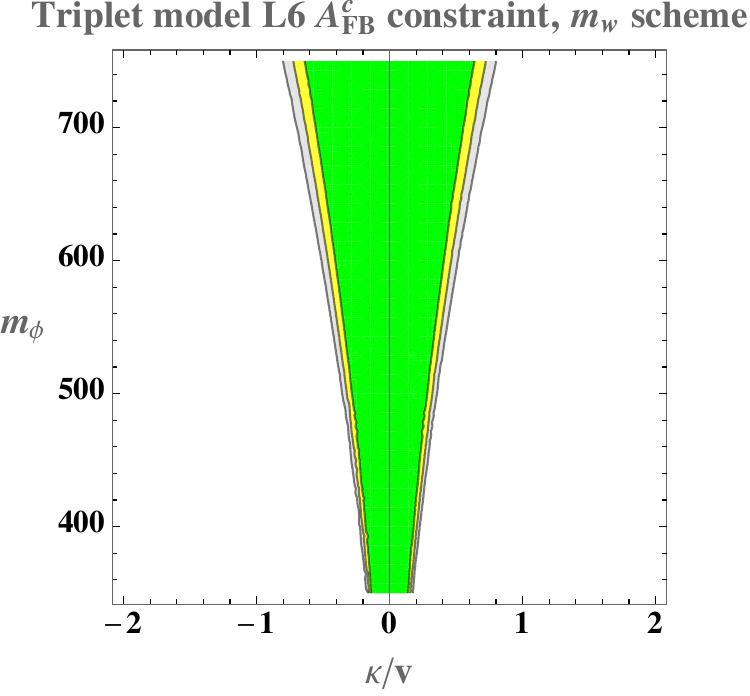}
\includegraphics[height=0.24\textheight,width=0.45\textwidth]{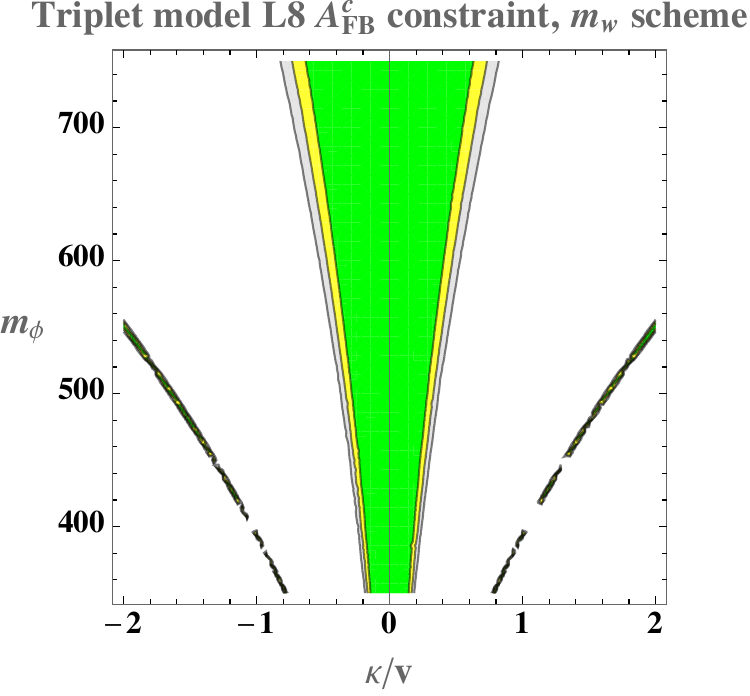}
\caption{Shown are the constraints on the model
space from the $A_{FB}^\ell$ and $A_{FB}^c$ EWPD observables.\label{scalarplotfig3}}
\end{figure}

\begin{figure}
\includegraphics[height=0.24\textheight,width=0.45\textwidth]{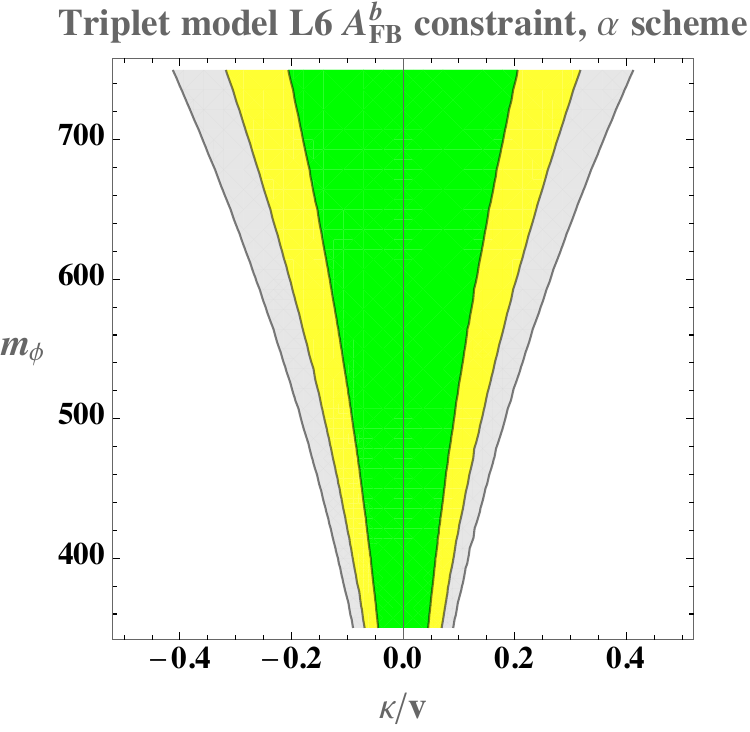}
\includegraphics[height=0.24\textheight,width=0.45\textwidth]{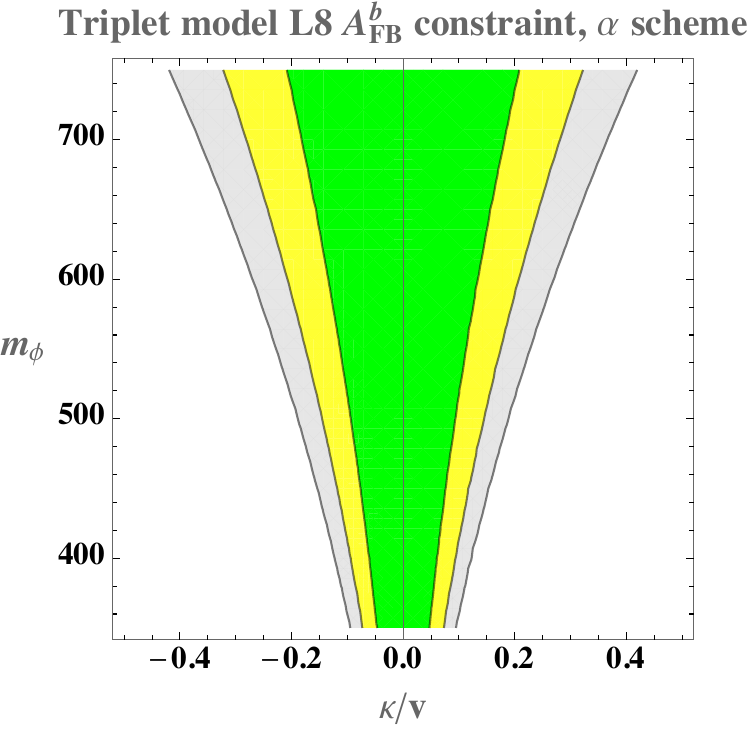}
\includegraphics[height=0.24\textheight,width=0.45\textwidth]{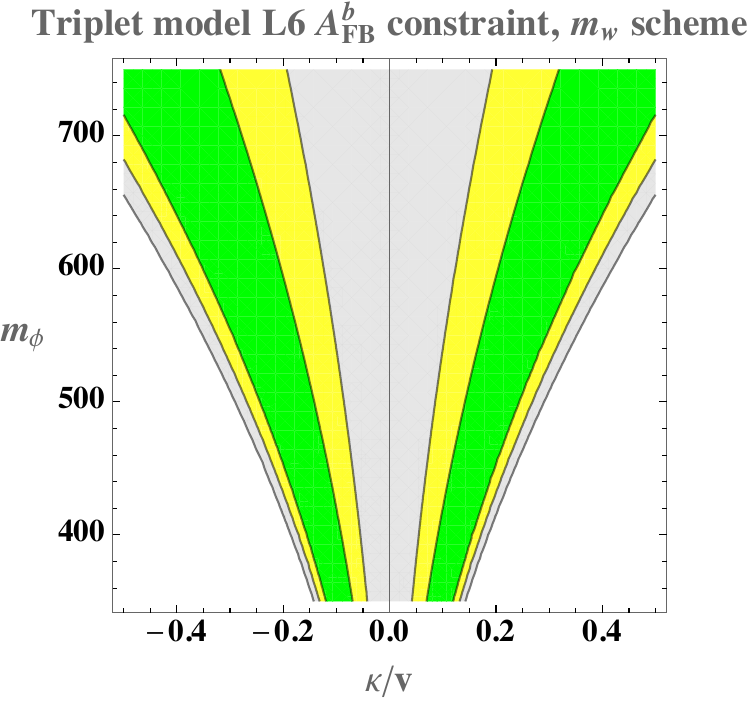}
\includegraphics[height=0.24\textheight,width=0.45\textwidth]{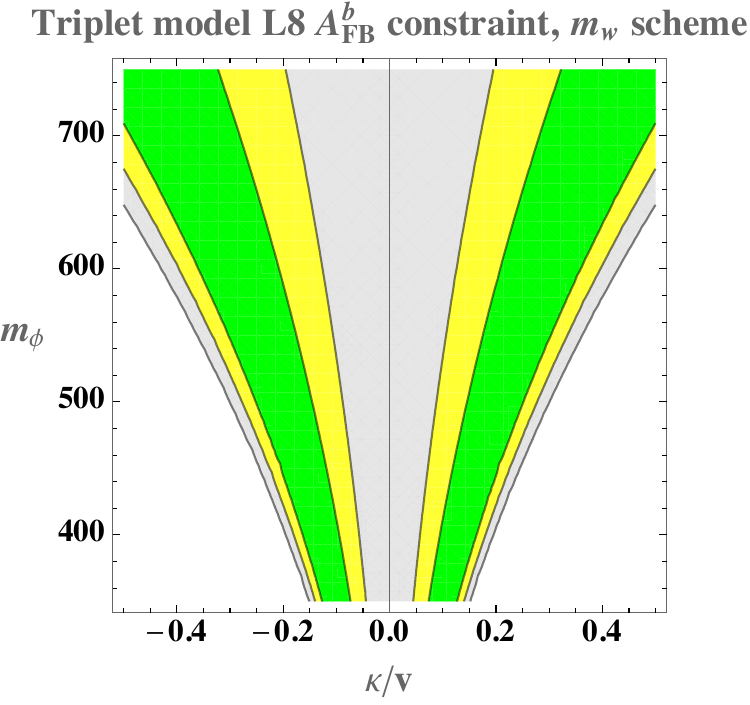}
\includegraphics[height=0.24\textheight,width=0.45\textwidth]{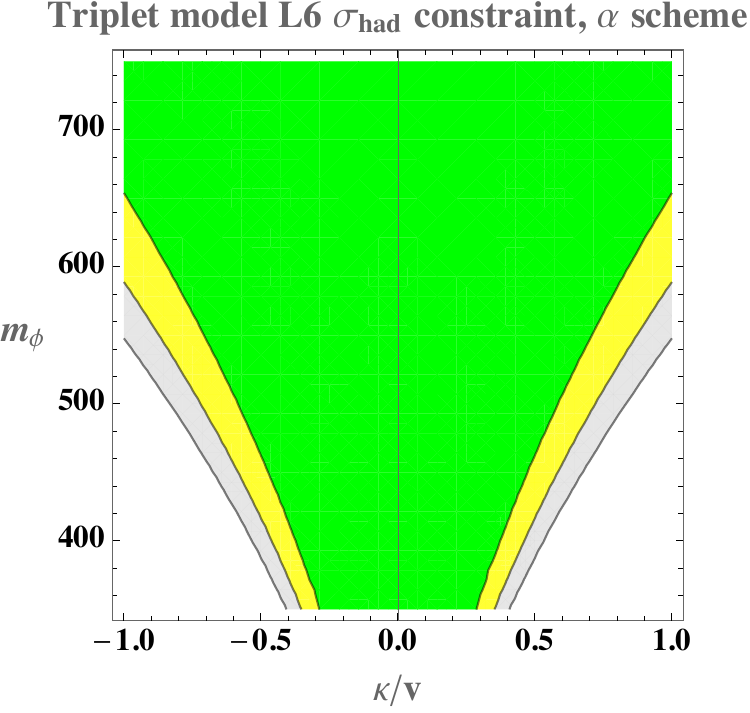}
\includegraphics[height=0.24\textheight,width=0.45\textwidth]{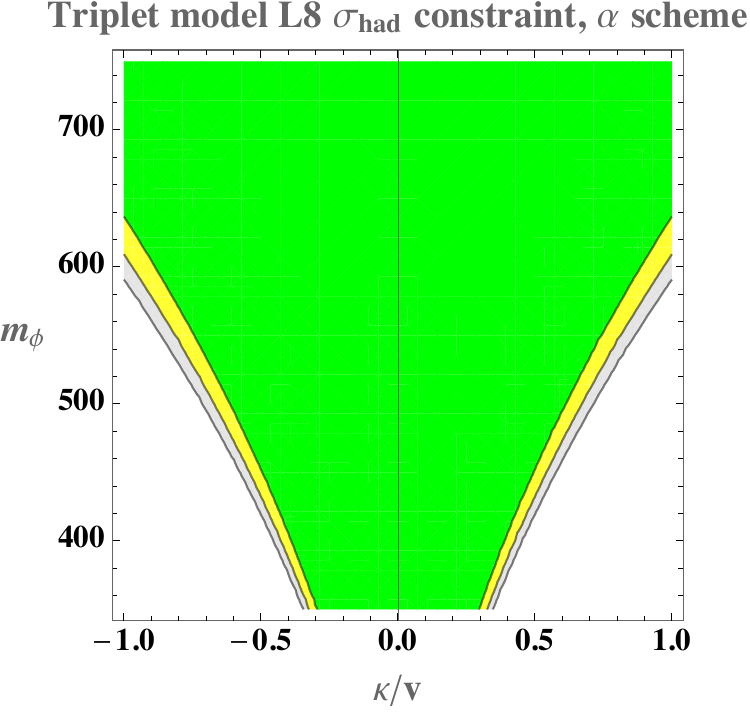}
\includegraphics[height=0.24\textheight,width=0.45\textwidth]{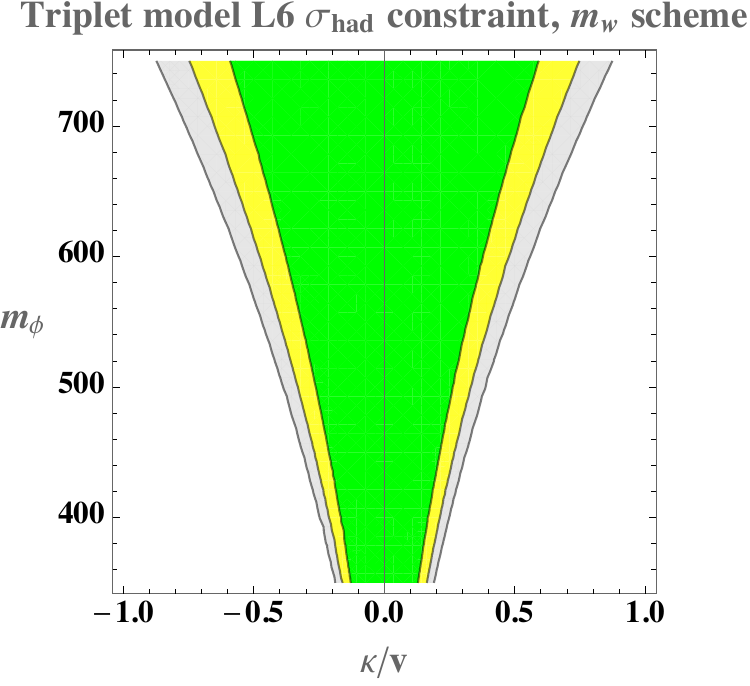}
\includegraphics[height=0.24\textheight,width=0.45\textwidth]{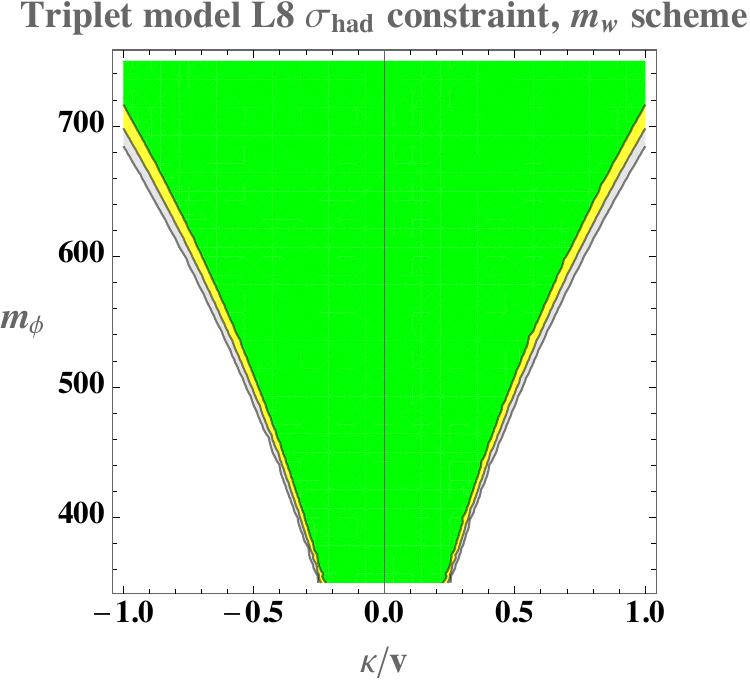}
\caption{Shown are the constraints on the model
space from the $A_{FB}^b$ and $\sigma_{had}^0$ EWPD observables.\label{scalarplotfig4}}
\end{figure}

The total width is
\bea
 \frac{\sum_{\psi} \bar{\Gamma}^{{\rm SMEFT},\hat{\alpha}_{ew}}_{\mathcal{Z} \rightarrow \bar{\psi}_p \psi_p}}{\sum_{\psi} \hat{\Gamma}^{{\rm SM}, \hat{\alpha}_{ew}}_{\mathcal{Z} \rightarrow \bar{\psi}_p \psi_p}}
&=& 1 + 1.3 \frac{\kappa^2\, \bar v^2_T}{m^4_\Phi} - 2.6\,\eta\frac{\kappa^2\, \bar v^4_T}{m^6_\Phi} - 3.8\,\frac{\kappa^4\, \bar v^4_T}{m^8_\Phi},
\eea
\bea
 \frac{\sum_{\psi} \bar{\Gamma}^{{\rm SMEFT}, \hat{m}_{W}}_{\mathcal{Z} \rightarrow \bar{\psi}_p \psi_p}}{\sum_{\psi} \hat{\Gamma}^{{\rm SM},\hat{m}_{W}}_{\mathcal{Z} \rightarrow \bar{\psi}_p \psi_p}}
&=& 1 + 0.14 \frac{\kappa^2\, \bar v^2_T}{m^4_\Phi} - 0.28\,\eta\frac{\kappa^2\, \bar v^4_T}{m^6_\Phi} + 1.7\,\frac{\kappa^4\, \bar v^4_T}{m^8_\Phi}.
\eea

The remaining EWPD observables, with numerical values ordered in the $[\hat{m}_W,\alpha_{EW}]$ schemes, are
\bea
\frac{\bar{R}_{c}^{\textrm{SMEFT}}}{\hat{R}_{c}^{\textrm{SM}}} = [-0.30/0.098]\frac{\kappa^2\, \bar v^2_T}{m^4_\Phi} + [0.60/-0.20]\eta\frac{\kappa^2\, \bar v^4_T}{m^6_\Phi} + [2.7/-0.16]\frac{\kappa^4\, \bar v^4_T}{m^8_\Phi},
\eea

\bea
\frac{\bar{R}_{b}^{\textrm{SMEFT}}}{\hat{R}_{b}^{\textrm{SM}}} = [0.15/-0.030]\frac{\kappa^2\, \bar v^2_T}{m^4_\Phi} + [-0.31/0.059]\eta\frac{\kappa^2\, \bar v^4_T}{m^6_\Phi} + [-1.4/-0.016]\frac{\kappa^4\, \bar v^4_T}{m^8_\Phi},
\eea

\bea
\frac{\bar{R}_{\ell}^{\textrm{SMEFT}}}{\hat{R}_{\ell}^{\textrm{SM}}} =  [-0.46/0.22]\frac{\kappa^2\, \bar v^2_T}{m^4_\Phi} + [0.92/-0.44]\eta\frac{\kappa^2\, \bar v^4_T}{m^6_\Phi} + [-6.3/-2.5]\frac{\kappa^4\, \bar v^4_T}{m^8_\Phi},
\eea

\bea
\frac{(\bar{\sigma}_{had}^{0})^{\textrm{SMEFT}}}{(\hat{\sigma}_{had}^{0})^{\textrm{SM}}} =  [-0.082/-0.019]\frac{\kappa^2\, \bar v^2_T}{m^4_\Phi} + [0.16/0.039]\eta\frac{\kappa^2\, \bar v^4_T}{m^6_\Phi} + [7.2/1.3]\frac{\kappa^4\, \bar v^4_T}{m^8_\Phi},
\eea

\bea
\frac{(\bar{A}_{FB}^{0,c})^{\textrm{SMEFT}}}{(\hat{A}_{FB}^{0,c})^{\textrm{SM}}} = [-32/19]\frac{\kappa^2\, \bar v^2_T}{m^4_\Phi} + [63/-38]\eta\frac{\kappa^2\, \bar v^4_T}{m^6_\Phi} + [190/-65]\frac{\kappa^4\, \bar v^4_T}{m^8_\Phi},
\eea

\bea
\frac{(\bar{A}_{FB}^{0,b})^{\textrm{SMEFT}}}{(\hat{A}_{FB}^{0,b})^{\textrm{SM}}} = [-28/17]\frac{\kappa^2\, \bar v^2_T}{m^4_\Phi} + [57/-35]\eta\frac{\kappa^2\, \bar v^4_T}{m^6_\Phi} + [85/-82]\frac{\kappa^4\, \bar v^4_T}{m^8_\Phi},
\eea

\bea
\frac{(\bar{A}_{FB}^{0,\ell})^{\textrm{SMEFT}}}{(\hat{A}_{FB}^{0,\ell})^{\textrm{SM}}} =  [-56/34]\frac{\kappa^2\, \bar v^2_T}{m^4_\Phi} + [110/-69]\eta\frac{\kappa^2\, \bar v^4_T}{m^6_\Phi} + [950/130]\frac{\kappa^4\, \bar v^4_T}{m^8_\Phi}.
\eea

\begin{figure}
\includegraphics[height=0.24\textheight,width=0.45\textwidth]{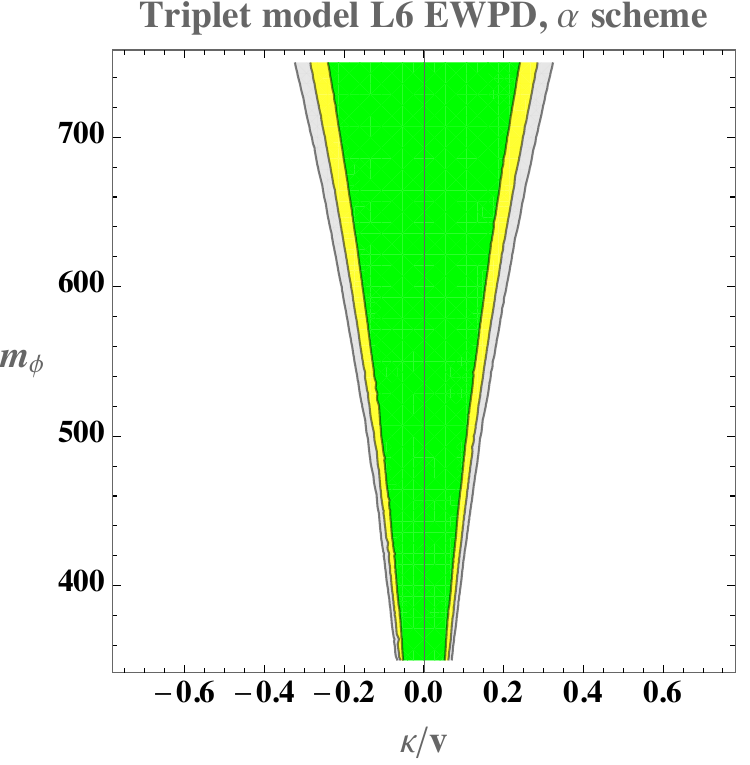}
\includegraphics[height=0.24\textheight,width=0.45\textwidth]{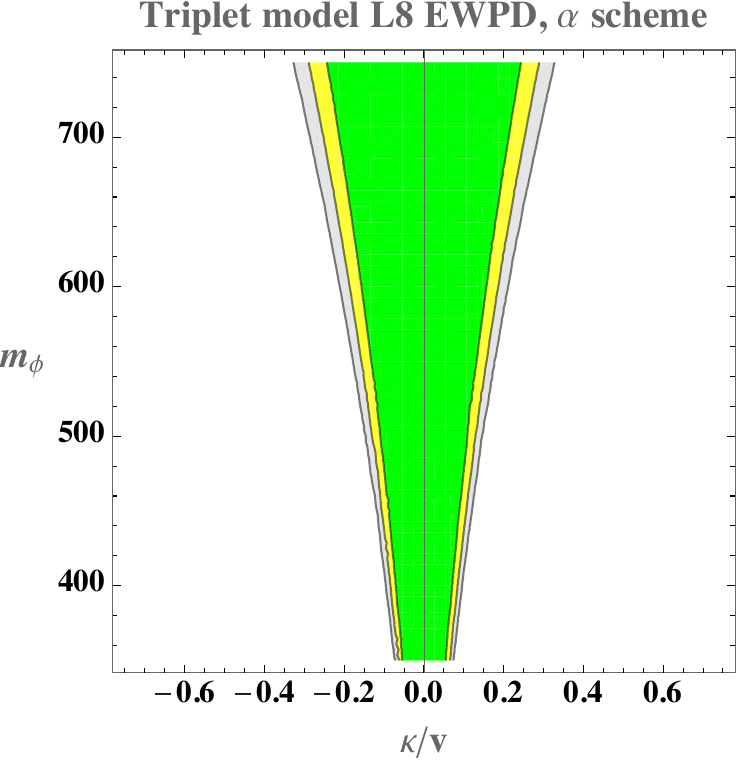}
\includegraphics[height=0.24\textheight,width=0.45\textwidth]{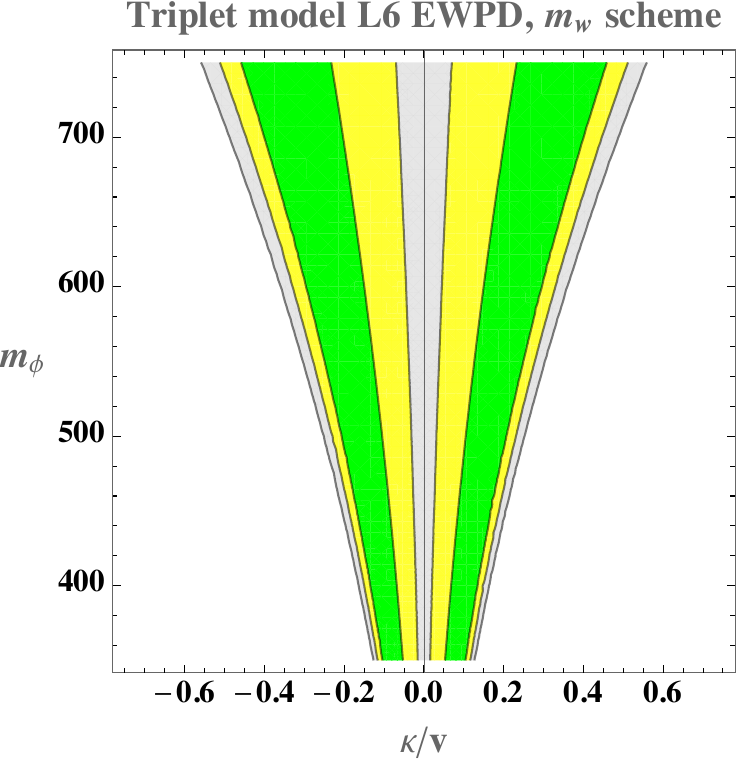}
\includegraphics[height=0.24\textheight,width=0.45\textwidth]{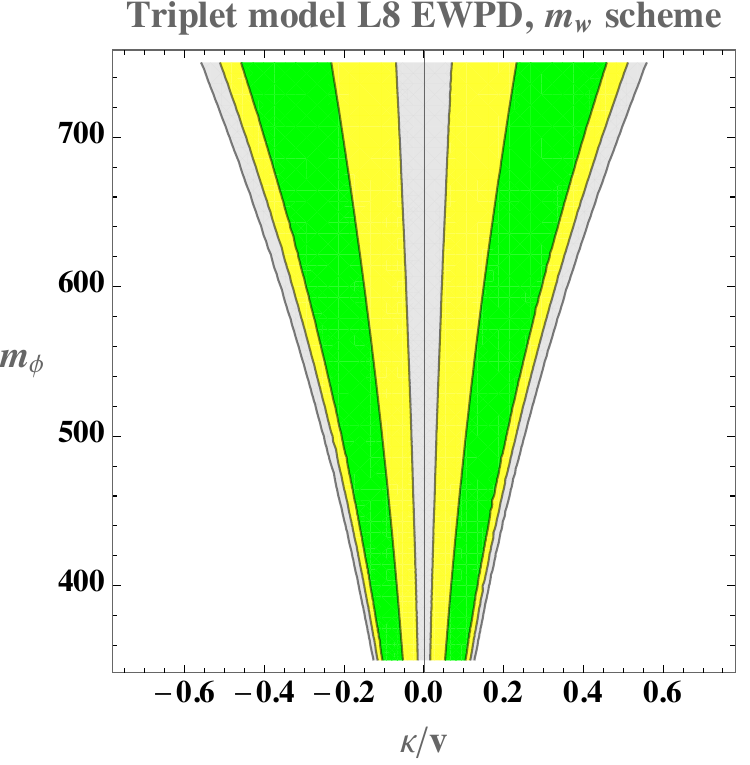}
\caption{Combined constraints from the full set of EWPD LEP observables in the scalar
triplet model with $\eta = 0.1$.\label{scalarplotfig5}}
\end{figure}

Again, the above expressions are the perturbations to the SM, as such there is an implied $1+$ in each equation. Figures~\ref{scalarplotfig1},~\ref{scalarplotfig2},~\ref{scalarplotfig3},~\ref{scalarplotfig4},~and~\ref{scalarplotfig5} show the effects for each observable and their combination.
The effect of the $\mathcal{L}^{(8)}$ corrections in the scalar triplet model also
damp in significance as more observables are consistently combined in the SMEFT into a global fit, as seen in Fig.~\ref{scalarplotfig5}.
Once again, significant input parameter scheme dependence remains in the global fit combination.
The results are shown for the value $\eta = 0.1$, and are simply illustrative. The same numerical
behavior is present for other values of $\eta$.

\section{Discussion and Conclusions}
In this paper we have developed and reported the first analysis of EWPD
in the SMEFT to dimension eight. This result was enabled by the geoSMEFT formulation of the
SMEFT reported in Refs.~\cite{Helset:2020yio,Hays:2020scx}. The interpretation of EWPD
in the SMEFT has been subject to significant controversy over the years.
A cautious interpretation of EWPD has been advocated in some works, when  determining constraints
on $\mathcal{L}^{(6)}$ parameters in the SMEFT, due to the neglect of loop corrections,
and dimension-eight operator effects in a leading order SMEFT analysis of LEP data.
More aggressive interpretations of LEP data in the SMEFT have also been advanced.
All of the unknown corrections leading to past differences of opinion
are calculable in a well-defined formulation of the SMEFT. Recently, loop corrections
for LEP observables in the SMEFT have been
reported in Refs.~\cite{Hartmann:2016pil,Dawson:2018liq,Dawson:2019clf}. In this paper we have
reported the dimension-eight corrections; see Appendix~\ref{loopvsdim8} for a comparative discussion of loop corrections versus $\mathcal O(v^4/\Lambda^4)$ effects, both from the bottom-up SMEFT perspective and for the two UV models discussed.
The key point is that such calculable corrections introduce
more parameters into the predictions of LEP observables, compared to the number of parameters
present in a naive $\mathcal{L}^{(6)}$ SMEFT analysis.
We encourage the reader to draw their own conclusion on how the interpretation of LEP EWPD
is impacted by the dimension-eight contributions that are now known, and reported in this work.

\acknowledgments
We thank Chris Hays for insightful discussions.
M.T. acknowledges support from the Villum Fund, project number 00010102.
T.C. acknowledges funding from European Union’s Horizon 2020 research and innovation programme under the Marie Sklodowska-Curie grant agreement No. 890787.
The work of A.M. is partially supported by the National Science Foundation under Grant No. Phy-1230860.
A.H. is supported by the U.S. Department of Energy (DOE) under Award Number DE-SC0011632 and by the Walter Burke Institute for Theoretical Physics.

\appendix


\section{\label{app:matching}Matching the electroweak triplet scalar model}

Matching to dimension six of the real scalar triplet model has been considered in Refs.~\cite{Henning:2014wua,Corbett:2017ieo,deBlas:2017xtg}. We follow the notation of Ref.~\cite{Henning:2014wua} and extend the matching results to dimension eight. We can write the Lagrangian in Eq.~\eqref{eq:tripletLagr} in a different form, using vector notation,
\begin{align}
	\mathcal{L} = \frac{1}{2} \vec{\Phi}^{T}
	\left( P^{2} - m^{2}_{\Phi} - U \right) \vec{\Phi}
	+ \vec{\Phi} \cdot \vec{B}
	- \frac{1}{4} \lambda_{\Phi} (\vec{\Phi} \cdot \vec{\Phi})^{2} ,
\end{align}
where $U = 2 \eta (H^{\dagger} H)$, $\vec{B} = 2 \kappa H^{\dagger} \vec{\tau} H$,
and $P_{\mu} = i D_{\mu}$.
The normalization of the $\rm SU(2)_{L}$ matrices is $\tau^{a} = \sigma^{a}/2$, where
$\sigma^{a}$ are the Pauli matrices.
The equation of motion is
\begin{align}
	\vec{\Phi}_{c} = - \frac{1}{P^{2} - m^{2}_{\Phi} - U} \vec{B}
	+ \frac{1}{P^{2} - m^{2}_{\Phi} - U} \lambda_{\Phi} (\vec{\Phi}_{c} \cdot \vec{\Phi}_{c}) \vec{\Phi}_{c} .
\end{align}
Here we have the equation of motion of the theory before the Taylor expansion leading
to the matching to dimension 6 and 8.
Plugging this back into the Lagrangian, we find that
\begin{align}
	\mathcal{L}
	=& \frac{1}{2} \vec{\Phi}^{T}_{c}
	\left( P^{2} - m^{2}_{\Phi} - U \right) \vec{\Phi}_{c}
	+ \vec{\Phi}_{c} \cdot \vec{B}
	- \frac{1}{4} \lambda_{\Phi} (\vec{\Phi}_{c} \cdot \vec{\Phi}_{c})^{2}
	\nonumber \\
	=& \frac{1}{2 m^{2}_{\Phi} }
	\vec{B} \cdot \vec{B}
	+ \frac{1}{2}
	\vec{B}^{T}
	\frac{1}{ m^{2}_{\Phi} }
	(P^{2}  - U)
	\frac{1}{ m^{2}_{\Phi} } \vec{B}
	+ \frac{1}{2}
	\vec{B}^{T}
	\frac{1}{ m^{2}_{\Phi} }
	(P^{2}  - U)
	\frac{1}{ m^{2}_{\Phi} }
	(P^{2}  - U)
	\frac{1}{ m^{2}_{\Phi} } \vec{B}
	\nonumber \\
	 &- \frac{1}{4 m^{8}_{\Phi}} \lambda_{\Phi} (\vec{B} \cdot \vec{B})^{2}
	 + \mathcal{O}(\kappa^{2}/m^{8}_{\Phi}) .
\end{align}
We simplify the operators in the expansion.
The Lagrangian involving the Higgs field is
\begin{align}
	\mathcal{L}_{H} =& (D_{\mu} H^{\dagger} ) ( D^{\mu} H )
	- \lambda \left( H^{\dagger} H - \frac{1}{2} v^{2} \right)^{2}
	- H^{\dagger} \mathcal{Y} - \mathcal{Y}^{\dagger} H
	+ a_{0} \mathcal{O}^{(4)}_{H}
	+ a_{1} \mathcal{O}^{(6)}_{H}
	+ a_{2} \mathcal{O}^{(6)}_{H\Box}
	+ a_{3} \mathcal{O}^{(6)}_{HD}
	\nonumber \\ &
	+ b_{1} (H^{\dagger} H) (\Box H^{\dagger} H + H^{\dagger} \Box H)
	+ a_{4} \mathcal{O}^{(8)}_{H}
	+ a_{5} \mathcal{O}^{(8)}_{HD2}
	+ a_{6} \mathcal{O}^{(8)}_{4DH,1}
	+ a_{7} \mathcal{O}^{(8)}_{4HD,2}
	\nonumber \\ &
	+ b_{2} (H^{\dagger} H)^{2} (\Box H^{\dagger} H + H^{\dagger} \Box H)
	+ b_{3} (\Box H^{\dagger} H + H^{\dagger} \Box H)^{2}
	\nonumber \\ &
	+ b_{4} (D^{\mu} H^{\dagger} D_{\mu} H) (\Box H^{\dagger} H + H^{\dagger} \Box H)
	+ b_{5} (H^{\dagger} H) (\Box H^{\dagger} \Box H)
	+ b_{6} (H^{\dagger} \Box H) (\Box H^{\dagger} H)
	\nonumber \\ &
	+ b_{7} \left[ (D^{\mu} H^{\dagger} \Box H) (H^{\dagger} D_{\mu} H)
	+ (D^{\mu} H^{\dagger} H) (\Box H^{\dagger} D_{\mu} H) \right] .
\end{align}
where
\begin{align}
	a_{0} =& \frac{\kappa^{2}}{2m^{2}_{\Phi}} , \qquad
	a_{1} = - \frac{\eta \kappa^{2}}{m^{4}_{\Phi}} , \qquad
	a_{2} = \frac{\kappa^{2}}{2m^{4}_{\Phi}} , \qquad
	a_{3} = - \frac{2\kappa^{2}}{m^{4}_{\Phi}} , \nonumber \\
	a_{4} =& \left( \frac{2 \eta^{2} \kappa^{2}}{ m^{6}_{\Phi}}
	- \frac{ \kappa^{4}}{4 m^{8}_{\Phi}} \lambda_{\Phi} \right) , \qquad
	a_{5} = \frac{4 \eta \kappa^{2}}{m^{6}_{\Phi}} , \qquad
	a_{6} = - \frac{2 \kappa^{2}}{m^{6}_{\Phi}} , \qquad
	a_{7} = \frac{4 \kappa^{2}}{m^{6}_{\Phi}} , \nonumber \\
		b_{1} =& - \frac{\kappa^{2}}{m^{4}_{\Phi}} , \qquad
		b_{2} = \frac{2 \eta \kappa^{2}}{m^{6}_{\Phi}} , \qquad
		b_{3} = \frac{\kappa^{2}}{2m^{6}_{\Phi}} , \qquad
		b_{4} = - \frac{2 \kappa^{2}}{m^{6}_{\Phi}} , \nonumber \\
		b_{5} =& \frac{2 \kappa^{2}}{m^{6}_{\Phi}} , \qquad
		b_{6} = - \frac{2 \kappa^{2}}{m^{6}_{\Phi}} , \qquad
		b_{7} = \frac{4 \kappa^{2}}{m^{6}_{\Phi}} ,
\end{align}
and
\begin{align}
	\mathcal{O}^{(8)}_{4DH,1} =& (D^{\mu} H^{\dagger} D_{\mu} H) (D^{\nu} H^{\dagger} D_{\nu} H) ,\\
	\mathcal{O}^{(8)}_{4DH,2} =& (D^{\mu} H^{\dagger} D_{\nu} H) (D^{\nu} H^{\dagger} D_{\mu} H) .
\end{align}
Here we have written the Yukawa terms compactly as $-H^{\dagger}\mathcal{Y} + \rm{h.c.}$

We need to remove the higher-derivative operators to find a suitable form
of the Lagrangian. This we achieve by redefining the Higgs field as
\begin{align}
	H &\rightarrow H + b_{1} (H^{\dagger} H) H + \mathcal{O}^{(8)} ,
\end{align}
where
\begin{align}
	\mathcal{O}^{(8)} =&
	b_{3} (\Box H^{\dagger} H + H^{\dagger} \Box H) H
	+ b_{4} (D^{\mu} H^{\dagger} D_{\mu} H ) H
	+ \frac{1}{2} b_{5} (H^{\dagger} H ) \Box H
	+ \frac{1}{2} b_{6} (H^{\dagger} \Box H )  H
	\nonumber \\ &
	+ b_{7} (D^{\mu} H^{\dagger}  H) D_{\mu} H
	+ d_{1} v^{2} \lambda (H^{\dagger} H) H
	+ d_{2} (H^{\dagger} H)^{2} H
	- b_{3} (\mathcal{Y}^{\dagger} H + H^{\dagger} \mathcal{Y}) H
	\nonumber \\ &
	- \frac{1}{2} b_{5} (H^{\dagger} H) \mathcal{Y}
	- \frac{1}{2} b_{6} (H^{\dagger} \mathcal{Y}) H ,
\end{align}
and
\begin{align}
	d_{1} =& \left( 2 b_{3} - b_{4} + \frac{1}{2} (b_{5} + b_{6}) \right) ,
	\\
	d_{2} =& \left( b_{2} + 4 b^{2}_{1} + 4 a_{2} b_{1} - \frac{1}{2} a_{3} b_{1}+ 2 (a_{0} - \lambda) \left( 2 b_{3} + \frac{1}{2} b_{5} + \frac{1}{2} b_{6} \right)  \right) .
\end{align}
The final result for the matching (contributing to three-point couplings) is given in Table \ref{tab:matchingTriplet}.


\section{\label{app:moreplots} Additional EWPD, coefficient scan}

\begin{figure}
\includegraphics[height=0.24\textheight,width=0.45\textwidth]{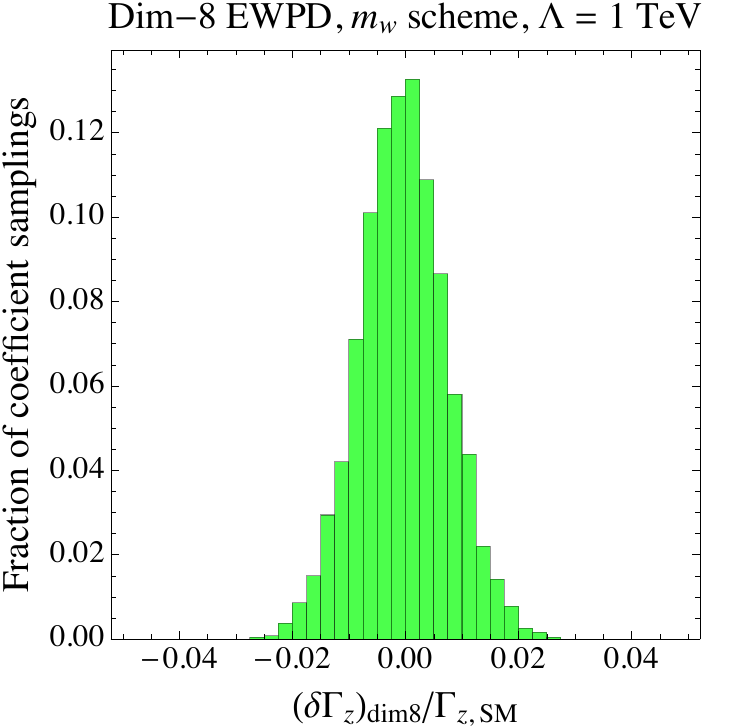}
\includegraphics[height=0.24\textheight,width=0.45\textwidth]{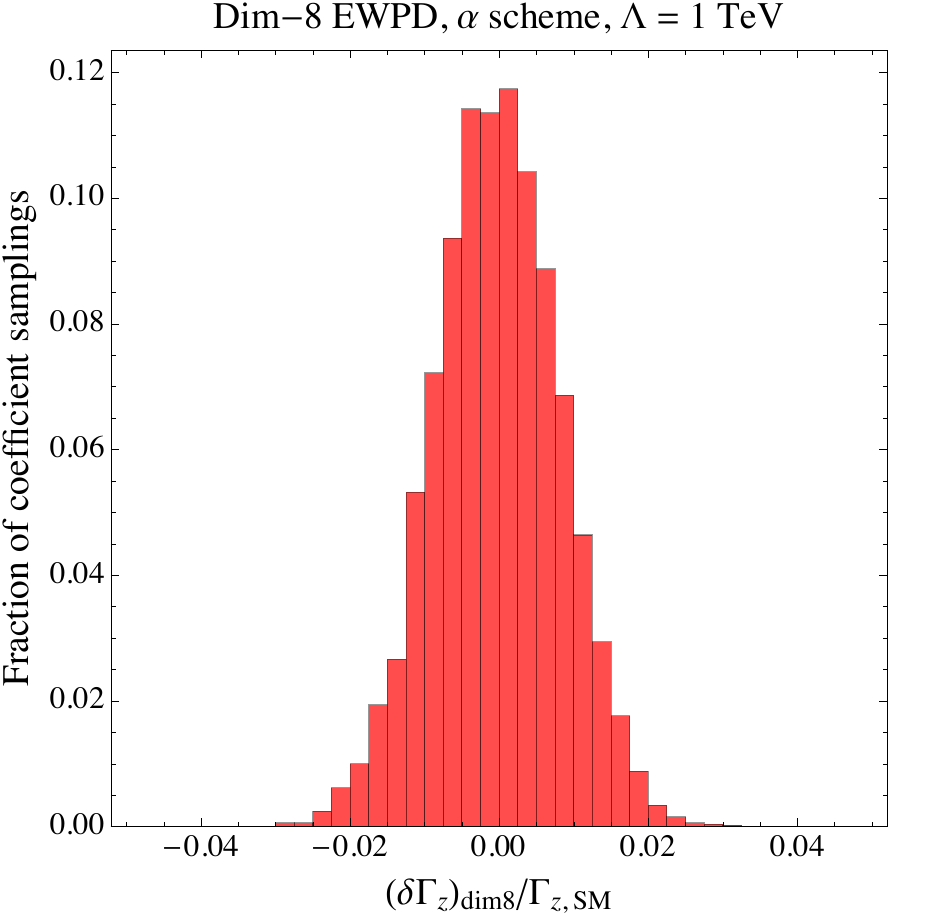}
\includegraphics[height=0.24\textheight,width=0.45\textwidth]{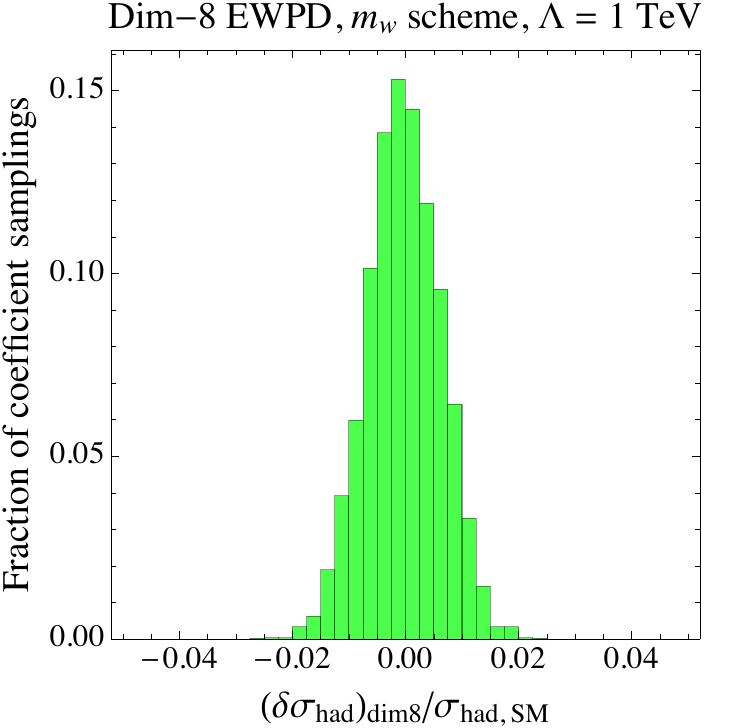}
\includegraphics[height=0.24\textheight,width=0.45\textwidth]{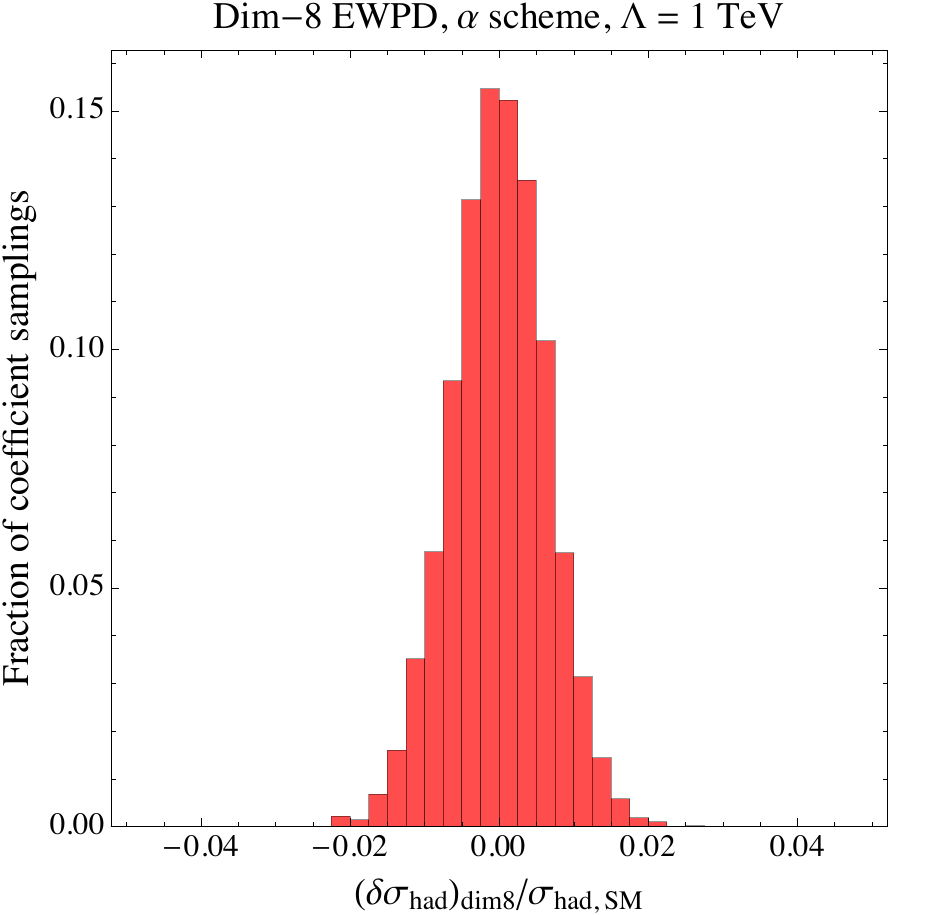}
\caption{Contributions to the $\Gamma_Z$ and  $\sigma_{had}^0$ from dimension-eight operators relative to the SM value. Here $\Lambda = 1\, $TeV. The histograms are formed from selecting random values for the coefficients 5000 times following the scheme described in the text. }
\label{fig:bottomupfigappB}
\end{figure}

For completeness, in figure \ref{fig:bottomupfigappB} we show the results of the bottom-up coefficient scan in Sec.~\ref{bottomup} for the remaining EWPD, $\Gamma_Z$ and $\sigma_{had}^0$.

\section{\label{loopvsdim8} $\mathcal O(v^4/\Lambda^4)$ EWPD, compared to loop corrections to $\mathcal{L}^{(6)}$}
A robust interpretation of EWPD at $\mathcal O(v^2/\Lambda^2)$ should define EWPD observables,
including loop correction to $\mathcal{L}^{(6)}$ effects, and also $\mathcal O(v^4/\Lambda^4)$ contributions.
When the size of sub-leading terms are known numerically, a conclusion drawn on leading order terms is more robust, and a theory error for the neglected higher order terms can be assigned. This is the case for both the bottom up (completely general) and top-down (UV model specific) SMEFT perspectives.

In this work, we have reported the first result
characterizing EWPD to $\mathcal O(v^4/\Lambda^4)$. These corrections modify observables via interference with the SM amplitude
for an EWPD observable $\mathcal{A}_{SM}^{o_1}$ as
\bea
\sim \mathcal{A}_{SM}^{o_1}  \frac{C^{(8)}_i \bar{v}_T^4}{\Lambda^4},
\eea
from the interference of a $\mathcal{L}^{(8)}$ correction with the SM, and via double insertions of $\mathcal{L}^{(6)}$ corrections in Feynman diagrams (or $\mathcal{L}^{(6)}$ cross terms) scaling as
\bea
\sim \frac{C_j^{(6)} \bar{v}_T^2}{\Lambda^2} \, \frac{C_k^{(6)} \bar{v}_T^2}{\Lambda^2}.
\eea
Conversely, loop corrections to EWPD scale as
\bea
\sim \mathcal{A}_{SM}^{o_1}  \frac{C^{(6)}_l \bar{v}_T^2}{16 \pi^2 \Lambda^2}.
\eea
Treating the SMEFT bottom up -- meaning we assume all operators are present and are agnostic about their respective Wilson coefficients -- the $\mathcal O(v^4/\Lambda^4)$ corrections dominate over loop corrections when
\bea
\mathcal{A}_{SM}^{o_1}  \frac{C^{(8)}_i \bar{v}_T^4}{\Lambda^4} \gtrsim  \mathcal{A}_{SM}^{o_1}  \frac{C^{(6)}_l \bar{v}_T^2}{16 \pi^2 \Lambda^2},
\eea
for corrections from $\mathcal{L}^{(8)}$ operators, or
\bea
\frac{C_j^{(6)} \bar{v}_T^2}{\Lambda^2} \, \frac{C_k^{(6)} \bar{v}_T^2}{\Lambda^2} \gtrsim \mathcal{A}_{SM}^{o_1}  \frac{C^{(6)}_l \bar{v}_T^2}{16 \pi^2 \Lambda^2}.
\eea
from double insertions of  $\mathcal{L}^{(6)}$ corrections. These conditions reduce to
\bea
\frac{C^{(8)}_i \bar{v}_T^2}{\Lambda^2} \gtrsim  \frac{C^{(6)}_l}{16 \pi^2} \quad \text{and}\quad  C_j^{(6)} \, C_k^{(6)}  \, \frac{\bar{v}_T^2}{\Lambda^2} \gtrsim \mathcal{A}_{SM}^{o_1}  \frac{C^{(6)}_l}{16 \pi^2}.
\label{eq:generalcond}
\eea
If one assumes the $C^{(6,8)}_{i,j,k,l}$ are similar numerically, and a factor of $\mathcal{A}_{SM}^{o_1}$
is not a significant numerical enhancement or suppression, these conditions can be reduced to $\sim 4 \pi \bar{v}_T \gtrsim \Lambda$
i.e. $3 \,  {\rm TeV} \gtrsim \Lambda$. For $\Lambda$ in the few $\rm TeV$ range, it is reasonable to expect that
dimension eight corrections will dominate over loop corrections.

In the case of the specific models, the inferred conditions scale with the model parameters. For the $U(1)$ kinetic mixing model we have $C_i^{(6)}\sim k^2 g_{SM}^2/m_K^2$ and $C_i^{(8)}\sim k^4 g_{SM}^4/m_K^4$, or $C_i^{(8)}\sim k^2 g_{SM}^4/m_K^4$.
Substituting these results into the reduced conditions for the $\mathcal O(v^4/\Lambda^4)$ corrections to dominate over the loop corrections to $\mathcal{L}^{(6)}$, Eq.~\eqref{eq:generalcond}, we find
\bea
\frac{k^2 g_{SM}^4 \bar{v}_T^2}{m_K^2} \gtrsim  \frac{g_{SM}^2}{16 \pi^2},
\eea
and
\bea
k^2 \, g_{SM}^4 \, \frac{\bar{v}_T^2}{m_K^2} \gtrsim \mathcal{A}_{SM}^{o_1}  \frac{g_{SM}^2}{16 \pi^2}.
\eea

For the scalar triplet model we have $C_i^{(6)}\sim \kappa^2/m_\phi^4$
and $C_i^{(8)}\sim \kappa^2 g_{SM}/m_\phi^6$, $C_i^{(8)}\sim \kappa^4 g_{SM}/m_\phi^8$
and $C_i^{(8)}\sim \kappa^6/m_\phi^{10}$.
Plugged into Eq.~\eqref{eq:generalcond}, these become
\bea
\frac{g_{SM} \bar{v}_T^2}{m_\phi^2} \gtrsim  \frac{1}{16 \pi^2}, \quad
\frac{g_{SM} \kappa^2 \bar{v}_T^2}{m_\phi^4} \gtrsim  \frac{1}{16 \pi^2}, \quad
\frac{g_{SM} \kappa^4 \bar{v}_T^2}{m_\phi^6} \gtrsim  \frac{1}{16 \pi^2},
\eea
and
\bea
\kappa^2  \, \frac{\bar{v}_T^2}{m_\phi^4} \gtrsim \mathcal{A}_{SM}^{o_1}  \frac{1}{16 \pi^2}.
\eea
Here we have neglected logs that can appear in the perturbative corrections, as we are restricted
to $\Lambda$ of a few {\rm TeV}.

While there is a regime where $\mathcal O(v^4/\Lambda^4)$ dominates over loop corrections, the scope of the parameter space depends on whether one takes the bottom-up or top-down approach.   An additional ingredient to consider when exploring different SMEFT perspectives is the number of operators present. 
The $U(1)$ model matches to operators which contribute to the Z-pole data; $3+5n_{f}$ at dimension six and $2+9n_{f}$ at dimension eight 
(treating $\delta G_F^{(8)}$ as one parameter for the purposes of this counting).
In contrast, only one operator at dimension six and two at dimension eight coming from matching in the triplet model contribute to Z-pole data.
In the bottom-up approach, a total of $6+7n_{f}$ dimension-six operators as well as $5+9n_{f}$ dimension-eight operators contribute to Z-pole data. 
Contrasting the number of operators present for the top-down models with the bottom-up approach, we see the importance of a global analysis using the bottom up approach.

%




\bibliographystyle{JHEP}
\bibliography{bibliography.bib}

\end{document}